\documentclass[twoside,twocolumn,9pt]{article}
\usepackage[utf8]{inputenc}
\usepackage[T1]{fontenc}
\usepackage[raggedright]{sidecap}

\usepackage{graphicx,caption,amsmath,amssymb,amsthm,amsfonts,epsfig,siunitx,chemformula,booktabs}
\usepackage[super,sort&compress,comma]{natbib}

\usepackage[left=1.5cm, right=1.5cm, top=1.785cm, bottom=2.0cm]{geometry}
\usepackage{fancyhdr}
\usepackage{fnpos}
\usepackage{setspace}
\usepackage[compact]{titlesec}

\usepackage[frozencache=true,cachedir=minted-cache]{minted}

\usepackage{times,pifont,colortbl,hyperref,mathptmx,authblk,mdframed,etoolbox,lineno,inconsolata,sectsty,doi}
\usepackage[labelfont={sf},labelsep=period,justification=raggedright]{caption}
\usepackage{xcolor}
\usepackage{orcidlink}
\allsectionsfont{\bfseries\sffamily}
\definecolor{boxgray}{gray}{0.95}
\definecolor{RoyalBlue}{RGB}{65,105,225}
\definecolor{Violet}{RGB}{61, 60, 159}
\definecolor{ForestGreen}{RGB}{18, 159, 87}
\definecolor{cream}{RGB}{222,217,201}

\newcommand{\optimadeprop}[1]{\colorbox{boxgray}{\texttt{\color{RoyalBlue}{#1}}}}
\newcommand{\optimadeendp}[1]{\colorbox{boxgray}{\texttt{\color{darkorchid}{#1}}}}
\newcommand{\optimadeprefix}[1]{\colorbox{boxgray}{\texttt{\color{RoyalBlue}{\_#1\_}}}}
\newcommand{\optimadefilter}[1]{\colorbox{boxgray}{\texttt{\color{RoyalBlue}{#1}}}}

\DeclareCaptionType{Listing}[Box]

\newcommand{\database}{\noindent\emph{Database}:~}
\newcommand{\implementation}{\noindent\emph{OPTIMADE implementation}:~}
\newcommand{\baseurl}[1]{\noindent\emph{Base URL}:\\\href{#1}{\colorbox{boxgray}
{\texttt{{\color{dogwoodrose}#1}}}}}
\newcommand{\indexbaseurl}[1]{\noindent\emph{Index base URL}:\\\href{#1}{\colorbox{boxgray}
{\texttt{{\color{dogwoodrose}#1}}}}}

\hypersetup{breaklinks=true,colorlinks=true,citecolor=ForestGreen,filecolor=ForestGreen,linkcolor=ForestGreen,urlcolor=ForestGreen}
\usepackage{minted, mdframed, listings}

\setminted[json]{
    numbers=left,
    breaklines,
    style=emacs,
    breakbefore=\&/?=,
    fontsize=\footnotesize, 
    linenos
}

\mdfdefinestyle{mintframe}{
    innertopmargin=2pt,
    innerbottommargin=2pt,
    innerleftmargin=2pt,
    innerrightmargin=2pt,
    backgroundcolor=gray!5,
    skipbelow=0,
    skipabove=0,
    linecolor=gray!50,
}

\definecolor{eclipseStrings}{RGB}{42,0.0,255}
\definecolor{eclipseKeywords}{RGB}{127,0,85}
\definecolor{darkpastelblue}{rgb}{0.47, 0.62, 0.8}
\definecolor{darkpastelgreen}{rgb}{0.01, 0.75, 0.24}
\definecolor{darkpastelpurple}{rgb}{0.59, 0.44, 0.84}
\definecolor{darkorchid}{rgb}{0.6, 0.2, 0.8}
\definecolor{darkpink}{rgb}{0.91, 0.33, 0.5}
\definecolor{deepskyblue}{rgb}{0.0, 0.75, 1.0}
\definecolor{dodgerblue}{rgb}{0.12, 0.56, 1.0}
\definecolor{dogwoodrose}{rgb}{0.84, 0.09, 0.41}
\definecolor{dollarbill}{rgb}{0.52, 0.73, 0.4}
\colorlet{numb}{darkpastelgreen}
\colorlet{keywords}{dodgerblue}
\colorlet{strings}{dogwoodrose}
\colorlet{punct}{black}
\lstdefinelanguage{json}{
    basicstyle=\footnotesize\ttfamily,
    commentstyle=\color{strings},
    stringstyle=\color{keywords},
    frame=none,
    numberstyle=\ttfamily,
    stepnumber=1,
    numbersep=8pt,
    upquote=true,
    showstringspaces=false,
    breaklines=true,
    backgroundcolor=gray!5,
    string=[b]",
    morestring=[b]",
    literate=
        *{0}{{{\color{numb}0}}}{1}
         {1}{{{\color{numb}1}}}{1}
         {2}{{{\color{numb}2}}}{1}
         {3}{{{\color{numb}3}}}{1}
         {4}{{{\color{numb}4}}}{1}
         {5}{{{\color{numb}5}}}{1}
         {6}{{{\color{numb}6}}}{1}
         {7}{{{\color{numb}7}}}{1}
         {8}{{{\color{numb}8}}}{1}
         {9}{{{\color{numb}9}}}{1}
         {.}{{{\color{numb}.}}}{1}
         {:}{{{\color{punct}{:}}}}{1}
         {,}{{{\color{punct}{,}}}}{1}
}
\lstdefinelanguage{html}{
    basicstyle=\footnotesize\ttfamily,
    stringstyle=\color{strings},
    commentstyle=\color{keywords},
    frame=none,
    stepnumber=1,
    numbersep=8pt,
    showstringspaces=false,
    breaklines=true,
    string=[s]{"}{"},
    comment=[l]{:\ "},
    morecomment=[l]{:"}
}

\begin{document}

\makeFNbottom
\makeatletter
\renewcommand\LARGE{\@setfontsize\LARGE{15pt}{17}}
\renewcommand\Large{\@setfontsize\Large{12pt}{14}}
\renewcommand\large{\@setfontsize\large{10pt}{12}}
\renewcommand\footnotesize{\@setfontsize\footnotesize{7pt}{10}}
\makeatother

\renewcommand{\thefootnote}{\fnsymbol{footnote}}
\renewcommand\footnoterule{\vspace*{1pt}%
\color{cream}\hrule width 3.5in height 0.4pt \color{black}\vspace*{5pt}} 
\setcounter{secnumdepth}{5}

\makeatletter 
\renewcommand\@biblabel[1]{#1}            
\renewcommand\@makefntext[1]%
{\noindent\makebox[0pt][r]{\@thefnmark\,}#1}
\makeatother 
\renewcommand{\figurename}{\small{Fig.}~}
\sectionfont{\sffamily\Large}
\subsectionfont{\normalsize}
\subsubsectionfont{\bf}
\setstretch{1.125} 
\setlength{\skip\footins}{0.8cm}
\setlength{\footnotesep}{0.25cm}
\setlength{\jot}{10pt}
\titlespacing*{\section}{0pt}{4pt}{4pt}
\titlespacing*{\subsection}{0pt}{15pt}{1pt}
\fancyfoot{}
\fancyfoot[LO,RE]{\vspace{-7.1pt}\includegraphics[height=9pt]{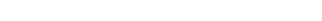}}
\fancyfoot[CO]{\vspace{-7.1pt}\hspace{13.2cm}\includegraphics{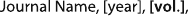}}
\fancyfoot[CE]{\vspace{-7.2pt}\hspace{-14.2cm}\includegraphics{head_foot/RF}}
\fancyfoot[RO]{\footnotesize{\sffamily{1--\pageref{LastPage} ~\textbar  \hspace{2pt}\thepage}}}
\fancyfoot[LE]{\footnotesize{\sffamily{\thepage~\textbar\hspace{3.45cm} 1--\pageref{LastPage}}}}
\fancyhead{}
\renewcommand{\headrulewidth}{0pt} 
\renewcommand{\footrulewidth}{0pt}
\setlength{\arrayrulewidth}{1pt}
\setlength{\columnsep}{6.5mm}
\setlength\bibsep{1pt}

\makeatletter 
\newlength{\figrulesep} 
\setlength{\figrulesep}{0.5\textfloatsep} 

\newcommand{\topfigrule}{\vspace*{-1pt}%
\noindent{\color{cream}\rule[-\figrulesep]{\columnwidth}{1.5pt}} }

\newcommand{\botfigrule}{\vspace*{-2pt}%
\noindent{\color{cream}\rule[\figrulesep]{\columnwidth}{1.5pt}} }

\newcommand{\dblfigrule}{\vspace*{-1pt}%
\noindent{\color{cream}\rule[-\figrulesep]{\textwidth}{1.5pt}} }

\makeatother

\twocolumn[
  \begin{@twocolumnfalse}
{\includegraphics[height=30pt]{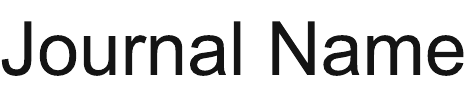}\hfill\raisebox{0pt}[0pt][0pt]{\includegraphics[height=55pt]{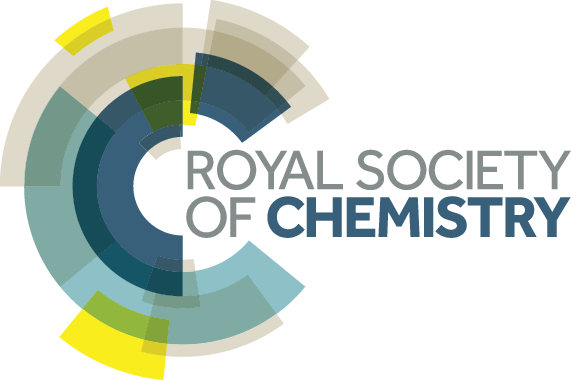}}\\[1ex]
\includegraphics[width=18.5cm]{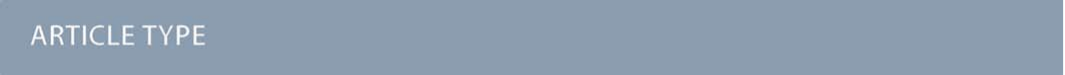}}\par
\vspace{1em}
\sffamily
\begin{tabular}{m{4.5cm} p{13.5cm} }

\includegraphics{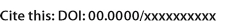} & \noindent\LARGE{\textbf{Developments and applications of the OPTIMADE API for materials discovery, design, and data exchange}} \\
\vspace{0.3cm} & \vspace{0.3cm} \\

 & \noindent\large{%
Matthew~L.~Evans,$^{1,2}$\,\orcidlink{0000-0002-1182-9098}
Johan~Bergsma,$^{3,\dag}$\,\orcidlink{0000-0003-3072-3798}
Andrius~Merkys,$^{4}$\,\orcidlink{0000-0002-7731-6236}
Casper~W.~Andersen,$^{5}$\,\orcidlink{0000-0002-2547-155X}
Oskar~B.~Andersson,$^{6}$\,\orcidlink{0009-0003-6158-1857}
Daniel~Beltrán,$^{7}$
Evgeny~Blokhin,$^{8,9}$\,\orcidlink{0000-0002-5333-3947}
Tara~M.~Boland,$^{10}$\,\orcidlink{0000-0002-2587-5677}
Rubén~Castañeda~Balderas,$^{11}$\,\orcidlink{0000-0002-0040-6321}
Kamal~Choudhary,$^{12}$\,\orcidlink{0000-0001-9737-8074}
Alberto~Díaz~Díaz,$^{11}$\,\orcidlink{0000-0002-4873-7947}
Rodrigo~Domínguez~García,$^{11}$\,\orcidlink{0000-0002-6805-705X}
Hagen~Eckert,$^{13,14}$\,\orcidlink{0000-0003-4771-1435}
Kristjan~Eimre,$^{15}$\,\orcidlink{0000-0002-3444-3286}
María~Elena~Fuentes~Montero,$^{16}$\,\orcidlink{0000-0002-3388-8224}
Adam~M.~Krajewski,$^{17}$\,\orcidlink{0000-0002-2266-0099}
Jens~Jørgen~Mortensen,$^{10}$\,\orcidlink{0000-0001-5090-6706}
José~Manuel~Nápoles~Duarte,$^{16}$\,\orcidlink{0000-0001-6823-4733}
Jacob~Pietryga,$^{18}$\,\orcidlink{0009-0009-6649-3665}
Ji~Qi,$^{19}$
Felipe~de~Jesús~Trejo~Carrillo,$^{11}$\,\orcidlink{0009-0005-1973-4615}
Antanas~Vaitkus,$^{4}$\,\orcidlink{0000-0002-5944-1391}
Jusong~Yu,$^{15}$
Adam~Zettel,$^{13,14}$\,\orcidlink{0000-0003-1645-9476}
Pedro~Baptista~de~Castro,$^{20}$\,\orcidlink{0000-0001-8673-2840}
Johan~Carlsson,$^{21}$\,\orcidlink{0000-0002-1596-0923}
Tiago~F.~T.~Cerqueira,$^{22}$\,\orcidlink{0000-0002-4147-8129}
Simon~Divilov,$^{13,14}$\,\orcidlink{0000-0002-4185-6150}
Hamidreza~Hajiyani,$^{21,\ddag}$
Felix~Hanke,$^{23}$\,\orcidlink{0000-0002-4155-5980}
Kevin~Jose,$^{24}$
Corey~Oses,$^{25}$\,\orcidlink{0000-0002-3790-1377}
Janosh~Riebesell,$^{24,26}$\,\orcidlink{0000-0001-5233-3462}
Jonathan~Schmidt,$^{27}$\,\orcidlink{0000-0001-5685-6404}
Donald~Winston,$^{28}$\,\orcidlink{0000-0002-8424-0604}
Christen~Xie,$^{19}$
Xiaoyu~Yang,$^{29,30,31}$
Sara~Bonella,$^{3}$\,\orcidlink{0000-0003-4131-2513}
Silvana~Botti,$^{32}$\,\orcidlink{0000-0002-4920-2370}
Stefano~Curtarolo,$^{13,14}$\,\orcidlink{0000-0003-0570-8238}
Claudia~Draxl,$^{33}$\,\orcidlink{0000-0003-3523-6657}
Luis~Edmundo~Fuentes~Cobas,$^{11}$\,\orcidlink{0000-0001-6063-3967}
Adam~Hospital,$^{7}$\,\orcidlink{0000-0002-8291-8071}
Zi-Kui~Liu,$^{17}$\,\orcidlink{0000-0003-3346-3696}
Miguel~A.~L.~Marques,$^{32}$\,\orcidlink{0000-0003-0170-8222}
Nicola~Marzari,$^{15,34}$\,\orcidlink{0000-0002-9764-0199}
Andrew~J.~Morris,$^{35}$\,\orcidlink{0000-0001-7453-5698}
Shyue~Ping~Ong,$^{19}$\,\orcidlink{0000-0001-5726-2587}
Modesto~Orozco,$^{7}$\,\orcidlink{0000-0002-8608-3278}
Kristin~A.~Persson,$^{26,36}$\,\orcidlink{0000-0003-2495-5509}
Kristian~S.~Thygesen,$^{10}$\,\orcidlink{0000-0001-5197-214X}
Chris~Wolverton,$^{18}$\,\orcidlink{0000-0003-2248-474X}
Markus~Scheidgen,$^{33}$\,\orcidlink{0000-0002-8038-2277}
Cormac~Toher,$^{37,14}$\,\orcidlink{0000-0001-7073-8690}
Gareth~J.~Conduit,$^{24}$\,\orcidlink{0000-0003-3807-6361}
Giovanni~Pizzi,$^{15,34}$\,\orcidlink{0000-0002-3583-4377}
Saulius~Gra\v{z}ulis,$^{4,38}$\,\orcidlink{0000-0002-7928-5218}
Gian-Marco~Rignanese,$^{1,2,39}$\,\orcidlink{0000-0002-1422-1205}
Rickard~Armiento$^{6}$\,\orcidlink{0000-0002-5571-0814}} \\

\includegraphics{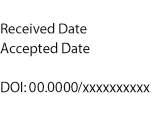} & \noindent\normalsize{The Open Databases Integration for Materials Design (OPTIMADE) application programming interface (API) empowers users with holistic access to a growing federation of databases, enhancing the accessibility and discoverability of materials and chemical data.
Since the first release of the OPTIMADE specification (v1.0), the API has undergone significant development, leading to the v1.2 release, and has underpinned multiple scientific studies.
In this work, we highlight the latest features of the API format, accompanying software tools, and provide an update on the implementation of OPTIMADE in contributing materials databases. We end by providing several use cases that demonstrate the utility of the OPTIMADE API in materials research that continue to drive its ongoing development.} \\

\end{tabular}

\end{@twocolumnfalse} \vspace{0.6cm}
]


\renewcommand*\rmdefault{bch}\normalfont\upshape
\rmfamily
\section*{}
\vspace{-1cm}

\footnotetext{\textit{$^{1}$~UCLouvain, Institut de la Mati\`ere Condens\'ee et des Nanosciences (IMCN), Chemin des \'Etoiles~8, Louvain-la-Neuve 1348, Belgium}}
\footnotetext{\textit{$^{2}$~Matgenix SRL, 185 Rue Armand Bury, 6534 Gozée, Belgium}}
\footnotetext{\textit{$^{3}$~Centre Européen de Calcul Atomique et Moléculaire (CECAM), \'Ecole Polytechnique F\'ed\'erale de Lausanne, Avenue de Forel 3, 1015 Lausanne, Switzerland}}
\footnotetext{\textit{$^{4}$~Institute of Biotechnology, Life Sciences Center, Vilnius University, Saul\.{e}tekio av.\ 7, LT-10257 Vilnius, Lithuania}}
\footnotetext{\textit{$^{5}$~SINTEF, P.O. Box 4760 Torgarden, NO-7465 Trondheim, Norway}}
\footnotetext{\textit{$^{6}$~Materials Design and Informatics unit, Department of Physics, Chemistry and Biology, Link\"oping University, Sweden}}
\footnotetext{\textit{$^{7}$~Institute for Research in Biomedicine (IRB Barcelona), Baldiri i Reixac 10-12, 08028 Barcelona, Spain}}
\footnotetext{\textit{$^{8}$~Tilde Materials Informatics, Stra{\ss}mannstra{\ss}e 25, 10249, Berlin, Germany}}
\footnotetext{\textit{$^{9}$~Materials Platform for Data Science, Sepapaja 6, 15551, Tallinn, Estonia}}
\footnotetext{\textit{$^{10}$~Computational Atomic-Scale Materials Design, Technical University of Denmark, Kgs. Lyngby, Denmark}}
\footnotetext{\textit{$^{11}$~Centro de Investigación en Materiales Avanzados, S.C. (CIMAV), Av. Miguel de Cervantes 120, Complejo Industrial Chihuahua, 31136, Chihuahua, Chih., México}}
\footnotetext{\textit{$^{12}$~Material Measurement Laboratory, National Institute of Standards and Technology, Gaithersburg, MD, 20899, USA}}
\footnotetext{\textit{$^{13}$~Department of Mechanical Engineering and Materials Science, Duke University, Durham, NC 27708, USA}}
\footnotetext{\textit{$^{14}$~Center for Extreme Materials, Duke University, Durham, NC 27708, USA}}
\footnotetext{\textit{$^{15}$~Theory and Simulation of Materials (THEOS), and National Centre for Computational Design and Discovery of Novel Materials (MARVEL), \'{E}cole Polytechnique F\'{e}d\'{e}rale de Lausanne, 1015 Lausanne, Switzerland}}
\footnotetext{\textit{$^{16}$~Universidad Autónoma de Chihuahua, Facultad de Ciencias Quimicas, 31125 Chihuahua, México}}
\footnotetext{\textit{$^{17}$~Department of Materials Science and Engineering, The Pennsylvania State University, University Park, PA 16802, USA}}
\footnotetext{\textit{$^{18}$~Department of Materials Science and Engineering, Northwestern University, Evanston, IL 60208, USA}}
\footnotetext{\textit{$^{19}$~Department of NanoEngineering, University of California, San Diego, 9500 Gilman Drive, La Jolla, California, 92093-0448, United States}}
\footnotetext{\textit{$^{20}$~National Institute for Materials Science, 1-2-1 Sengen, Tsukuba, Ibaraki, 305-0047 Japan}}
\footnotetext{\textit{$^{21}$~Dassault Systèmes Germany GmbH, Am Kabellager 11-13, 51063 Cologne, Germany}}
\footnotetext{\textit{$^{22}$~CFisUC, Department of Physics, University of Coimbra, Rua Larga, 3004-516 Coimbra, Portugal}}
\footnotetext{\textit{$^{23}$~Dassault Systèmes, 22 Science Park, CB4 0FJ, United Kingdom.}}
\footnotetext{\textit{$^{24}$~Theory of Condensed Matter, Cavendish Laboratory, Cambridge, United Kingdom}}
\footnotetext{\textit{$^{25}$~Department of Materials Science and Engineering, Johns Hopkins University, Baltimore, MD 21218, USA}}
\footnotetext{\textit{$^{26}$~Lawrence Berkeley National Lab, Berkeley, CA, USA}}
\footnotetext{\textit{$^{27}$~Materials Theory, ETH Z\"{u}rich, Wolfgang-Pauli-Strasse 27, 8093 Zurich, Switzerland}}
\footnotetext{\textit{$^{28}$~Polyneme LLC, New York, NY 10038, USA}}
\footnotetext{\textit{$^{29}$~Computer Network Information Center,  Chinese Academy of Sciences, Beijing, 100083, China}}
\footnotetext{\textit{$^{30}$~University of Chinese Academy of Sciences,  Beijing, 101408, China}}
\footnotetext{\textit{$^{31}$~Beijing MaiGao MatCloud Technology Co. Ltd., Beijing, 100149, China}}
\footnotetext{\textit{$^{32}$~Research Center Future Energy Materials and Systems of the University Alliance Ruhr and Interdisciplinary Centre for Advanced Materials Simulation, Ruhr University Bochum, Universitätsstraße 150, D-44801 Bochum, Germany}}
\footnotetext{\textit{$^{33}$~Humboldt-Universit\"at zu Berlin, Institut f\"ur Physik and IRIS Adlershof, 12489 Berlin, Germany}}
\footnotetext{\textit{$^{34}$~Laboratory for Materials Simulations (LMS), Paul Scherrer Institute (PSI), 5232 Villigen PSI, Switzerland}}
\footnotetext{\textit{$^{35}$~School of Metallurgy and Materials, University of Birmingham, Edgbaston, Birmingham B15 2TT, United Kingdom}}
\footnotetext{\textit{$^{36}$~Department of Materials Science and Engineering, UC Berkeley, Hearst Mining Memorial Building, Berkeley, 94720 CA, USA}}
\footnotetext{\textit{$^{37}$~Department of Materials Science and Engineering and Department of Chemistry and Biochemistry, The University of Texas at Dallas, Richardson, TX 75080, USA}}
\footnotetext{\textit{$^{38}$~Institute of Computer Science, Faculty of Mathematics and Informatics, Vilnius University, Naugarduko g. 24, LT-03225 Vilnius, Lithuania}}
\footnotetext{\textit{$^{39}$~School of Materials Science and Engineering, Northwestern Polytechnical University, Xi’an, Shaanxi, 710072, China}}
\footnotetext{\dag~Present address: Van ’t Hoff Institute for Molecular Sciences, University of Amsterdam, PO Box 94157, 1090 GD Amsterdam, Netherlands}
\footnotetext{\ddag~Present address: Henkel, AID/Digital Twins \& Data Analytics, 40589 D\"usseldorf, Germany}




\newpage

\section{Introduction}

Industrial chemicals and materials underpin the global economy: for example, chemicals alone contribute \$6.4trillion~\cite{ChemicalsReport2022} annually to the global economy. Industrial chemicals and materials companies are under significant pressure to improve the environmental, social, and governance impact of their business~\cite{Baratta2023}, with a particular focus on reducing the carbon footprint. 
Unfortunately, the discovery of chemicals and drugs is traditionally a time-consuming and expensive process driven by experiment-led trial-and-improvement, delaying the response to the climate crisis. 
However, in the last few years, high-throughput calculations have led to an explosion in the volume of available materials data~\cite{Himanen2019, Ghiringhelli2023, Liu2014}. Machine learning has emerged as a pivotal tool to exploit this data~\cite{Suh2020}, accelerating the discovery of chemicals and drugs that meet the challenges faced today.

A core requirement for the data and machine-learning revolution is data availability and interoperability.
Therefore, the Open Databases Integration for Materials Design (OPTIMADE) universal application programming interface (API) was created to empower users with programmatic access to many leading materials databases.
By organising under an open federation, and emphasising the interoperability of search as well as access, OPTIMADE improves the discoverability of materials data, especially from smaller, less known databases.
As we move into the era of autonomous laboratories (both computational and experimental), the technical approach taken renders all OPTIMADE APIs machine actionable, allowing for automated serendipitous discovery of newly added data entries in a given materials space without needing to specify which databases to access.
Such an extended data availability requires explicit clarification of data permissions and ownership, which can be different for each database.

Since the first release of the OPTIMADE specification (v1.0)~\cite{OPTIMADE_spec_2020} with accompanying article~\cite{OPTIMADE2021}, the OPTIMADE API format has enjoyed significant adoption, with 22 registered providers~\cite{OPTIMADE_providers_dashboard, OPTIMADE_providers_list} of 25 interoperable databases serving over 22 million crystal structures with associated properties.
In the v1.2 release~\cite{OPTIMADE_spec_1.2}, the specification has undergone significant extensions and enhancements that enable novel use cases whilst making the format accessible to both users and developers. This gives users access to data from both large and well-known sources, and many specialist datasets focused on a family of materials of particular interest. The combination of a general overview of all possible materials and detailed knowledge of particular materials enables novel discovery and deep insights with for example machine learning.

In this paper, we highlight recent developments to and uses of the OPTIMADE API.
First, in Section~\ref{sec:OPTIMADEintro}, we provide an overview of the OPTIMADE API format, and of the latest features.
Next, in Section~\ref{sec:UpdateOnDatabases}, we highlight the efforts of leading materials databases to provide access through the OPTIMADE API format.
In Section~\ref{sec:ApplicationOfOPTIMADE}, we show how the OPTIMADE API has been used in computational screening and for the creation of machine learning datasets.
Section~\ref{sec:FutureOfOPTIMADE} discusses future plans for OPTIMADE, both in terms of new technical frontiers, and of the sustainability of the ecosystem and community.
Finally, in Section~\ref{sec:Conclusion}, we summarise and look ahead to the future of materials databases.

\section{Overview of the OPTIMADE API}\label{sec:OPTIMADEintro}

The OPTIMADE API is well-positioned to set the standard to enable search, retrieval and annotation in a common way for all databases.
Crystal structure data has benefited from decades of standardisation work in the form of the Crystallographic Information File (CIF)~\cite{Hall_IUCr_CIF_1991,cif_2016} and related initiatives, which heavily inspired the crystal structure representation employed by the OPTIMADE API format.
By building a standardised, open format, both proprietary and open data can be aggregated and used on the same footing.
 
Building on these seminal standardisation efforts, OPTIMADE goes considerably further than standardising the representation of crystal structure data, by including: the means for filtering entries (via the OPTIMADE filter grammar), a standard for laying out resources on the web (by providing rules and expectations of URL formats), a means for introspectively defining additional properties and entry types per-database, and the creation of a decentralised federation of compatible databases.
These additional aspects are what maximises the impact of the OPTIMADE API and enable new scientific applications.
The OPTIMADE API is registered with FAIRsharing.org as a data standard~\cite{fairsharing}, and releases are archived on Zenodo~\cite{OPTIMADE_spec_2020}, with ongoing development occurring openly under the \href{https://github.com/Materials-Consortia}{Materials-Consortia} banner on GitHub~\cite{OPTIMADE_github}. The initial motivation for OPTIMADE, and a discussion of the previously existing materials API formats and filter mechanisms can be found in Ref.~\cite{OPTIMADE2021} which described the first release.

The process for a user to access OPTIMADE compliant data is as follows.
The user starts with the base URL defined by the database provider (available from the federated OPTIMADE provider list~\cite{OPTIMADE_providers_dashboard, OPTIMADE_providers_list}), and then appends a common string describing the entry type to query, plus any filter or implementation parameters, which is submitted as an HTTP GET request.
For example, to probe the Materials Project~\cite{Materials_Project} for materials containing SiO$_\text{2}$, we make a GET request to the following URL:
\begin{center}
\vspace*{-1mm}
\begin{minipage}{\textwidth}
\colorbox{boxgray}{
$
\overbrace{
\texttt{\color{dogwoodrose}https://optimade.materialsproject.org}
}^{\textbf{\color{dogwoodrose}base URL}}
\overbrace{
\texttt{\color{darkorchid}/v1/structures}}^{\textbf{\color{darkorchid}endpoint}}$
}

\colorbox{boxgray}{
${\texttt{?filter=}}
\underbrace{
\texttt{\color{RoyalBlue}chemical\_formula\_reduced="O2Si"}}_{\textbf{\color{RoyalBlue}OPTIMADE filter}}
$
}
\end{minipage}
\end{center}
This delivers the response in Box~\ref{listing:O2Si} that contains entries where oxygen and silicon occur in a 2:1 ratio.
The power of the OPTIMADE API is that the same generic request can be appended to the base URL of any other database, and its matching entries will be returned.
This is a non-trivial step for database providers, who must convert the OPTIMADE filter grammar into the corresponding query for their own database engine.
The benefit is that client code can then be written to unify the results from multiple databases, allowing users to receive the most comprehensive results for the query. Such an extended data availability requires explicit clarification of data permissions and ownership, which can be different for each database or even each entry.

\begin{Listing}[ht!]
\begin{center}
\begin{mdframed}[style=mintframe]
\begin{minted}{json}
{
 "data": [
  {
   "id": "mp-7000",
   "type": "structures",
   "attributes": {
    "immutable_id": "645d2ba4bcd30f748b475981",
    "last_modified": "2023-03-11T14:56:30Z",
    "elements": ["O", "Si"],
    "nelements": 2,
    "elements_ratios": [0.3333333333333333, 0.6666666666666666],
    "chemical_formula_descriptive": "O6Si3",
    "chemical_formula_reduced": "O2Si",
    "chemical_formula_hill": "O6Si3",
    "chemical_formula_anonymous": "A2B",
    "dimension_types": [1, 1, 1],
    "nperiodic_dimensions": 3,
    "lattice_vectors": [
      [4.914966, -1e-8, 0],
      [-2.45748252, 4.2564861, 0],
      [0, 0, 5.43130114]
    ],
    "nsites": 9,
    "species_at_sites": ["Si", "Si", "Si", "O", "O", "O", "O", "O", "O"],
\end{minted}
\end{mdframed}
\end{center}
 \caption{An excerpt of the JSON response showing the material attributes for one of the returned entries.\label{listing:O2Si}}
\end{Listing}

\subsection{OPTIMADE core design principles}\label{sec:CorePrinciples}

\begin{figure}
 \begin{center}
 \includegraphics[width=0.8\linewidth]{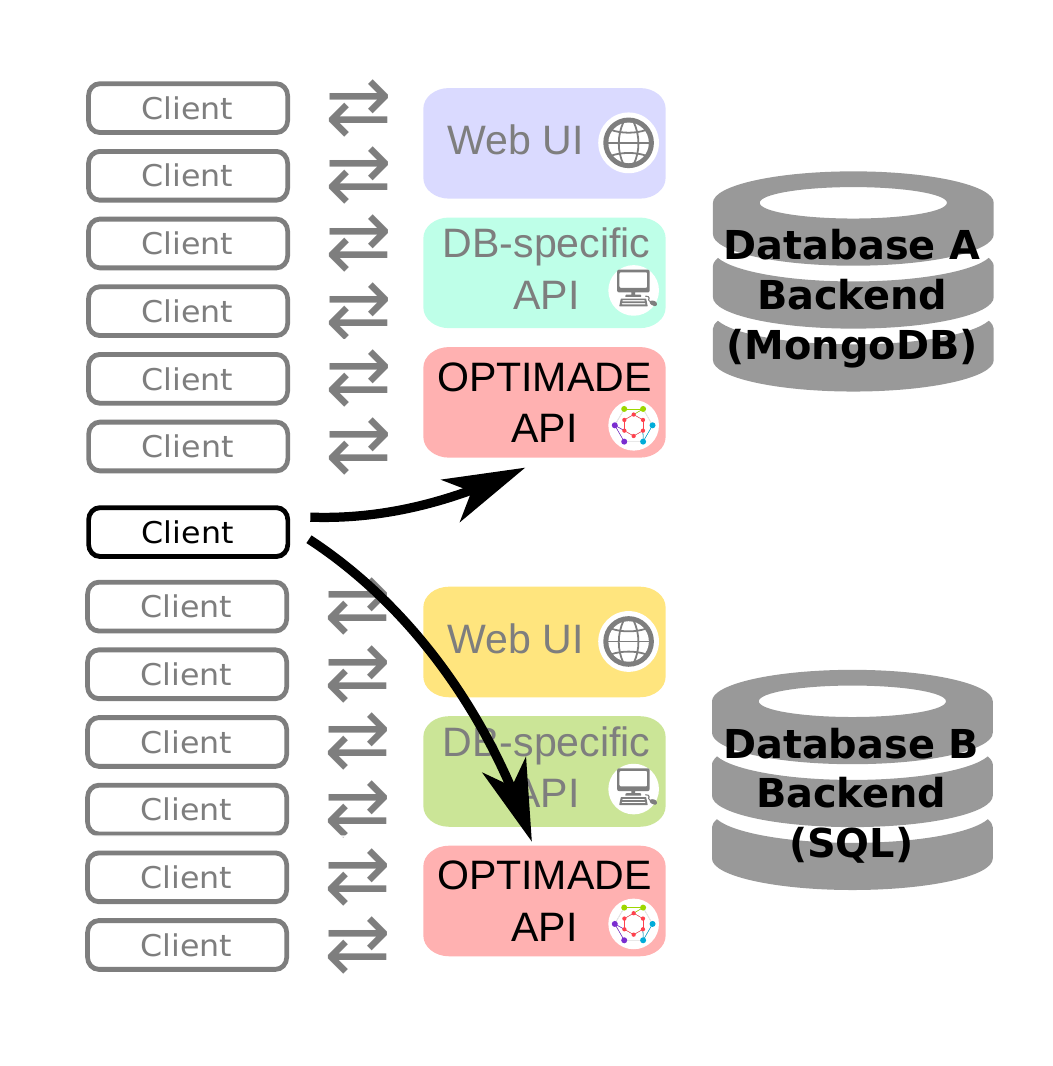}
 \end{center}
 \caption{Schematic for how databases with different backends may provide their own web-based user interfaces and database-specific APIs alongside the OPTIMADE API. A single client can then interact with the databases via the OPTIMADE API without having to be aware of these differences.}
 \label{fig:api_schematic}
\end{figure}

A materials database provider that implements the OPTIMADE API will have a database backend and one or more interfaces available to clients.
These interfaces include the OPTIMADE API, but can also provide access by other means, e.g., a database-specific API or web-based graphical user interface.
Figure~\ref{fig:api_schematic} serves as a schematic illustration of this point.

The providers that presently support the OPTIMADE API represent a wide range of underlying backends.
The primary backend component is, usually, a database engine, but the backend covers all parts of the system that manages the stored data. Relevant backends for OPTIMADE implementations range from simple flat files in a filesystem, to sophisticated setups with load-balanced distributed cloud hosting of relational database engines (e.g., Structured Query Language, SQL-based) or non-relational key-value, document or graph database engines  (so-called NoSQL).

There exists a multitude of APIs for data access and storage (both for materials databases and more generally) that have been designed for a specific database backend.
These APIs are typically designed around specific features the backend provides in terms of, e.g., browse, search, and retrieval, and, crucially, \emph{how} these features are implemented by the backend.
The API then typically becomes a thin wrapper for these features, which exposes the functionality of the backend implementation.
There is thus a crucial need for a generic database interface, which should be based on the central principles:

\begin{itemize}
 \item A core feature set that any reasonable materials database backend can implement via cheap, single-pass, on-the-fly translations in the API layer. Furthermore, the translations should be possible \emph{without} modifying the underlying backend -- i.e., participating databases should not be required to reformat or amend the stored data, software features, etc.\ to provide these core OPTIMADE features.
 \item Extended features beyond the core features that are shared among multiple (but not all) backends should be standardised as \emph{optional} features. It may appear that this design works against interoperability, as it may lead to query $Q$ working for databases $A$ and $B$, but not for database $C$. However, if the database backend for $C$ \emph{cannot} support the type of query that $Q$ represents (for example, a database containing molecular dynamics calculations of proteins cannot sensibly be searched on the chemical formula in the simulation cell), it means that there exists no way to make that query interoperable across $A$, $B$, and $C$ (without altering the backend of $C$). Hence, the highest level of interoperability is achieved by standardising $Q$ as an optional feature.
 \item For standardised optional features, multiple overlapping representations of the same features and/or data should be avoided. The reason is that, for two different ways of providing feature $Q$ as $Q_1$ and $Q_2$, we may end up with database $A$ supporting only $Q_1$ and database $B$ supporting only $Q_2$. Hence, the client either has to tailor the query differently depending on the destination or provide multiple versions of the query in the request, which is an undesirable burden to put on clients and works against an interoperable design. What is strictly standardised is how a database that does not implement a certain feature should respond should that feature be requested.
 \item Those features and data that only occur in a particular database should be provided in isolated database-specific namespaces that other databases can recognise and handle appropriately. The OPTIMADE specification describes how references to such fields should be announced by the particular database and also how clients making multi-provider queries should handle in the response in various situations for optimal interoperability. The latest version of OPTIMADE even outlines a mechanism for creating sub-specifications that allow multiple database providers to collaborate on custom communal definitions, as shall be discussed later.
\end{itemize}

\subsection{Recent OPTIMADE improvements}\label{sec:OPTIMADEimprovements}

Since the initial v1.0 release~\cite{OPTIMADE_spec_2020} of the OPTIMADE API in July 2020, many features have been added, driven by user feedback and use cases. The major enhancements to the specification introduced in versions 1.1~\cite{OPTIMADE_spec_2021} and 1.2 are discussed below, with the full \href{https://github.com/Materials-Consortia/OPTIMADE/blob/master/CHANGELOG.md}{changelog} and specification text available online on GitHub at \href{https://github.com/Materials-Consortia/OPTIMADE}{Materials-Consortia/OPTIMADE}~\cite{OPTIMADE_github}.

\subsubsection{Property definitions}

In previous versions of OPTIMADE, the information served through introspection for a provided property was limited to a single specification of an OPTIMADE type, which is far from sufficient for a client to use the data. For standard fields, clients were referred to the human readable descriptions in the OPTIMADE specification, and for database-specific fields to side channel information, e.g., at the website of the database. 

\sloppypar
OPTIMADE v1.2 includes a full schema format, based on -- and compatible with -- JSONSchema~\cite{jsonschema}, capable of fully describing in a machine-readable way what a property is (including definition of sophisticated data structures with multiple layers of lists and dictionary subfields). All properties are given clear versioned stable identifiers (URIs) that can be used to identify that multiple databases refer to the exact same property.
A simple example is provided in Box~\ref{listing:nsites}, which displays the definition of the \optimadeprop{nsites} field. 
Furthermore, a browsable interface to these definitions for the OPTIMADE standard properties is available at
\href{https://schemas.optimade.org/}{https://schemas.optimade.org/}.

There is related work outside of the specification on providing an enlarged set of shared definitions that can be incrementally adopted by the subsections of the community. This will allow databases to point to and share the same property definition without requiring the slow consensus-building step for the relevant fields to be promoted to the main specification. This is especially useful in cases where this shared information is the defining feature of a given database; for example, the first target namespace is that of stability predictions arising from density-functional theory calculations, the key property required to truly enable materials discovery applications with OPTIMADE (see \ref{sec:MaterialsDiscovery} later for more information).

\begin{Listing}[ht!]
\begin{center}
\begin{mdframed}[style=mintframe]
\begin{minted}{json}
{
    "$id": "https://schemas.optimade.org/defs/v1.2/properties/optimade/structures/nsites",
    "$schema": "https://schemas.optimade.org/meta/v1.2/optimade/property_definition.json",
    "title": "number of sites",
    "x-optimade-type": "integer",
    "x-optimade-definition": {
        "label": "nsites_optimade_structures",
        "kind": "property",
        "version": "1.2.0",
        "format": "1.2",
        "name": "nsites"
    },
    "type": [
        "integer",
        "null"
    ],
    "description": "An integer specifying the length of the `cartesian_site_positions` property.\n\n**Requirements/Conventions**:\n\n- MUST be equal to the lengths of the list properties elements and elements_ratios, if they are provided.\n\n**Query examples**:\n\n- Match only structures with exactly 4 sites: `nsites=4`\n- Match structures that have between 2 and 7 sites: `nsites>=2 AND nsites<=7`",
    "examples": [
        42
    ],
    "x-optimade-unit": "dimensionless"
}
\end{minted}
\end{mdframed}
\end{center}
 \caption{An example OPTIMADE property definition for the \href{https://schemas.optimade.org/defs/v1.2/properties/optimade/structures/nsites}{\optimadeprop{nsites}} property.\label{listing:nsites}}
\end{Listing}


\subsubsection{Streaming of partial data}

So far, OPTIMADE has been based on the JSON format, which has the advantage that it is relatively human readable and well supported by most programming languages. Most tools, however, can only process an entire JSON file and thus do not support streaming processing, which makes it difficult to handle large JSON files. In addition, JSON does not support binary data: a response with large amounts of numerical data needs to be encoded (e.g., in \texttt{base64} or similar encodings), significantly increasing the size of the data to be transferred over the network. Adding support for a format like BSON, which does support binary data, or JSON variants which do support streamed processing (such as JSON Lines) could therefore improve the data transfer rates for OPTIMADE requests, and allow for immediate visualisation of partial results.

In previous versions of OPTIMADE, the size of property data was limited so that a single database entry could fit in a single HTTP response. In the most recent version of OPTIMADE, the implementation can decide to defer large properties to be communicated over a separate simplified streamable protocol, which in practice can be implemented by serving large static files over HTTP. This addition was driven by the future application of serving trajectory data with OPTIMADE (see \ref{sec:FutureOfOPTIMADE} for more information) where individual ``properties'', such as atomic positions in each frame of a trajectory, can individually be prohibitively large.

\subsubsection{Other technical and scientific enhancements}

There have been several other extensions of the format since version 1.0:

\begin{description}
 \item[Symmetry information for structures]The latest OPTIMADE release includes a set of standardised and comprehensive descriptions of structural symmetry. This has been achieved with 5 fields that can be variously used for filtering by symmetry and for reconstructing atomic site positions, building on existing standards laid out by the IUCr and others:
 \begin{itemize}
 \item{\optimadeprop{space\_group\_symmetry\_operations\_xyz},}\item{\optimadeprop{space\_group\_symbol\_hall},}\item{\optimadeprop{space\_group\_symbol\_hermann\_mauguin},} \item{\optimadeprop{space\_group\_symbol\_hermann\_mauguin\_extended},}
 \item{\optimadeprop{space\_group\_it\_number}.}
 \end{itemize}
 \item[\optimadeendp{/files} endpoint]The OPTIMADE specification has been extended to enable the description of files and their precise relationships to other OPTIMADE entries. To implement this, a new \optimadeendp{/files} entry endpoint has been defined. This addition allows, for example, linking \optimadeendp{/structures} entries with their representations in widely used structural data formats such as CIF, POSCAR, and SDF, or linking directly to the raw input or output files associated with a calculation involving a structure.
 
 \item[Per property and per entry metadata]We plan to create a mechanism to provide metadata per property for each individual entry. This metadata could, for example, be confidence intervals, or information about how a property was calculated.

 \item[Request delay]\sloppypar In order not to overload a particular OPTIMADE server, the metadata property \optimadeprop{request\_\-delay} appears in the latest release to allow an implementation to suggest a specific back-off time delay between subsequent requests. It is up to the given server implementation to decide what to do with clients violating the requested delay: refuse to serve, intentionally delay the response, or ignore.
 
 \item[Licensing]The possibility of unsupervised database harvesting raises a need for machine-readable definition of data licenses. Whilst OPTIMADE is an open format, it can be used to serve proprietary or otherwise restricted data, the usage terms of which must be described. The latest OPTIMADE release addresses this issue by introducing the metadata fields \optimadeprop{license}, \optimadeprop{available\_licenses}, and \optimadeprop{available\_licenses\_for\_entries}. The property \optimadeprop{license} is intended for databases to link to human-readable licensing terms, while \optimadeprop{available\_licenses} and  \optimadeprop{available\_licenses\_for\_entries} allow specifying machine-readable license identifiers following the Software Package Data Exchange (SPDX) standard, making interpreting requirements easier for automated clients and web crawlers. This standardised way to announce that either the whole database, or the individual entries, are explicitly available under certain licenses enables machine-actionable licensing decisions, e.g. for commercial re-use and republishing of data.
 
 \item[Substring comparisons on list elements]The OPTIMADE filter grammar defines substring comparison operators to match either the start, end or any part of a property value. However, until version 1.2, such comparisons could not be carried out on list elements. The latest version of OPTIMADE now explicitly supports such substring queries on elements of list properties.

 \item[Boolean values]Although there are no Boolean fields in the main OPTIMADE specification, the latest version of the OPTIMADE specification includes support in the filter grammar for defining and filtering on such custom fields defined by data providers.

\end{description}

\subsection{Associated software tools}

\subsubsection{optimade-python-tools}\label{sec:PythonTools}

\texttt{optimade-python-tools} is an open-source software (MIT license) package that provides tooling for serving, validating and consuming OPTIMADE APIs in Python~\cite{Evans2021}, available on GitHub at \href{https://github.com/Materials-Consortia/optimade-python-tools}{Materials-Consortia/optimade-python-tools}.
Now in version 1.0, it provides a highly extensible reference server implementation of an OPTIMADE API, with support for different database backends.
This server is provided as a Docker container for easy deployment and can be configured to use an existing database, or generate one from scratch in the OPTIMADE format.
Existing databases wanting to make use of the library need to provide mappings to and from their existing data format and query mechanisms.
The package contains isolated modules for various OPTIMADE-related functionalities, for example, a grammar and parser for the OPTIMADE filter language, mappers for querying different database backends, and a fuzzy validator that can dynamically generate requests to an OPTIMADE server to assess its compliance with the standard (which is also used to generate the OPTIMADE provider dashboard: \href{https://optimade.org/providers-dashboard}{optimade.org/providers-dashboard})~\cite{OPTIMADE_providers_dashboard}.
In addition to server-focused functionalities, the package includes reusable code that can help OPTIMADE consumers and clients, including adapters for converting OPTIMADE entries into common formats used in the community, such as ASE \texttt{Atoms}~\cite{ase}, pymatgen \texttt{Structure}s~\cite{Ong_pymatgen_2013} and AiiDA nodes~\cite{AiiDA2}.

The package also contains an advanced asynchronous HTTP client that can be used within Python code or at the command line to concurrently filter multiple OPTIMADE databases with multiple queries, paginating and validating the results, as well as searching across databases for supported properties.
Users can provide callbacks to store the results in local secondary databases for re-use in other projects.

\texttt{optimade-python-tools} is fully documented online at \href{https://www.optimade.org/optimade-python-tools/}{optimade.org/optimade-python-tools} with guides for setting up and validating an API, deploying a server and using the client.
In this way, \texttt{optimade-python-tools} significantly lowers the barrier to retrieving data from OPTIMADE APIs, and to the development of new server implementations.
Future developments will focus on extending the use cases of \texttt{optimade-python-tools} to operating on static data, so that it can be embedded within archival infrastructure, such as Materials Cloud~\cite{MaterialsCloud}, to serve user data without any additional input.

\subsubsection{optimade-gateway}\label{sec:Gateway}

\texttt{optimade-gateway} is an open-source software (MIT license) package implementing a RESTful API in Python for querying OPTIMADE databases, available on GitHub at \href{https://github.com/Materials-Consortia/optimade-gateway}{Materials-Consortia/optimade-gateway}. This provides a powerful yet straightforward tool to allow users of the Python programming language (that is widespread in machine learning and other branches of data science) to access material database results. \texttt{optimade-gateway} supports both synchronous and asynchronous searches via HTTP GET and POST requests, respectively. It can return search results in the standard OPTIMADE data format, as well as a custom OPTIMADE-inspired data format. The main purpose and goal of the deployed service is to be a client backend. A gateway to version 0.4 is running at \href{https://mmp-optimade-gateway.materialscloud.io}{mmp-optimade-gateway.materialscloud.io} as well as at \href{https://optimade-gateway.fly.dev}{optimade-gateway.fly.dev}. Both services utilise a MongoDB database for time-limited caching of the query results, increasing response speeds for common queries.

\subsubsection{Optimade.Science}\label{sec:OptimadeDotScience}

\href{https://optimade.science}{Optimade.Science} is a minimalist in-browser OPTIMADE aggregator, written in the TypeScript language on top of the Svelte frontend framework \cite{svelte_2024}.
It fetches the official OPTIMADE providers list, looks for the structure endpoints, and allows simultaneous querying against all of them, collecting the results together on a single webpage.
Technically, this is just the single file \texttt{index.html} and is thus highly-portable, can be opened from anywhere, on any environment (e.g., on a smartphone or locally from a USB stick).
To increase ease-of-use, a simple pattern-matching library was developed (\href{https://github.com/mpds-io/optimade-mpds-nlp}{mpds-io/optimade-mpds-nlp}), transforming the free-text user input into a standard OPTIMADE query (e.g., a keyword \texttt{ternary} is transformed into \texttt{nelements=3}).
A separate Svelte user interface kit (\href{https://github.com/basf/svelte-spectre}{basf/svelte-spectre}) was developed offering a range of the modular GUI components, willingly accepted by the frontend community and already re-used in many other web-projects, including commercial ones.
A standalone OPTIMADE client written in TypeScript was employed, being fully isomorphic (that is, the same code can be used inside the web browser and on the web server).

\section{Contributing databases}\label{sec:UpdateOnDatabases}

The burgeoning community of materials databases are the core that underpins the OPTIMADE consortium.
Therefore, below we discuss the key features of the major materials databases that make data available through the OPTIMADE API.
We first briefly introduce each database and its offering, and we then discuss its particular OPTIMADE implementation. Finally, we provide an updated table from our previous work~\cite{OPTIMADE2021} that compares the amount of compliant data available in different databases.

\subsection{AFLOW}

\database One of the largest open-access databases for inorganic materials, with 4 million compounds and 800 million associated properties~\cite{afloworg2023,aflowpp2023}, which also includes the 2000+ entries of the AFLOW Encyclopedia of Crystallographic Prototypes
~\cite{AFLOW_PROTO1,AFLOW_PROTO2,AFLOW_PROTO3}.
The data has been employed for the discovery of new permanent magnets~\cite{aflowmagnets2017}, superalloys~\cite{aflowsuperalloys2017,ReyesTirado_superalloys_ActaMat_2018}, high-entropy high-hardness plasmonic carbides~\cite{aflowhep2022},  super-hard disordered carbides~\cite{aflowhec2018}, borides and carbo-nitrides~\cite{DEED}, and phase-change memory compositions~\cite{aflowcameo2020}, and has also been used to study bulk metallic glasses~\cite{aflowbmg2016,aflowbmg2019}, superconductors~\cite{aflowcartography2015,aflowsc2018}, and thermoelectrics~\cite{aflowte2016}. The data can be retrieved conveniently through the AFLUX search API~\cite{AFLUX} with a minimal, flexible, and human-readable query language.\\

\implementation The AFLOW OPTIMADE API builds on AFLUX to offer a common query syntax across multiple materials databases, mapping AFLOW property labels to that of OPTIMADE while still offering access to AFLOW-specific properties with the \optimadeprefix{aflow} prefix. A full list of keywords available to use with OPTIMADE to query AFLOW are available at the info endpoint \optimadeendp{/info/structures}.\\

\baseurl{https://aflow.org/API/optimade}

\subsection{Alexandria}\label{Alexandria}

\database Comprises both hypothetical and existing compounds, which have been relaxed using density functional theory (DFT).
Currently, the database contains 5,062,521 entries, spanning nearly the entire periodic table with 89 elements.
The database was primarily generated by scanning binary, ternary, and quaternary prototypes to identify stable compounds.
This process employed crystal graph attention networks~\cite{CGAT, CGATHT, garnet} to predict the stability of all potential compositions for each prototype. Compounds that were found to be close to stability were subsequently confirmed using DFT.
Additionally, the database includes compounds obtained from traditional high-throughput searches conducted previously~\cite{jonathan2018, schmidt2017, Wang2021b}.

Most entries were calculated using the PBE functional with parameters mostly consistent with those of the Materials Project~\cite{Materials_Project}, and includes 3D (4\,489\,295 entries), 2D (137\,833) and 1D (13\,295 entries) compounds.
Additionally, a total of 422\,098 materials were computed using the PBEsol and SCAN functionals to yield more precise geometries, formation energies, and bandgaps~\cite{scan_dataset}.
The PBE version of the Alexandria database, which comprises 115\,535 potentially stable materials, represents the most extensive publicly available DFT convex hull of thermodynamic stability in our knowledge.
Furthermore, 771\,696 materials lie within a distance of less than 50\,meV/atom from the convex hull.
The database encompasses various properties, including structure (lattice and atomic positions), energy distance to the convex hull, formation energy and direct as well as indirect bandgaps.
Continuous expansion of the database is underway through further ongoing high-throughput searches.\\

\implementation Uses the \texttt{optimade-python-tools}~\cite{Evans2021} reference implementation and provides a list of extra properties with the \optimadeprefix{alexandria} prefix.
Prior to this development, the Alexandria database was made available solely as a static archive that users had to download in its entirety to explore, but now the OPTIMADE format supports filtering on both composition and the predicted stability of database entries.\\

\baseurl{https://alexandria.icams.rub.de}

\subsection{BioExcel COVID-19}\label{BioExcel}

\database A platform designed to provide online access to atomistic molecular dynamics trajectories for biological macromolecules related to the COVID-19 disease \cite{Beltran2024}. The project is part of the open access initiatives promoted by the world-wide scientific community to share information about COVID-19 research and integrate technology developed in previous biology related projects \cite{Hospital2015, Zivanovic2020, Andrio2019, Hospital2020}.\\

\implementation A web-server interface \href{https://bioexcel-cv19.bsc.es}{https://bioexcel-cv19.bsc.es} presents the MD trajectories, with a set of quality control analyses and system information. Using an extension of version 1.1.0 of the OPTIMADE specification, a basic OPTIMADE server based on the \texttt{optimade-python-tools} has been set up, which provides the trajectory data at the \optimadeendp{/trajectories} endpoint. This server also provides protein specific properties and metadata under database specific fields with the \optimadeprefix{bioxl} prefix.
Querying has been partially implemented, but is not yet available for all fields.
No atomistic structures are shared, so the \optimadeendp{/structures} endpoint is not available.\\

\baseurl{https://bioexcel-cv19.bsc.es/optimade/}

\subsection{Computational Materials Repository (CMR)}\label{CMR}

\database CMR is a repository of databases containing calculated atomic structures and basic properties of a broad set of materials. The CMR databases can be browsed online using a simple querying system or downloaded in various formats (\href{https://cmr.fysik.dtu.dk}{https://cmr.fysik.dtu.dk}). Currently, the CMR holds more than 30 different databases.


The flagship database in CMR is the Computational Two-Dimensional Materials Database (C2DB)~\cite{moustafa2022computational,haastrup2018computational,gjerding2021recent} that contains structural, thermodynamic, elastic, electronic, magnetic, and optical properties of more than 15,000 two-dimensional (2D) monolayer materials computed using the GPAW~\cite{GPAW,mortensen2023gpaw} package. The core set of materials in C2DB have been obtained by extracting monolayers from experimentally known layered van der Waals crystals. Subsequently, new monolayers have been generated by systematic atom-substitution applied to the core materials, or using deep generative AI models~\cite{lyngby2022data}. Recently, the C2DB has been complemented by the BiDB database containing homobilayers formed by stacking 1,000 of the most stable monolayers from the C2DB in all possible commensurate configurations~\cite{pakdel2023emergent}. \\


\implementation CMR implements the OPTIMADE API through the \href{https://gitlab.com/jensj/ase-optimade/}{OASE} package, utilising the \texttt{optimade-python-tools} library. At present only a subset of C2DB is available via an OPTIMADE API, but in the future other CMR databases will also be available via OPTIMADE.\\

\baseurl{https://cmr-optimade.fysik.dtu.dk}

\subsection{Crystallography Open Database (COD)}

\database The largest open access collection of \emph{experimental} crystal structures~\cite{Grazulis_COD_2012}.
It is widely used by the scientific community to explore different material categories such as superconductors~\cite{Choudhary_COD_superconductors_2022}, metal-organic frameworks~\cite{Barjasteh_COD_MOF_2023}, high entropy alloys~\cite{Schwarz_CODalloys_2022}, organic molecules~\cite{Chan_CODRingPuckering_2021} as well as for conformer sampling~\cite{Mendenhall2020} or custom force field generation~\cite{Gomzi_CODff_2021}.
Having a set of experimental structures readily available under the same format as is required for computational materials research is highly beneficial since these structures serve as initial points for material property calculations~\cite{MaterialsCloud} or for the search of new materials~\cite{Pizzi_COD_2D_2021}.
They also serve as experimental points that theoretical computations can be checked against~\cite{Wang_COD_model_validation_2022}.\\

\implementation The COD database currently implements version v1.1.0 of the OPTIMADE standard.
The new OPTIMADE version will allow this implementation to be enriched with new features that are required for the faithful representation of experimental data, thus making computations from these data and comparisons of theory and experiments more accurate.
Being an experimental structure database, the COD requires a slightly different data presentation than computational material databases.
Structures from the COD are evaluated using a set of experimental data quality criteria, established by the IUCr and the chemical crystallography community~\cite{IUCr_quality_criteria_2023}.
An OPTIMADE response within the core features does not contain all of the necessary fields to convey these additional data elements; however, the OPTIMADE standard allows introducing database-specific fields in a regular way.
As a result, all established crystallographic quality criteria are included into a COD response as COD-specific fields with the \optimadeprefix{cod} prefix.
This allows OPTIMADE to include experimental position and composition disorder information following the Crystallographic Information Framework~\cite{Hall_IUCr_CIF_1991}.\\

\baseurl{https://www.crystallography.net/cod/optimade}

\subsection{Joint Automated Repository for Various Integrated Simulations (JARVIS)}

\database A repository designed to automate materials design using classical force-field, density functional theory (DFT), machine learning calculations and experiments. The JARVIS-DFT originated about 5 years ago and contains millions of properties materials with carefully converged atomic structures as well as tight convergence parameters and various exchange-correlation functionals. The JARVIS-DFT contains metallic, semiconducting, insulator, superconductor, high-strength, topological, solar, thermoelectric, piezoelectric, dielectric, two-dimensional, magnetic, porous, defect and various other classes of materials \cite{winesreview,choudhary2020joint}.\\

\implementation Based on the Django Rest Framework and the JARVIS-Tools packages to follow OPTIMADE protocols of filtering and curating data. JARVIS-DFT specific fields are included in the results with the\optimadeprefix{jarvis} prefix.\\

\baseurl{https://jarvis.nist.gov/optimade/jarvisdft}

\subsection{Materials Cloud}

\database A platform created to enable sharing and dissemination of resources in computational materials science~\cite{MaterialsCloud}.
A major service offered is the archiving and publishing of research data for the community via the open Materials Cloud Archive service (\href{https://archive.materialscloud.org}{https://archive.materialscloud.org}).
Moreover, several databases that are generated within the AiiDA~\cite{AiiDA2,AiiDA,AiiDA3} framework are published in the Materials Cloud Explore section, which enables users to interactively browse the data and its provenance. Curated visualisations of these databases are also provided in the Materials Cloud Discover section.
These databases are accessible via the OPTIMADE RESTful API. The databases are divided into flagship and contributed databases.
The current flagship databases are MC3D and MC2D~\cite{mc2d_1,mc2d_2} hosting over 34,000 3D crystals and 3,000 2D crystals, respectively, providing properties of experimentally-known inorganic compounds obtained via DFT simulations.
The contributed databases include 2D topological insulators, pyrene-based metal organic frameworks, high-throughput Wannierisation, \ch{SrTiO3}-\ch{CeO2} interfaces, tail-corrections in the molecular simulations of porous materials, hidden spontaneous polarisation in the chalcohalide photovoltaic \ch{Sn2SbS2I3}, and the CURATED covalent organic frameworks database.
The data has been used to investigate transport properties such as mobility~\cite{Sohier2018,Sohier2019}, to search for $\mathbb{Z}_2$ topological order~\cite{Marrazzo2019}, to screen and discover quantum Spin-Hall insulators~\cite{Marrazzo2018, Grassano2023} and Weyl semimetals~\cite{Grassano2023a}, to obtain tight-binding-like Wannier Hamiltonians in a fully automated fashion~\cite{Qiao2023}, and to develop machine-learning methods for fast identification of low-dimensional materials~\cite{Vahdat2022}.

Current developments are focusing on a second ``Materials Cloud Archive'' provider, allowing submissions to Materials Cloud Archive to specify whether and how data contributed by users should be served via an OPTIMADE API, to enable advanced federated search over archived data. The first proof-of-concept of this integration is a recently published dataset of novel electride materials~\cite{WangElectrides2023, wangElectridesMCA2023}.
The full list of OPTIMADE databases served by the Materials Cloud can be explored at the \href{https://materialscloud.org/optimade/main/v1/links}{\optimadeendp{/main/v1/links}} endpoint, with a similar list for the Materials Cloud Archive available at \href{https://materialscloud.org/optimade/archive/v1/links}{\optimadeendp{/archive/v1/links}}. A landing page for both OPTIMADE providers is available at \href{https://www.materialscloud.org/optimade}{https://www.materialscloud.org/optimade}.
\\

\implementation The data on the backend of Materials Cloud is managed via AiiDA. Along with a custom REST API, AiiDA can serve data in the OPTIMADE format thanks to the AiiDA-OPTIMADE (\href{https://github.com/aiidateam/aiida-optimade}{https://github.com/aiidateam/aiida-optimade}) plugin, that is thus also used to serve the main Materials Cloud data. The OPTIMADE implementation of the Materials Cloud Archive provider is instead based directly on the \texttt{optimade-python-tools}~\cite{Evans2021} package.
In addition to the server implementations, Materials Cloud also offers users several web applications that include clients of the OPTIMADE API, as we describe in more detail in Section~\ref{sec:optimade-client}.\\

\indexbaseurl{https://www.materialscloud.org/optimade/main}

\indexbaseurl{https://www.materialscloud.org/optimade/archive}

\subsection{Materials Platform for Data Science}

\database Materials Platform for Data Science (MPDS) serves the Pauling File dataset~\cite{PaulingFile_2024}. Started in 1993, Pauling File is the oldest privately funded initiative for the curation and standardisation of the published inorganic chemistry data\footnote{The double awarded Nobel laureate Linus Pauling personally endorsed this project and gave an explicit written permission to use his name. In 2019, the Pauling File's founder Pierre Villars was acknowledged with the NIMS Award (Tsukuba, Japan) for the fundamental research for data-driven materials development.}.
Data is drawn from nearly 400 thousand publications and backs up such commercial products as Springer Materials, ICDD PDF 4+, ASM's Alloy Phase Diagram Database and Pearson's Crystal Data, MedeA, and AtomWork Advanced.\\

\implementation MPDS presents curated experimental data of three types: crystalline structures (\optimadeendp{/structures} endpoint), physical properties (\optimadeendp{/extensions/properties} endpoint), and phase diagrams (\optimadeendp{/extensions/phase\_diagrams} endpoint).
These three data types are inter-linked into about 200 thousand distinct phases (\optimadeendp{/extensions/phases} endpoint).
Any distinct phase is uniquely determined by the chemical formula, space group, and Pearson symbol.
Furthermore, each distinct phase has the permanent integer identifier \texttt{phase\_id}, e.g., see \href{https://mpds.io/phase\_id/27712}{brookite}. The MPDS OPTIMADE implementation is specifically designed for the low response time and high retrieval speed, therefore some expensive operators (\optimadefilter{ANY}, \optimadefilter{OR}) are currently not supported (cf. Table~\ref{tab:table_redux}).\\

\baseurl{https://api.mpds.io}

\subsection{Materials Project}

\database This multi-institution, multi-national effort~\cite{Materials_Project} aims at computing the properties of all inorganic materials and providing the data and associated analysis algorithms for every materials researcher free of charge. Currently, over 172k molecules and over 154k inorganic compounds are included in the database. The project was established in 2011 with an emphasis on battery research, but includes property calculations for many areas of clean energy systems such as photovoltaics, thermoelectric materials, and catalysts.\\

\implementation The Materials Project (MP)~\cite{Materials_Project} makes use of the reference server implementation provided by the \texttt{optimade-python-tools}~\cite{Evans2021}.
Since May 2022, Materials Project has been serving formation energy data via its OPTIMADE API.
In June 2023, MP started exposing additional thermodynamic stability in the form of energy distance to the convex hull via OPTIMADE for all 154k materials in its core database. The convex hull distance to the Materials Project is one of the most important properties of theoretical structures for experimentalists and simulators alike, as it indicates whether a postulated material is potentially synthesisable.\\

The MP OPTIMADE integration is complemented by a convenient open-source pymatgen~\cite{Ong_pymatgen_2013} interface in the form of the \href{https://github.com/materialsproject/pymatgen/blob/ec750ca15d02cdd51b0c0a7a4408af8e0d259223/pymatgen/ext/optimade.py#L31}{\texttt{OptimadeRester}} class, designed to streamline access to these resources for existing users of pymatgen and the Materials Project.
Additionally, efforts are underway to further expose the full set of MP \texttt{summary} data via the OPTIMADE endpoint under the \optimadeprefix{mp} namespace, mirroring the complete set of data recorded in the \texttt{emmet} \href{https://github.com/materialsproject/emmet/blob/bf8a4ef09a0d9f91bb6e9fe3e2fca0acd3582306/emmet-core/emmet/core/summary.py#L137}{\texttt{SummaryDoc}}.\\

\baseurl{https://optimade.materialsproject.org}

\subsection{Material-Property-Descriptor Database}

\database Material-Property-Descriptor Database (MPDD) is an extensive database (4M+) of \emph{ab initio} relaxations of 3D crystal structures,  combined with an infrastructure of tools allowing efficient descriptor calculation (featurization), as well as the deployment of ML models.~\cite{Krajewski2024pySIPFENN}
The most critical feature of the MPDD is the retention of intermediate modelling data, including structure-informed descriptors, which typically cost orders of magnitude more computational time than any of the other steps performed during ML model deployment~\cite{Krajewski2022ExtensibleNetworks}.
Thus, many ML models can be run at a small fraction of the original cost if the same descriptor (or, more commonly, a subset chosen through feature selection) is used.
This benefit applies regardless of whether a model is just another iteration, e.g., fine-tuned to a specific class of materials like perovskites, or an entirely new model for a different property.
Furthermore, MPDD's access to stored atomic structures and associated metadata has been shown to be useful, for instance, in the fully data-driven prediction of atomic structures (validated with DFT and experiments), allowing quick identification of unknown structures in Nd-Bi~\cite{Im2022ThermodynamicModelingNdBi} and Al-Fe~\cite{Shang2021FormingAlFeJoints} systems.\\

\implementation MPDD has a stable OPTIMADE API that serves the entire core MPDD dataset, fully implementing v1.1.0 of the OPTIMADE standard through a server based on \texttt{optimade-python-tools}~\cite{Evans2021}.
Making the MPDD available via OPTIMADE was initially challenging, as MPDD stores and exchanges data in a way that prioritises high throughput and low storage requirements, including binary data, making it difficult or slow to make MPDD queryable as an OPTIMADE API on-the-fly.
However, issues have been resolved by establishing a self-updating mirror of the dataset where structures are made OPTIMADE-compliant during transfer and with most associated MPDD-specific data available under the \optimadeprefix{mpdd} namespace, including dictionaries of metadata, properties, and descriptors.\\

\baseurl{http://optimade.mpdd.org}\\

\subsection{Materials Properties Open Database}

\database The Material Properties Open Database (MPOD) \cite{MPOD2012, MPOD2017} is a web-based, open access repository of experimentally determined quantitative information about the physical properties of crystalline materials. MPOD is oriented at design engineers, scientists, science teachers and students. Properties are generally treated as tensor magnitudes. In MPOD the compact matrix notation is applied. To bring an intuitive view of tensor properties, so-called longitudinal properties surfaces are displayed. 3D printing of properties surfaces is implemented via creation of STL files. A dictionary of properties definitions is included. Eventually, comments are added. Syntax and notation in MPOD files are oriented towards matching IUCr standards and so tries to comply with CIF format.\\

\implementation The integration of OPTIMADE with MPOD encompassed two distinct phases. Initially, the process entailed migrating all data from the MySQL database to MongoDB. This was followed by the mapping of MPOD objects to OPTIMADE, utilizing \texttt{optimade-python-tools}~\cite{Evans2021}. In this context, the prefix \optimadeprefix{mpod} was employed to delineate specific database fields.\\

\baseurl{http://mpod\_optimade.cimav.edu.mx}

\subsection{Materials Resource Registry}

\database A federated, decentralised registry of resources in the domain of materials science. It exposes these resources to users and machines via XML and OAI-PMH APIs~\cite{Plante2021}.\\

\implementation OPTIMADE has been added to the Materials Resource Registry as an API format that other services and datasets can link to, to indicate their own compliance.
Materials Resource Registry's rich semantic description of databases, with regards to their scientific content, techniques, and material focus~\cite{Medina-Smith2021}, as curated by the provider, enables users to make expressive queries over OPTIMADE providers, to narrow down which databases may be of interest to them.
This makes it much easier to discover data and direct clients to the resources that are scientifically the most relevant to them.

\subsection{Matterverse}

\database A database of yet-to-be-synthesised materials predicted using state-of-the-art ML models, currently comprised of 31,664,858 hypothetical materials. The current structures were generated by combinatorial isovalent ionic substitutions on 5,283 binary, ternary, and quaternary structural prototypes from the 2019 version of the ICSD database.
A critical enabler for this database is the Materials 3-body Graph Network (M3GNet) universal interatomic potential encompassing 89 elements of the periodic table~\cite{Chen2022b}.
Along with the information of lattice parameters, atom coordinates and $E_\text{hull}$, \href{https://matterverse.ai}{matterverse.ai} also provides the predicted formation energies, bandgaps (of multiple fidelities, including PBE, HSE and experimental), and bulk and shear moduli. As an ongoing effort, matterverse.ai is growing in two directions, (i) increasing the number of hypothetical materials via various structure generation strategies, and (ii) increasing the number of ML-predicted properties.\\

\implementation Support of the OPTIMADE API is under active development, with so far successful mapping of data to the OPTIMADE format using the \texttt{optimade-python-tools}~\cite{Evans2021}.\\


\subsection{NOMAD}

\database An open-source software and free service for managing and publishing FAIR~\cite{NOMAD_fair2022} materials science data. 
NOMAD~\cite{NOMAD_repository1,NOMAD_repository2} was made publicly available in 2014; it provides over 12 million data entries from over 500 researchers~\cite{NOMAD_2023}.
Originally, NOMAD focused on ab initio codes based on density-functional theory (DFT), automatically extracting data and metadata from input and output files.
Meanwhile, NOMAD was significantly expanded in scope by the consortium FAIRmat (\href{https://fairmat-nfdi.eu}{fairmat-nfdi.eu}).
It now supports file types from over 60 simulation codes, it encompasses advanced many-body calculations, including GW, the Bethe-Salpeter equation (BSE), and dynamical mean-field theory (DMFT), and classical molecular dynamics simulations.
It can cope with different types of experimental data. For instance, it provides support for electronic lab notebooks and the NeXus format. 
NOMAD can track data provenance in complex simulation and experiment workflows. 

NOMAD enables individual researchers to make their data available to a wide range of possible clients and applications. 
All data is formally described through rich metadata schema~\cite{Ghiringhelli2023} and can be analyzed with build in containerised tools and notebooks~\cite{NOMAD_toolkit2022}.
Data in NOMAD is provided through the OPTIMADE API, NOMAD specific APIs, and a rich graphical user interface with faceted search, (meta)data explorer, and visualisations for material properties.
API access is particularly important for re-use of the data, e.g. with artificial-intelligence (AI) methods. A collection of AI tools is available in the NOMAD AI Toolkit~\cite{NOMAD_toolkit2022}.
Besides supporting the community with the central data infrastructure, NOMAD offers the same software~\cite{NOMAD_2023} for local installation through NOMAD Oasis, which allows research groups to manage and provide their own research data individually and customise the software accordingly. \\

\implementation NOMAD supports a full OPTIMADE API implementation based on the ~\cite{Evans2021} using the Elasticsearch database engine, and a web-based search interface that allows users to formulate queries based on the standardised OPTIMADE query strings.
Furthermore, NOMAD users can search for related resources from all other OPTIMADE database providers in the OPTIMADE provider list.\\

\baseurl{https://nomad-lab.eu/prod/rae/optimade}

\subsection{Open Database of Xtals (\emph{odbx})}

\database A small database serving selected phase diagrams studied with \emph{ab initio} crystal structure prediction techniques~\cite{Harper2020, Evans2020}. Recently, \emph{odbx} has been used to ingest new materials discovery datasets into the OPTIMADE ecosystem as part of the \href{https://optimade-misc.odbx.science}{optimade-misc.odbx.science} sub-database~\cite{Goodall2022, Law2023, Ye2022}, as well as the GNome dataset~\cite{Merchant2023} at \href{https://optimade-gnome.odbx.science}{optimade-gnome.odbx.science}, as will be discussed in~\ref{sec:MaterialsDiscovery}.\\

\implementation \emph{odbx} was created using the matador~\cite{Evans2020} and \texttt{optimade-python-tools}~\cite{Evans2021} packages. As well as serving the standard OPTIMADE properties, \emph{odbx} also serves stability data (hull distances, formation energies) and the DFT parameters used to relax the structures under the \optimadeprefix{odbx} namespace. \emph{odbx} serves multiple distinct datasets; the \optimadeendp{links} endpoint of the index base URL below can be used to retrieve them.\\

\indexbaseurl{https://optimade-index.odbx.science}

\subsection{Open Materials Database (\emph{omdb})}
\database \emph{omdb} provides materials properties and is maintained by the developers of the High-Throughput Toolkit \emph{(httk)} \cite{HTTKOMDB}. It contains 205,264 structures for access via programmatic interaction using this toolkit. The structures are also accessible via a web interface. Recently, it is being integrated in the broader database effort Anyterial (\url{https://www.anyterial.se/}), which also includes the ADAQ database of point defects \cite{ADAQ}.\\

\implementation \emph{omdb} uses the built-in implementation of the OPTIMADE API provided in \emph{httk}. This implementation is written in Python using no dependencies beyond the Python standard library. Work is currently ongoing to extend the implementation to fully support version v1.2.0.\\

\baseurl{https://optimade.openmaterialsdb.se}

\subsection{Open Quantum Materials Database}
\database The Open Quantum Materials Database (OQMD) holds over 1 million materials, consisting of both experimental and hypothetical compounds~\cite{Saal_OQMD2013,Shen_ReflectionOQMD}.
The overarching interest of OQMD is to understand the competing stability between known and unknown compounds by generating large-scale convex hulls -- a stable yet-to-be-synthesised material should fall along or close to this convex hull.
In addition, OQMD grows and develops organically with interests in Wolverton group, including calculations targeting thermoelectrics~\cite{He_OQMD_Thermoelectrics}, battery materials~\cite{Aykol_OQMD_Batteries}, and high-strength alloys~\cite{Kirklin_OQMD_alloy}.\\

\implementation OQMD currently utilises v1.0.0 of the OPTIMADE standard and will adopt newer versions of OPTIMADE to replace OQMD’s current qmpy API. OQMD offers database specific properties through the \optimadeprefix{oqmd} prefix, including formation energies, bandgap, and stabilities of compounds.\\

\baseurl{https://oqmd.org/optimade}

\subsection{Comparison of data available}

\begin{table*}[t]
\centering
\ttfamily
\small{
\begin{tabular}{l|rrrr}
\bf{\texttt{PROVIDER}}&\bf{N$_1$}&\bf{N$_2$}&\bf{N$_3$}&\bf{N\textsubscript{tot}}\\
\hline
AFLOW & \href{http://aflow.org/API/optimade/v1/structures?filter=elements%20HAS%20ANY%20%22C%22%2C%22Si%22%2C%22Ge%22%2C%22Sn%22%2C%22Pb%22}{704,302} ({700,192}) & \href{http://aflow.org/API/optimade/v1/structures?filter=elements%20HAS%20ANY%20%22C%22%2C%22Si%22%2C%22Ge%22%2C%22Sn%22%2C%22Pb%22%20AND%20nelements%3D2}{63,017} ({62,293}) & \href{http://aflow.org/API/optimade/v1/structures?filter=elements%20HAS%20ANY%20%22C%22%2C%22Si%22%2C%22Ge%22%2C%22Sn%22%20AND%20NOT%20elements%20HAS%20%22Pb%22%20AND%20elements%20LENGTH%203}{413,797} ({382,554}) & \href{http://aflow.org/API/optimade/v1/structures?filter=nelements>=1}{3,530,330} \\
Alexandria$^\star$ & \href{https://alexandria.icams.rub.de/pbe/v1/structures?filter=elements%20HAS%20ANY%20%22C%22%2C%22Si%22%2C%22Ge%22%2C%22Sn%22%2C%22Pb%22}{939,084} & \href{https://alexandria.icams.rub.de/pbe/v1/structures?filter=elements%20HAS%20ANY%20%22C%22%2C%22Si%22%2C%22Ge%22%2C%22Sn%22%2C%22Pb%22%20AND%20nelements%3D2}{48,510} & \href{https://alexandria.icams.rub.de/pbe/v1/structures?filter=elements%20HAS%20ANY%20%22C%22%2C%22Si%22%2C%22Ge%22%2C%22Sn%22%20AND%20NOT%20elements%20HAS%20%22Pb%22%20AND%20elements%20LENGTH%203}{437,768} & \href{https://alexandria.icams.rub.de/pbe/v1/structures?filter=nelements>=1}{5,055,842} \\
COD & \href{https://www.crystallography.net/cod/optimade/v1/structures?filter=elements%20HAS%20ANY%20%22C%22%2C%22Si%22%2C%22Ge%22%2C%22Sn%22%2C%22Pb%22}{458,249} ({416,314}) & \href{https://www.crystallography.net/cod/optimade/v1/structures?filter=elements%20HAS%20ANY%20%22C%22%2C%22Si%22%2C%22Ge%22%2C%22Sn%22%2C%22Pb%22%20AND%20nelements%3D2}{4,082} ({3,896}) & \href{https://www.crystallography.net/cod/optimade/v1/structures?filter=elements%20HAS%20ANY%20%22C%22%2C%22Si%22%2C%22Ge%22%2C%22Sn%22%20AND%20NOT%20elements%20HAS%20%22Pb%22%20AND%20elements%20LENGTH%203}{34,739} ({32,420}) & \href{https://www.crystallography.net/cod/optimade/v1/structures?filter=nelements>=1}{512,282} \\
CMR & \href{https://cmr-optimade.fysik.dtu.dk/v1/structures?filter=elements%20HAS%20ANY%20%22C%22%2C%22Si%22%2C%22Ge%22%2C%22Sn%22%2C%22Pb%22}{147} & \href{https://cmr-optimade.fysik.dtu.dk/v1/structures?filter=elements%20HAS%20ANY%20%22C%22%2C%22Si%22%2C%22Ge%22%2C%22Sn%22%2C%22Pb%22%20AND%20nelements%3D2}{147} & \href{https://cmr-optimade.fysik.dtu.dk/v1/structures?filter=elements%20HAS%20ANY%20%22C%22%2C%22Si%22%2C%22Ge%22%2C%22Sn%22%20AND%20NOT%20elements%20HAS%20%22Pb%22%20AND%20elements%20LENGTH%203}{0} & \href{https://cmr-optimade.fysik.dtu.dk/v1/structures?filter=nelements>=1}{1,536} \\

JARVIS-DFT & \href{https://jarvis.nist.gov/optimade/jarvisdft/v1/structures/?filter=elements HAS ANY "C", "Si", "Ge", "Sn", "Pb"}{9,017} & \href{https://jarvis.nist.gov/optimade/jarvisdft/v1/structures/?filter=elements HAS ANY "C", "Si", "Ge", "Sn", "Pb" AND nelements=2}{1,426} & \href{
https://jarvis.nist.gov/optimade/jarvisdft/v1/structures/?filter=elements HAS ANY "C", "Si", "Ge", "Sn", AND NOT elements HAS "Pb" AND elements LENGTH 3}{8,084}  & \href{https://jarvis.nist.gov/optimade/jarvisdft/v1/structures/?filter=nelements%3E=1}{77,096} \\

Materials Cloud$^\star$ & \href{https://aiida.materialscloud.org/optimade-sample/optimade/v1/structures?filter=elements%20HAS%20ANY%20%22C%22%2C%22Si%22%2C%22Ge%22%2C%22Sn%22%2C%22Pb%22}{961,564} & \href{https://aiida.materialscloud.org/optimade-sample/optimade/v1/structures?filter=elements%20HAS%20ANY%20%22C%22%2C%22Si%22%2C%22Ge%22%2C%22Sn%22%2C%22Pb%22%20AND%20nelements%3D2}{4,218} & \href{https://aiida.materialscloud.org/optimade-sample/optimade/v1/structures?filter=elements%20HAS%20ANY%20%22C%22%2C%22Si%22%2C%22Ge%22%2C%22Sn%22%20AND%20NOT%20elements%20HAS%20%22Pb%22%20AND%20elements%20LENGTH%203}{136,176} & \href{https://aiida.materialscloud.org/optimade-sample/optimade/v1/structures?filter=nelements>=1}{4,515,120} \\
Materials Project & \href{https://optimade.materialsproject.org/v1/structures/?filter=elements HAS ANY "C", "Si", "Ge", "Sn", "Pb"}{34,424} (27,309) & \href{https://optimade.materialsproject.org/v1/structures/?filter=elements HAS ANY "C", "Si", "Ge", "Sn", "Pb" AND nelements=2}{3,750} (3,545) & \href{https://optimade.materialsproject.org/v1/structures/?filter=elements HAS ANY "C", "Si", "Ge", "Sn", AND NOT elements HAS "Pb" AND elements LENGTH 3}{11,861} (10,501) & \href{https://optimade.materialsproject.org/v1/structures}{154,387} \\
MPDD & \href{http://mpddoptimade.phaseslab.org/v1/structures?filter=elements%20HAS%20ANY%20%22C%22%2C%22Si%22%2C%22Ge%22%2C%22Sn%22%2C%22Pb%22}{811,136} & \href{http://mpddoptimade.phaseslab.org/v1/structures?filter=elements%20HAS%20ANY%20%22C%22%2C%22Si%22%2C%22Ge%22%2C%22Sn%22%2C%22Pb%22%20AND%20nelements%3D2}{80,195} & \href{http://mpddoptimade.phaseslab.org/v1/structures?filter=elements%20HAS%20ANY%20%22C%22%2C%22Si%22%2C%22Ge%22%2C%22Sn%22%20AND%20NOT%20elements%20HAS%20%22Pb%22%20AND%20elements%20LENGTH%203}{490,900} & \href{http://mpddoptimade.phaseslab.org/v1/structures?filter=nelements>=1}{3,975,666} \\
MPOD & \href{http://mpod_optimade.cimav.edu.mx/v1/structures?filter=elements%20HAS%20ANY%20%22C%22%2C%22Si%22%2C%22Ge%22%2C%22Sn%22%2C%22Pb%22}{91} & \href{http://mpod_optimade.cimav.edu.mx/v1/structures?filter=elements%20HAS%20ANY%20%22C%22%2C%22Si%22%2C%22Ge%22%2C%22Sn%22%2C%22Pb%22%20AND%20nelements%3D2}{8} & \href{http://mpod_optimade.cimav.edu.mx/v1/structures?filter=elements%20HAS%20ANY%20%22C%22%2C%22Si%22%2C%22Ge%22%2C%22Sn%22%20AND%20NOT%20elements%20HAS%20%22Pb%22%20AND%20elements%20LENGTH%203}{16} & \href{http://mpod_optimade.cimav.edu.mx/v1/structures?filter=nelements>=1}{401} \\
MPDS & \href{https://api.mpds.io/v1/structures?filter=elements%20HAS%20ANY%20%22C%22%2C%22Si%22%2C%22Ge%22%2C%22Sn%22%2C%22Pb%22}{-} & \href{https://api.mpds.io/v1/structures?filter=elements%20HAS%20ANY%20%22C%22%2C%22Si%22%2C%22Ge%22%2C%22Sn%22%2C%22Pb%22%20AND%20nelements%3D2}{-} & \href{https://api.mpds.io/v1/structures?filter=elements%20HAS%20ANY%20%22C%22%2C%22Si%22%2C%22Ge%22%2C%22Sn%22%20AND%20NOT%20elements%20HAS%20%22Pb%22%20AND%20elements%20LENGTH%203}{-} & \href{https://api.mpds.io/v1/structures?filter=nelements>=1}{507,178} \\
NOMAD & \href{https://nomad-lab.eu/prod/rae/optimade/v1/structures?filter=elements%20HAS%20ANY%20%22C%22%2C%22Si%22%2C%22Ge%22%2C%22Sn%22%2C%22Pb%22}{4,451,056} ({3,359,594}) & \href{https://nomad-lab.eu/prod/rae/optimade/v1/structures?filter=elements%20HAS%20ANY%20%22C%22%2C%22Si%22%2C%22Ge%22%2C%22Sn%22%2C%22Pb%22%20AND%20nelements%3D2}{587,923} ({532,123}) & \href{https://nomad-lab.eu/prod/rae/optimade/v1/structures?filter=elements%20HAS%20ANY%20%22C%22%2C%22Si%22%2C%22Ge%22%2C%22Sn%22%20AND%20NOT%20elements%20HAS%20%22Pb%22%20AND%20elements%20LENGTH%203}{2,092,989} ({1,611,302}) & \href{https://nomad-lab.eu/prod/rae/optimade/v1/structures?filter=nelements>=1}{12,116,021} \\
\emph{odbx}$^\star$ & \href{https://optimade-misc.odbx.science/v1/structures?filter=elements%20HAS%20ANY%20%22C%22%2C%22Si%22%2C%22Ge%22%2C%22Sn%22%2C%22Pb%22}{125,648} ({55}) & \href{https://optimade-misc.odbx.science/v1/structures?filter=elements%20HAS%20ANY%20%22C%22%2C%22Si%22%2C%22Ge%22%2C%22Sn%22%2C%22Pb%22%20AND%20nelements%3D2}{3,179} ({54}) & \href{https://optimade-misc.odbx.science/v1/structures?filter=elements%20HAS%20ANY%20%22C%22%2C%22Si%22%2C%22Ge%22%2C%22Sn%22%20AND%20NOT%20elements%20HAS%20%22Pb%22%20AND%20elements%20LENGTH%203}{17,009} ({0}) & \href{https://optimade-misc.odbx.science/v1/structures?filter=nelements>=1}{523,216} \\
\emph{omdb} & \href{http://optimade.openmaterialsdb.se/v1/structures?filter=elements%20HAS%20ANY%20%22C%22%2C%22Si%22%2C%22Ge%22%2C%22Sn%22%2C%22Pb%22}{58,718} ({58,718}) & \href{http://optimade.openmaterialsdb.se/v1/structures?filter=elements%20HAS%20ANY%20%22C%22%2C%22Si%22%2C%22Ge%22%2C%22Sn%22%2C%22Pb%22%20AND%20nelements%3D2}{690} ({690}) & \href{http://optimade.openmaterialsdb.se/v1/structures?filter=elements%20HAS%20ANY%20%22C%22%2C%22Si%22%2C%22Ge%22%2C%22Sn%22%20AND%20NOT%20elements%20HAS%20%22Pb%22%20AND%20elements%20LENGTH%203}{7,428} ({7,428}) & \href{http://optimade.openmaterialsdb.se/v1/structures?filter=nelements>=1}{68,566} \\
OQMD & \href{http://oqmd.org/optimade/v1/structures?filter=elements%20HAS%20ANY%20%22C%22%2C%22Si%22%2C%22Ge%22%2C%22Sn%22%2C%22Pb%22}{261,400} ({153,113}) & \href{http://oqmd.org/optimade/v1/structures?filter=elements%20HAS%20ANY%20%22C%22%2C%22Si%22%2C%22Ge%22%2C%22Sn%22%2C%22Pb%22%20AND%20nelements%3D2}{15,375} ({11,011}) & \href{http://oqmd.org/optimade/v1/structures?filter=elements%20HAS%20ANY%20%22C%22%2C%22Si%22%2C%22Ge%22%2C%22Sn%22%20AND%20NOT%20elements%20HAS%20%22Pb%22%20AND%20elements%20LENGTH%203}{81,673} ({70,252}) & \href{http://oqmd.org/optimade/v1/structures?filter=nelements>=1}{1,226,781} \\
TCOD & \href{https://www.crystallography.net/tcod/optimade/v1/structures?filter=elements%20HAS%20ANY%20%22C%22%2C%22Si%22%2C%22Ge%22%2C%22Sn%22%2C%22Pb%22}{7,161} ({2,631}) & \href{https://www.crystallography.net/tcod/optimade/v1/structures?filter=elements%20HAS%20ANY%20%22C%22%2C%22Si%22%2C%22Ge%22%2C%22Sn%22%2C%22Pb%22%20AND%20nelements%3D2}{296} ({296}) & \href{https://www.crystallography.net/tcod/optimade/v1/structures?filter=elements%20HAS%20ANY%20%22C%22%2C%22Si%22%2C%22Ge%22%2C%22Sn%22%20AND%20NOT%20elements%20HAS%20%22Pb%22%20AND%20elements%20LENGTH%203}{662} ({660}) & \href{https://www.crystallography.net/tcod/optimade/v1/structures?filter=nelements>=1}{7,452} \\
2DMatpedia & \href{http://optimade.2dmatpedia.org/v1/structures?filter=elements%20HAS%20ANY%20%22C%22%2C%22Si%22%2C%22Ge%22%2C%22Sn%22%2C%22Pb%22}{1,172} & \href{http://optimade.2dmatpedia.org/v1/structures?filter=elements%20HAS%20ANY%20%22C%22%2C%22Si%22%2C%22Ge%22%2C%22Sn%22%2C%22Pb%22%20AND%20nelements%3D2}{739} & \href{http://optimade.2dmatpedia.org/v1/structures?filter=elements%20HAS%20ANY%20%22C%22%2C%22Si%22%2C%22Ge%22%2C%22Sn%22%20AND%20NOT%20elements%20HAS%20%22Pb%22%20AND%20elements%20LENGTH%203}{255} & \href{http://optimade.2dmatpedia.org/v1/structures?filter=nelements>=1}{6,351} \\
\end{tabular}
}
\caption{The table from Ref.~\cite{OPTIMADE2021} recreated in March 2024, with new providers. The final column indicates the total number of structures served by each OPTIMADE API.  Providers that serve multiple databases are indicated with $\star$. Results for Materials Cloud and Materials Cloud Archive have been aggregated under the same title. The corresponding values from the 2021 paper \cite{OPTIMADE2021} are provided in brackets, where appropriate.}
\label{tab:table_redux}
\end{table*}

An important benefit of a universal API format such as OPTIMADE is the ability to simultaneously request and unify results from different databases.
While the key features of the databases are highlighted under the subsections dedicated to the respective providers in Sec.~\ref{sec:UpdateOnDatabases}, a summarizing list of the implementations tested and confirmed to support the OPTIMADE API is shown in Table~\ref{tab:table_redux}.
These databases are all openly accessible and provide users with a broad range of materials classes, applications and modalities.

The table shows the return from three requests that explore materials that contain at least one element from Group 14 ({\bf N$_{1}$}), and then constrains the search to cover only binary materials ({\bf N$_{2}$}), and only ternary materials without toxic lead ({\bf N$_{3}$}):
\begin{description}
 \item[N$_{1}$]\;\colorbox{boxgray}{\parbox{0.43\textwidth}{\texttt{{\color{darkorchid}/v1/structures?filter=\color{RoyalBlue}elements
  \color{dodgerblue}HAS ANY
  \color{RoyalBlue}"C", "Si","Ge","Sn","Pb"}}}}
 \item[N$_{2}$]\;\colorbox{boxgray}{\parbox{0.43\textwidth}{\texttt{{\color{darkorchid}/v1/structures?filter=\color{RoyalBlue}elements
  \color{dodgerblue}HAS ANY
  \color{RoyalBlue}"C", "Si","Ge","Sn","Pb"
  \color{dodgerblue}AND
  \color{RoyalBlue}nelements=2}}}}
 \item[N$_{3}$]\;\colorbox{boxgray}{\parbox{0.43\textwidth}{\texttt{\color{darkorchid}/v1/structures?filter=\color{RoyalBlue}elements
  \color{dodgerblue}HAS ANY
  \color{RoyalBlue}"C", "Si","Ge","Sn"
  \color{dodgerblue}AND NOT
  \color{RoyalBlue}elements
  \color{dodgerblue}HAS
  \color{RoyalBlue}"Pb"
  \color{dodgerblue}AND
  \color{RoyalBlue}elements
  \color{dodgerblue}LENGTH
  \color{RoyalBlue}3}}}
\end{description}
These queries directly duplicate the three detailed in the 2021 OPTIMADE paper~\cite{OPTIMADE2021}. The ability to repeat the query attests to how the OPTIMADE API helps with reproducibility in research. In addition, we provide the total number of structures served by each OPTIMADE API in the {\bf N$_\text{tot}$} column.

Comparison of this table to that in Ref.~\cite{OPTIMADE2021} bears witness to the growth and impact of OPTIMADE.
We see several additional providers (Alexandria, BioExcel, CMR, JARVIS, MPOD, MPDD and 2DMatpedia) that now support OPTIMADE, and several new databases hosted by pre-existing providers.
Furthermore, among the databases that did support OPTIMADE in 2021, there has been an impressive growth in the volume of returned data, reflecting their continued efforts to assimilate further data.


\section{Application of OPTIMADE to real-life problems}\label{sec:ApplicationOfOPTIMADE}

A key goal for the OPTIMADE API is for it to act as an enabling technology for materials discovery, design and other new research avenues.
Feedback from users is crucial to motivate the future development of the API.
Therefore, in the following sections, we spotlight several use cases of the application of the OPTIMADE API to real-life systems, firstly in Section~\ref{sec:MLApplications} by supplying data for machine learning, and secondly in Section~\ref{sec:DataProvision} by providing data for screening and other studies. We highlight examples which benefit from access to the wealth of data available in large databases (e.g., the hard-coating alloys database discussed immediately below), and examples that benefit from access to specialist data available only in the small and focused databases (e.g., Section~\ref{sec:HighEntropyAlloys}).

Furthermore, there has been additional use of OPTIMADE in the literature: firstly how OPTIMADE has been a central tool to access materials data for materials discovery, secondly as a template for materials data curation and access, and thirdly through online web-based interfaces:

\begin{description}
    \item[Discovery] The hard-coating alloys database (HADB)~\cite{HADB_2023} exploited the OPTIMADE API to rapidly and easily provide the browser-based graphical web interface as well as a RESTful API. A second application of the OPTIMADE API was to query and retrieve an unprecedented volume of data to train an attention-based crystal graph convolutional neural network to accurately predict the formation energy, total energy, bandgap, and Fermi energy of a broad range of crystals~\cite{CGCNN_2022}. Finally, OPTIMADE has found application in materials discovery, where it was used in Ref.~\cite{Henkel2023} to assess the novelty of a predicted structures in a high-throughput study on quaternary mixed metal chalcohalide perovskites using the \texttt{optimade-python-tools} client~\cite{Evans2021}. As these structures were ingested into an OPTIMADE-compliant database, in this case NOMAD~\cite{NOMAD_2023}, any future OPTIMADE queries for novelty in this chemical space will yield the results of this study.
 \item[Template] Development of the OPTIMADE API has motivated and guided data access in other ongoing projects. For example, firstly OPTIMADE API collaborates with, and is being used in, the development of the FAIRmat metadata, dictionaries, and materials ontology~\cite{NOMAD_fair2022}. The inclusion in other community efforts reflects the maturity and uptake of the OPTIMADE API. A second example is the BIG-MAP Project, where the consortium plans to use the OPTIMADE API to guide the access of the data gathered in the Battery Interface Genome~\cite{BIG-MAP_2021}.
 \item[Interfaces and integrations] The MarketPlace Project~\cite{MarketPlace} has integrated the OPTIMADE Gateway (Section~\ref{sec:Gateway}) into the platform, which will make it possible to perform OPTIMADE queries through its global search functionality. OntoTrans~\cite{OntoTrans} has developed the Open Translation Environment to perform ontology-driven data pipelines to retrieve, parse, map, and transform data. As part of the project, an OPTIMADE plugin has been developed for the system, making it possible to request and digest OPTIMADE resources. The next steps include semantic mappings for the OPTIMADE data models for true semantic data interoperability.
\end{description}

\subsection{Machine learning}\label{sec:MLApplications}

Machine learning is a promising tool that is already having a significant impact in the materials sciences. Machine learning starts from already computed data about a system and trains a model to capture trends. The machine learning model can then make predictions and design materials quicker and more cost effectively than performing additional experiments.

Machine learning relies on having a pool of historical data available. This is where the OPTIMADE API offers a significant boost, by opening access to a wide range of materials databases that hold complementary data. We highlight the help offered by the OPTIMADE API to machine learning with two case studies.

\subsubsection{High-entropy alloys}\label{sec:HighEntropyAlloys}

High-entropy alloys are comprised of roughly equal parts of five or more elements. This endows the alloy with a high entropy of mixing, which in turn delivers excellent high-temperature properties, such as strength-to-weight ratio, corrosion resistance, and fracture resistance. These favourable properties have driven an acceleration in research into high-entropy alloys over the last decade, but this means that there is still relatively little historical data available for methods such as machine learning.

We use the OPTIMADE API to retrieve high-entropy alloy materials using the filter query
\begin{center}
\colorbox{boxgray}{\parbox{0.47\textwidth}{
{\texttt{%
\color{RoyalBlue}elements
\color{dodgerblue}HAS ANY
\color{RoyalBlue}"W","Al","Cd","Zn"
\color{dodgerblue}AND NOT\\
\color{RoyalBlue}elements
\color{dodgerblue}HAS ANY
\color{RoyalBlue}"B","Cl","F","H","N","O","S"
\color{dodgerblue}AND\\
\color{RoyalBlue}nelements$>$=5
}}}}
\end{center}
from three different providers (P1, P2, P3).

The complete dataset obtained using OPTIMADE is split into training and testing sets (4:1 ratio) while ensuring that ratio of entries from each provider is the same in both training and testing sets. The training set is used to train the ``combined'' model M-C. Data points from the providers P1, P2, and P3 that appear in the training set are used to train the models M-P1, M-P2, M-P3, respectively. The predictive power of the models is assessed by calculating the $R^2$ on the same test set.

\begin{figure}
 \centering
 \includegraphics[width=0.9\columnwidth, trim=0 0 100 0, clip]{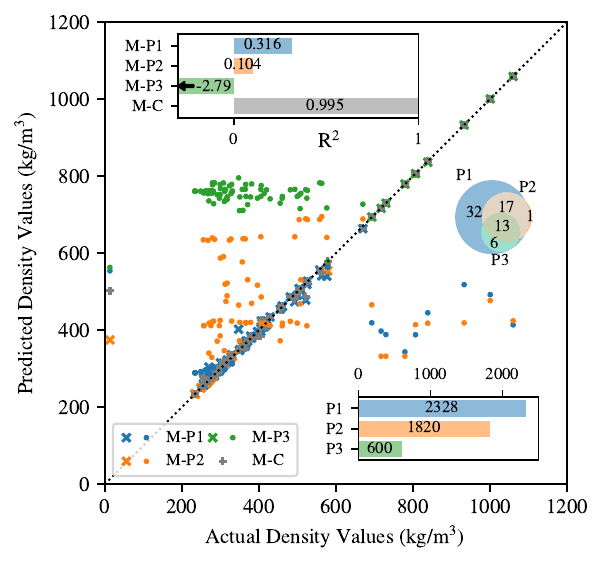}
 \caption{Scatter plot showing the comparison between actual density values and those predicted by the models trained on data obtained using OPTIMADE (M-P1, M-P2, M-P3 for three providers, and M-C trained on all data). For a particular provider, the $\times$ symbols are those points validated against blind data from the same provider, the {\textbullet} symbols those validated against blind data from a different provider. The top inset shows the $R^2$ values for each of the models. The bottom inset shows the number of entries returned by each provider. The inset on the centre-right shows the Venn diagram of unique elements in the entries returned by each of the three providers.}
 \label{fig:HighEntropyMachineLearning}
\end{figure}

We choose Random Forest Regressor (with default parameters, from the \texttt{scikit-learn} 1.2 package in Python~\cite{scikit-learn}) to construct the machine learning models for our example. Standard structural entries of the OPTIMADE specification, \optimadeprop{species\_at\_sites} and \optimadeprop{lattice\_vectors}, are used to construct vectors codifying the composition of each material and also to calculate the density of each material. The `composition vectors', described above, are used as input to machine learning models that are trained to predict the densities (output). Models that are trained on data from only one provider (M-P1, M-P2, and M-P3) perform poorly when tested on data from all the providers ($R^2=0.316$, $0.104$, $-2.79$ for M-P1, M-P2, and M-P3 respectively). Meanwhile, the ``combined'' model that is trained on data from all providers (M-C) performs very well ($R^2 = 0.995$). A comparison of the $R^2$ values is shown in the top-left of Fig. \ref{fig:HighEntropyMachineLearning}.

We can get a better insight into the benefits of leveraging data from multiple providers by looking at comparison of actual density values and those predicted by the machine learning models for small random sampling of materials (scatter plot in Fig. \ref{fig:HighEntropyMachineLearning}). For models trained only on data from a single provider (M-P1, M-P2, and M-P3), the prediction is quite accurate when tested on data from the same provider (indicted by `cross' markers). However, most of the predictive power is lost when tested on data from another provider (`dot' markers). This explains their poor $R^2$ values. Meanwhile, the model which is trained on data from all providers (M-C) retains its predictive power when tested on data from all the providers. The number of materials returned by each provider is shown in the bar-graph on the bottom right of Fig. \ref{fig:HighEntropyMachineLearning}. A Venn diagram of unique elements that appear in the materials from each provider are shown in the centre right in Fig. \ref{fig:HighEntropyMachineLearning}. Therefore, the OPTIMADE API offers the significant benefit to merge the information from the datasets together.

\subsubsection{Materials discovery and accelerated design}\label{sec:MaterialsDiscovery}

Recent advances in AI-driven materials discovery have created an abundance of hypothetical crystal structures that are expected to be stable~\cite{Wang2021b, Chen2022b, Ye2022, Goodall2022, Law2023, Merchant2023}. New datasets targeted towards materials discovery have been ingested and made available as OPTIMADE APIs within the \emph{odbx} provider~\cite{Evans2020, Harper2020}. Typically, these datasets would only be explored by other materials discovery specialists, at least until a more established DFT database ran them through their pipelines. With OPTIMADE, this dissemination process can be automated and greatly accelerated. Anyone can register as a provider and have their novel crystal structures appear in searches by other data-driven applications, such as X-ray diffraction phase identification to be discussed below. As OPTIMADE is not limited to purely theoretical crystal structures, any future experimental confirmation of a structure could also be served through OPTIMADE and used to further improve generative models for materials discovery.

Another possible application is to repeat previous high-throughput materials design campaigns on the wider set of structures now available through OPTIMADE. As each structure is time-stamped, active or ongoing workflows can be implemented to constantly monitor and screen new crystal structures against the search criteria of the campaign, to avoid having to redo such searches from scratch. Structure-based property prediction models, e.g., MODNet (for small structure-property datasets)~\cite{DeBreuck2021, DeBreuck2021a, Liu2022} or graph-based models (where larger structure-property datasets are available)~\cite{Chen2019, Xie2017, Choudhary2021a}, can be leveraged to sift through huge swathes of available crystal structures, with the results being used to prioritise future calculations, attempts at synthesis, characterisation experiments or model retraining, by selecting for structures with combinations of particular target properties.

\subsection{Data provision}\label{sec:DataProvision}

The OPTIMADE API has also been used to provide data for a rich variety of other scientific analysis approaches, below we highlight three projects that take advantage of the comprehensive range of data offered by OPTIMADE.

\subsubsection{OPTIMADE client: a web-based GUI to find and import structures}
\label{sec:optimade-client}

\begin{figure*}
 \centering
 \includegraphics[width=0.9\textwidth]{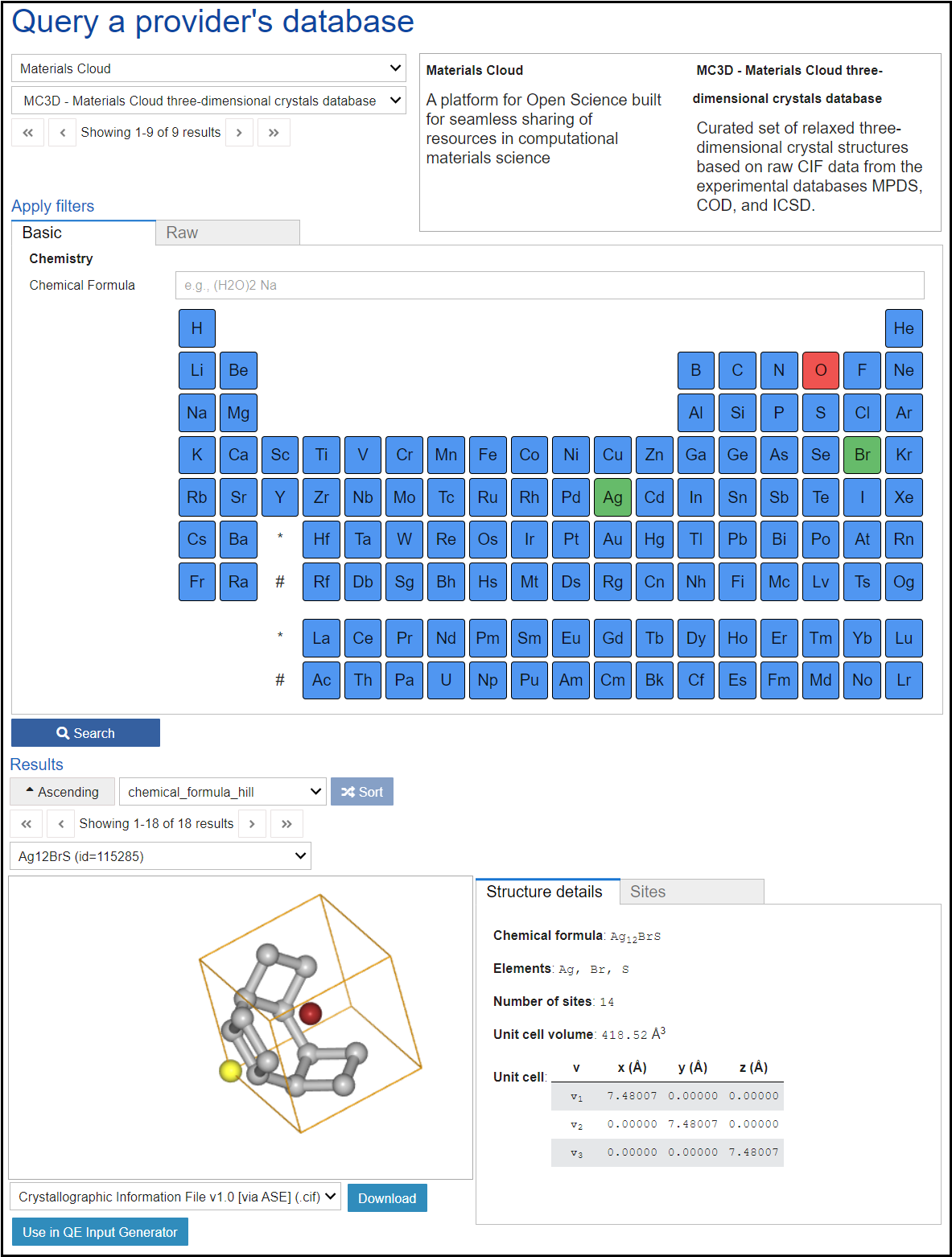}
 \caption{The search interface of the standalone OPTIMADE client. The user can select any OPTIMADE-compliant database and make a query based on filters specified through GUI elements.}
 \label{fig:optimadeclient}
\end{figure*}

The primary input to a first-principles materials calculation is the structure of the system. Experimental or computational crystal databases are commonly used as sources for the input structures of first-principles simulation software. To find the target structure in these databases, a query with filter conditions needs to be prepared, and then the structure needs to be downloaded, inspected, possibly converted into a different format, and finally used in simulations.

To facilitate this goal, Materials Cloud~\cite{MaterialsCloud} provides the OPTIMADE client, a web application to perform the structure search task via a unified and user-friendly GUI, empowering users to not only generate and execute complex OPTIMADE queries, but also to provide immediate graphical access and visualisation of the resulting structures. The OPTIMADE client can be embedded in other applications or used as a standalone tool, which is hosted on Materials Cloud at \href{https://optimadeclient.materialscloud.io}{optimadeclient.materialscloud.io}. The filtering section of the GUI is shown in Fig.~\ref{fig:optimadeclient}. A dropdown (at the top) is provided to select any of the known and automatically discovered OPTIMADE database providers. A periodic table widget allows users to select which elements need to be included (green) or excluded (red) in the compounds; additional filtering tools are also provided, such as for the number of elements and of sites, and for the dimensionality. An OPTIMADE query string is then produced (which can be optionally manually modified). After the search button is clicked, the OPTIMADE query is sent to the selected database provider. The results are curated and shown in the results widget. Here, the structures are visualised and can be downloaded.

Materials Cloud also provides various tools that leverage the power of the OPTIMADE client. One example is the Quantum ESPRESSO input generator, available as a tool at \href{https://www.materialscloud.org/work/tools/qeinputgenerator}{materialscloud.org/work/tools/qeinputgenerator}. A structure can be sent directly to this tool using a button in the OPTIMADE client (shown at the bottom of Fig.~\ref{fig:optimadeclient}). The Quantum ESPRESSO input generator enables any user to obtain a working input file for the Quantum ESPRESSO DFT code~\cite{Giannozzi2009,Giannozzi2017}, including an automated selection of all numerical parameters, by just specifying a crystal structure (either by uploading it, or by selecting it from OPTIMADE). A second example is the Quantum ESPRESSO app (\href{https://aiidalab-qe.readthedocs.io}{https://aiidalab-qe.readthedocs.io}) developed within the AiiDAlab platform~\cite{aiidalab}. It allows users to run complex computational workflows from the web browser, using straightforward graphical user interfaces for structure selection (including via OPTIMADE), parameter selection and inspection of the results.

\subsubsection{Automatic phase identification from X-ray diffraction}

The Xerus (X-ray Estimation and Refinement Using Similarity)~\cite{BaptistadeCastro, Toby2013, Iwasaki2017, Ozaki2020} software package implements procedures to refine and screen measured X-ray diffraction patterns of inorganic crystals against databases of crystal structures reported in the literature and beyond.
By querying for all possible structures in a given chemical space, Xerus excels at multiphase fits and performs competitively against more specialised and compute-intensive models constructed with machine learning.

Xerus uses a straightforward OPTIMADE interface to connect to the multiple databases hosted by members of the OPTIMADE consortia. The dynamic OPTIMADE providers list allows new databases to be automatically included in Xerus search results. Additional filtering parameters can be used to refine the searches towards materials stable (or predicted to be stable) at the experimental conditions (e.g., low temperature or high pressure).

\subsubsection{Workflows for automated and simultaneous queries of different databases}

\begin{figure*}
 \centering
 \includegraphics[width=0.9\textwidth]{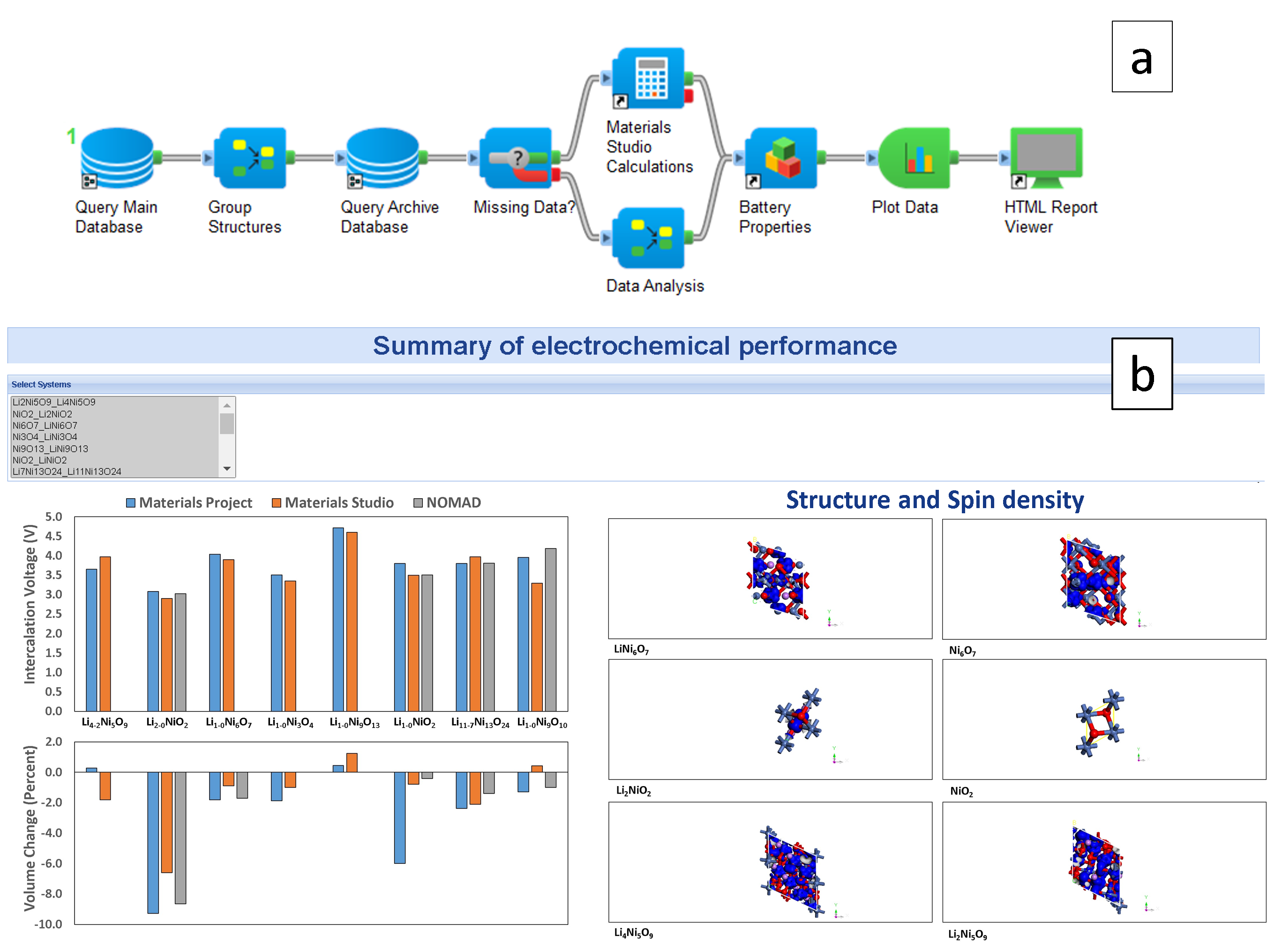}
 \caption{(a) Workflow architecture for automated simulations using BIOVIA Pipeline Pilot. The workflow queries the structure and energies from the main and archive databases, analyses the available data, and performs CASTEP calculations for the missing data. (b)  Screenshot of the app that reports the corresponding electrochemical properties, such as Intercalation Voltage and Crystal structures.}
 \label{fig:3DS}
\end{figure*}

\textbf{BIOVIA Pipeline Pilot}~\cite{PipelinePilot2022} is a scientific workflow system that allows users to automate calculations and visualise and report research results by graphically composing a protocol from hundreds of different configurable components. As a technology demonstrator, we investigate the intercalation voltage of a series of cathode materials with the workflow in Figure~\ref{fig:3DS}(a). It uses OPTIMADE to adopt the available structures and energetics from different providers and performs complementary calculations using the CASTEP DFT code~\cite{Clark2009}.

Results are shown in Fig.~\ref{fig:3DS}(b), where we plot the intercalation voltage of Li-Ni-O materials from the extracted VASP~\cite{Kresse1996} data in Materials Project and NOMAD and compare predictions to those from CASTEP. The comparability of results illustrates the functionality of the workflow. Since databases contain different structures, the OPTIMADE API facilitates the process of materials investigations by aggregating the query of all of them.

\textbf{MatCloud}~\cite{MatCloudURL} is a cloud-based integrated high-throughput computational materials infrastructure, which is directly connected to computing clusters and material property databases~\cite{Yang2018,MatCloud2018}. Users worldwide can visually design structures, create and run simulation jobs through workflows, and retrieve crystal structures from multiple databases using OPTIMADE; all the user needs is a web browser. MatCloud provides a Graphical User Interface (GUI)-based environment for users to intuitively create, enact and monitor a workflow. The MatCloud workflow system includes a front-end workflow designer and a back-end workflow engine, and supports the creation of workflows by a drag-and-drop approach.    

\begin{figure*}
    \centering
    \includegraphics[width=1\linewidth]{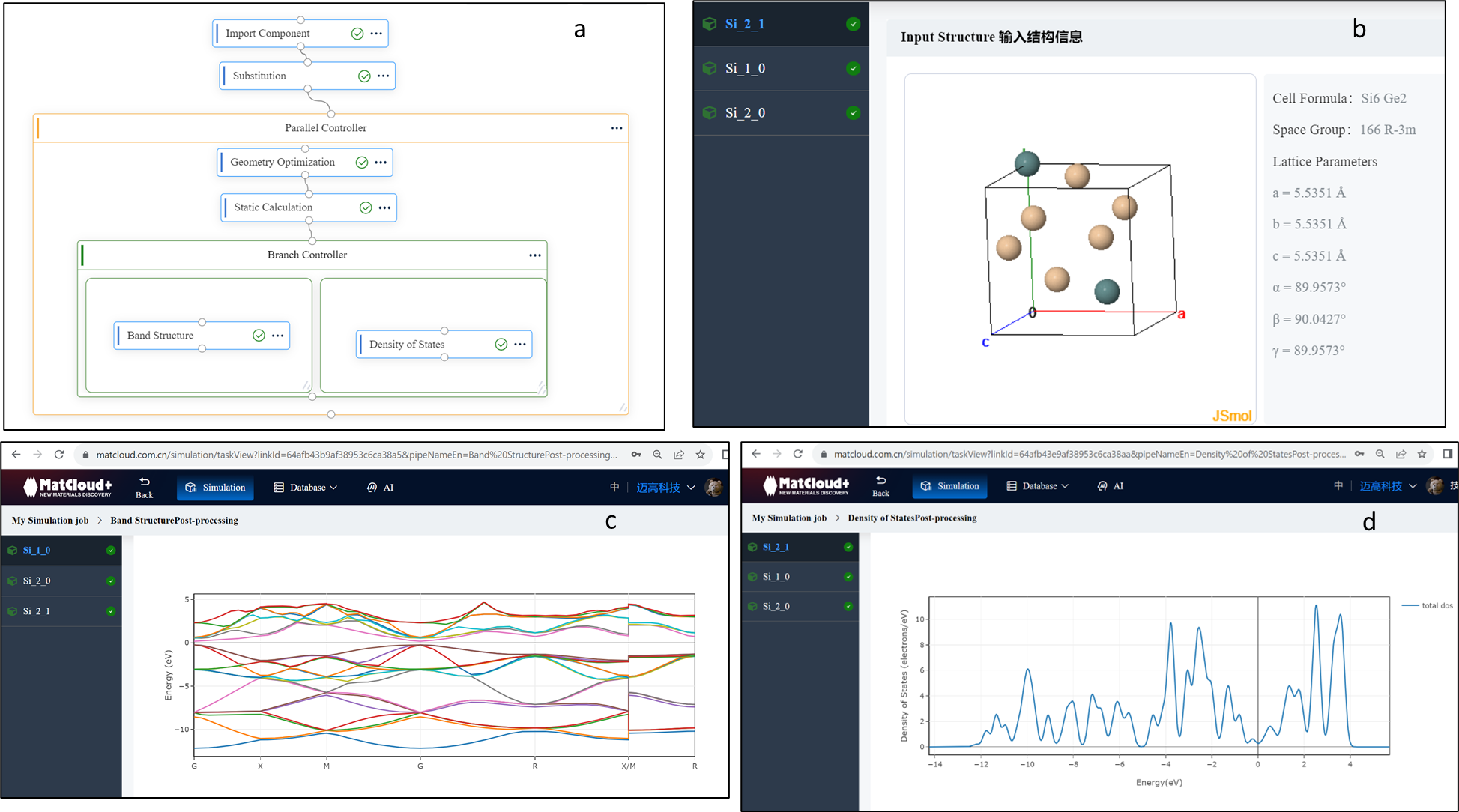}
    \caption{(a) MatCloud workflow of crystal structure retrieval through OPTIMADE, and job setup. (b) Visualisation of one of the Si-Ge structures. (c) Computed band structure for the Si-Ge structure. (d) Total DOS of the Si-Ge structure.}
    \label{fig:matcloud-sige}
\end{figure*}

Fig.~\ref{fig:matcloud-sige}a shows a workflow that retrieves a Si$_8$ crystal structure from the MPDS database through OPTIMADE, and replaces two Si atoms with Ge to produce structures in the Si-Ge chemical space (Fig.~\ref{fig:matcloud-sige}b). The band structure and density of states (DOS) are then simulated respectively over the chemical space in a high-throughput manner. Figs.~\ref{fig:matcloud-sige}c and \ref{fig:matcloud-sige}d show the visualisation results for the band structure and total DOS of the structures.       

\section{Future of OPTIMADE}\label{sec:FutureOfOPTIMADE}

The rapid expansion of materials databases, the use cases presented, and the adoption of machine learning motivate the continued work on the OPTIMADE API. We describe below firstly the ongoing tutorials that introduce new users to OPTIMADE, secondly the workshops that enable the development and promulgation of the OPTIMADE API, and finally the features currently under development coming out of those recent workshops.

\subsection{Workshops \& tutorials}

The OPTIMADE consortium originated from the workshop ``Open Databases Integration for Materials Design'', held at the Lorentz Center in Leiden, Netherlands in October 2016. There were follow-up workshops held at CECAM in Lausanne, Switzerland annually from 2018 to 2023, with events since 2020 also supporting remote attendees. The workshops in 2022 and 2023 were accompanied by on-site tutorials during the first two days. There was also a partner workshop, ``Ontologies for machine learning driven materials design'' held at Linköping University, Sweden in 2021. Going forward, the OPTIMADE API will undergo continued development through further annual workshops and publicly advertised monthly video calls. There are continual efforts to reach out to new databases to help accelerate their adoption of the format.

Several tutorial exercises have been developed and delivered at a variety of conferences and topical workshops, outside of the OPTIMADE annual meetings. Eight different exercises developed by the community are now hosted on the GitHub repository \href{https://github.com/Materials-Consortia/optimade-tutorial-exercises}{Materials-Consortia/optimade-tutorial-exercises}, ranging from the basics of the OPTIMADE filter syntax and URL structure, to machine-learning pipelines operating on database-specific properties, all the way up to hosting an OPTIMADE API for a new database.

\subsection{Upcoming features for future OPTIMADE releases}

The ongoing real-life use cases of the OPTIMADE API have identified a series of opportunities to extend the API and make it more applicable to a wide range of materials systems. The features currently under development include:
\begin{description}

  \item[\optimadeendp{/trajectories} endpoint]So far, OPTIMADE can only be used to describe static structures. We are, however, working to expand the OPTIMADE specification, so that OPTIMADE can be used for sharing trajectory data as well. Such data can originate from structural optimisations, or from Monte Carlo and molecular dynamics simulations. These trajectories could be used as a starting point for new simulations, to train machine learning potentials or to extract dynamical properties.
 \item[\optimadeendp{/collections} endpoint]It has been suggested several times whether it would be possible to have a method to create groups of entries. For example, all the structures that were used to generate a certain machine-learning potential, or all the structures that pertain to a certain research project. We therefore plan to introduce a \optimadeendp{/collections} endpoint, which would contain metadata for the collection as a whole, as well as references to all the individual entries that belong to this collection.

 \item[Ontologies and semantics]Building on the expanded property definitions outlined above, semantic mappings can be developed to existing and in-development materials and crystallography domain ontologies. For example, properties of an OPTIMADE structure (such as the specification of periodicity or types of disordered occupation) can be mapped into concepts in the crystallography domain ontology~\cite{crystallographydomainontology} under development within the Elementary Multiperspective Material Ontology (EMMO) ecosystem.~\cite{EMMO} This ontology is being created as a collaborative effort that includes members of the OPTIMADE consortium. These mappings of OPTIMADE properties into ontologies facilitate the alignment with other semantic data interoperability frameworks. Examples of such use include the ability to reference properties standardized by OPTIMADE in, e.g., future EMMO-aligned domain ontologies and giving access to property data via ontology-based GraphQL server generation.~\cite{Li2024}

 \item[SMILES property]So far, OPTIMADE has mostly been designed around describing inorganic crystal structures. There are, however, plenty of materials that are (at least partially) comprised of organic constituents. To make it easier to find and select these, we plan to implement a SMILES field~\cite{Weininger1988} and a SMARTS filter~\cite{smarts} for the organic parts of a structure.

 \item[Biomolecular fields]One major use case of the upcoming \optimadeendp{/trajectories} endpoint is to handle biomolecular structures, which can only be described statistically as a trajectory with multiple configurations. The OPTIMADE specification will therefore be extended to standardise the various dynamical fields and order parameters that are used to describe biomolecular structures. This covers fields that are typically stored in PDB files such as insertion codes, Chain IDs and sequence information of proteins, DNA and RNA. 

 \item[Proliferation of domain-specific namespaces]As mentioned in \ref{sec:OPTIMADEimprovements}, we have already added a mechanism for multiple providers to collaborate on subsets of shared property definitions. It is hoped that many such namespaces will arise to improve coverage of OPTIMADE data across various domains of atomistic science, potentially including the cheminformatics and biomolecular-focused fields above. Other immediate targets include stability information from density-functional theory calculations and magnetic properties.

 \item[Large language models (LLMs)] LLMs have emerged as an exciting frontier for data science and machine learning \cite{Jablonka2023b, Brown2020}. We are now considering two uses of LLMs within OPTIMADE. First, a large language model can help a non-expert formulate a query for OPTIMADE, for example the query in section 2 could be found by requesting ``tell me the structure of an oxide of silicon''; this could be readily performed by providing the LLM with the specification text or the machine-readable schemas, then constructing the relevant query with in-context learning \cite{Brown2020}. A second use is to pass the large language model either textual data or a scan of a page of historical data, which can then be readily parsed to extract out relevant numbers for an OPTIMADE database. The value provided by OPTIMADE here is to give a machine-actionable scaffold that an LLM can be validated and evaluated against, in such a way that the data produced is automatically compatible with other initiatives.

\end{description}

\section{Conclusion}\label{sec:Conclusion}

The OPTIMADE API provides users with easy access to many of the world leading materials databases. Since the initial release, the OPTIMADE API has not only been adopted by scientists as a tool to drive innovation, but furthermore served as a template for data curation. In this paper, we have provided use cases for how the breadth of data made available through OPTIMADE enables discovery in both academia and industry.
The development of the OPTIMADE API has continued apace. Major new support for enhanced property definitions and partial data formats have recently been added and will underpin future work on trajectories and biomolecular data. The concept that sub-consortia of databases are responsible for the definition of new sets of shared properties will accelerate the extension of the OPTIMADE API to other disciplines and fields.

Through monthly meetings, and with the continuing support of CECAM, the developers are continuing to extend the range of properties accessible via OPTIMADE APIs.
Plans to both expand the format to cover challenges arising from dealing with molecular dynamics data, and continued outreach to support the adoption of the API by additional databases, will further expand the range of scientific use cases that OPTIMADE enables.
Increasingly, use cases will take advantage of the unique advantages OPTIMADE has to offer; namely the robust and straightforward aggregation and data unification from the multitude of growing and federated data sources.

\section*{Acknowledgments}

The authors acknowledge support from CECAM in Lausanne (Switzerland) and the Lorentz Center in Leiden (Netherlands) for hosting OPTIMADE workshops; and for the partner workshop at Linköping University in 2021, support from Psi-k, NCCR MARVEL (a National Centre of Competence in Research, funded by the Swiss National Science Foundation, grant No. 205602), and the Swedish e-Science Research Centre (SeRC). 

G.J.C. would like to acknowledge financial support from the Royal Society.
M.L.E. thanks the BEWARE scheme of the Wallonia-Brussels Federation for funding under the European Commission's Marie Curie-Skłodowska Action (COFUND 847587).
CT thanks NSF (DMR-2219788). M.S. and C.D. acknowledge support from the German Research Foundation (DFG) through the NFDI consortium FAIRmat (project 460197019).
K.E., J.Y., S.B., N.M., and G.P. acknowledge funding by the NCCR MARVEL (a National Centre of Competence in Research, funded by the Swiss National Science Foundation, grant No. 205602). G.P. acknowledge funding by the Open Research Data Program of the ETH Board (project ``PREMISE'': Open and Reproducible Materials Science Research).
R.A. and O.B.A. acknowledges funding from Vetenskapsrådet Grant No. 2020-05402 and the Swedish e-Science Research Centre.
A.M., A.V. and S.G. acknowledge funding under the Programme "University Excellence Initiatives" of the Ministry of
Education, Science and Sports of the Republic of Lithuania (Measure No. 12-001-01-01-01 "Improving the Research
and Study Environment").
A.J.M. acknowledges support from EPSRC via CCP-NC (EP/T026642/1), CCP9 (EP/T026375/1), UKCP (EP/P022561/1) and Baskeville (EP/T022221/1).
The work by J.S, D. W. and K.A.P were supported by the U.S. Department of Energy, Office of Science, Office of Basic Energy Sciences, Materials Sciences and Engineering Division under contract no. DE-AC02-05-CH11231; the Materials Project program (KC23MP). 
D.B, A.H. and M.O. acknowledge funding by the European Union's Horizon 2020 programme under grant agreements No. 675728 (BioExcel), No. 823830 (BioExcel-2), No. 720270 (Human Brain Project SGA1), No. 785907 (Human Brain Project SGA2), No. 945539 (Human Brain Project SGA3) and by the Horizon Europe Programme under No. 101093290 (BioExcel-3) and No. 101094651 (Molecular Dynamics Data Bank). K.C. acknowledges funding support from the CHIPS Metrology Program, part of CHIPS for America, National Institute of Standards and Technology, U.S. Department of Commerce. X.Y. acknowledges the funding support from National Natural Science Foundation of China (NSFC) under the grant No. 62376258.

\section*{Author information}

These authors were active developers and reviewers of the specification: Johan Bergsma, Matthew L. Evans, and Andrius Merkys.

These authors were active developers of implementations for database providers and/or made contributions to specification: Johan Bergsma, Matthew L. Evans, Andrius Merkys, Oskar B.\ Andersson, Casper W. Andersen, Daniel Beltrán, Evgeny Blokhin, Tara Boland, Rubén Castañeda Balderas, Kamal Choudary, Alberto Díaz Díaz, Rodrigo Domínguez García, Hagen Eckert, Kristjan Eimre, María Elena Fuentes Montero, Adam M. Krajewski, Jens Jørgen Mortensen, José Manuel Nápoles Duarte, Jacob Pietryga, Ji Qi, Felipe de Jesús Trejo Carrillo, Antanas Vaitkus, Jusong Yu, Adam Zettel.

These authors primarily contributed to the paper (along with all other authors): Pedro Baptista de Castro, Johan Carlsson, Tiago F. T. Cerqueira, Simon Divilov, Hamidreza Hajiyani, Felix Hanke, Kevin Jose, Corey Oses, Janosh Riebesell, Jonathan Schmidt, Donald Winston, Christen Xie, Xiaoyu Yang.

These authors managed individual databases that have implemented the OPTIMADE API: Sara Bonella, Silvana Botti, Stefano Curtarolo, Claudia Draxl, Luis Edmundo Fuentes Cobas, Adam Hospital, Zi-Kui Liu, Miguel Marques, Nicola Marzari, Andrew Morris, Shyue Ping Ong, Modesto Orozco, Kristin A. Persson, Kristian Thygesen, Chris Wolverton.

These authors are organisers of the OPTIMADE API and are also senior developers who contributed to code and/or to the specification: Rickard Armiento, Gareth J. Conduit, Saulius Gra\v{z}ulis, Giovanni Pizzi, Gian-Marco Rignanese, Markus Scheidgen, Cormac Toher.

\section*{Competing interests}
G.J.C. is a shareholder and Director of Intellegens Ltd.\newline
G.-M.R. is a shareholder and Chief Innovation Officer of Matgenix SRL.\newline
E.B. is a shareholder and Director of Materials Platform for Data Science O{\"U}.

\bibliography{main}

\begin{thebibliography}{148}
\providecommand{\natexlab}[1]{#1}
\providecommand{\url}[1]{\texttt{#1}}
\expandafter\ifx\csname urlstyle\endcsname\relax
  \providecommand{\doi}[1]{doi: #1}\else
  \providecommand{\doi}{doi: \begingroup \urlstyle{rm}\Url}\fi

\bibitem[{A. Calzolari} et~al.(2022){A. Calzolari}, {C. Oses}, {C. Toher}, {M. Esters}, {X. Campilongo}, {S.~P. Stepanoff}, {D.~E. Wolfe}, and {S. Curtarolo}]{aflowhep2022}
{A. Calzolari}, {C. Oses}, {C. Toher}, {M. Esters}, {X. Campilongo}, {S.~P. Stepanoff}, {D.~E. Wolfe}, and {S. Curtarolo}.
\newblock {Plasmonic high-entropy carbides}.
\newblock \emph{Nat.\ Commun.}, 13:\penalty0 5993, 2022.
\newblock \doi{10.1038/s41467-022-33497-1}.

\bibitem[Andersen et~al.(2020)]{OPTIMADE_spec_2020}
C.~W. Andersen et~al.
\newblock {The OPTIMADE Specification (Version 1.0.0)}.
\newblock \emph{Zenodo}, page~1, 2020.
\newblock \doi{10.5281/zenodo.4195051}.

\bibitem[Andersen et~al.(2021{\natexlab{a}})]{OPTIMADE2021}
C.~W. Andersen et~al.
\newblock {OPTIMADE: an API for exchanging materials data}.
\newblock \emph{Scientific Data}, 8:\penalty0 217, 2021{\natexlab{a}}.
\newblock \doi{10.1038/s41597-021-00974-z}.

\bibitem[Andersen et~al.(2021{\natexlab{b}})]{OPTIMADE_spec_2021}
C.~W. Andersen et~al.
\newblock {The OPTIMADE Specification (Version 1.1.0)}.
\newblock \emph{Zenodo}, page~1, 2021{\natexlab{b}}.
\newblock \doi{10.5281/zenodo.4251947}.

\bibitem[Andersen et~al.(2024)]{OPTIMADE_spec_1.2}
C.~W. Andersen et~al.
\newblock {The OPTIMADE Specification (Version 1.2.0)}.
\newblock \url{https://github.com/Materials-Consortia/OPTIMADE/blob/v1.2.0/optimade.rst}, 2024.

\bibitem[Andrio et~al.(2019)Andrio, Hospital, Conejero, Jord{\`{a}}, del Pino, Codo, Soiland-Reyes, Goble, Lezzi, Badia, Orozco, and Gelpi]{Andrio2019}
P.~Andrio, A.~Hospital, J.~Conejero, L.~Jord{\`{a}}, M.~del Pino, L.~Codo, S.~Soiland-Reyes, C.~Goble, D.~Lezzi, R.~Badia, M.~Orozco, and J.~L. Gelpi.
\newblock {{BioExcel Building Blocks}}, a software library for interoperable biomolecular simulation workflows.
\newblock \emph{Scientific Data}, 6:\penalty0 169--179, 2019.
\newblock \doi{10.1038/s41597-019-0177-4}.

\bibitem[Armiento(2020)]{HTTKOMDB}
R.~Armiento.
\newblock Database-{Driven} {High}-{Throughput} {Calculations} and {Machine} {Learning} {Models} for {Materials} {Design}.
\newblock In K.~T. Sch{\"u}tt, S.~Chmiela, O.~A. von Lilienfeld, A.~Tkatchenko, K.~Tsuda, and K.-R. M{\"u}ller, editors, \emph{Machine {Learning} {Meets} {Quantum} {Physics}}, Lecture {Notes} in {Physics}, pages 377--395. Springer International Publishing, Cham, 2020.
\newblock ISBN 978-3-030-40245-7.
\newblock \doi{10.1007/978-3-030-40245-7_17}.

\bibitem[Aykol et~al.(2014)Aykol, Kirklin, and Wolverton]{Aykol_OQMD_Batteries}
M.~Aykol, S.~Kirklin, and C.~Wolverton.
\newblock Thermodynamic aspects of cathode coatings for lithium-ion batteries.
\newblock \emph{Advanced Energy Materials}, 4\penalty0 (17):\penalty0 1400690, 2014.
\newblock ISSN 1614-6832.
\newblock \doi{10.1002/aenm.201400690}.

\bibitem[{Baptista de Castro} et~al.(2022){Baptista de Castro}, Terashima, Esparza~Echevarria, Takeya, and Takano]{BaptistadeCastro}
P.~{Baptista de Castro}, K.~Terashima, M.~G. Esparza~Echevarria, H.~Takeya, and Y.~Takano.
\newblock {{XERUS}}: {{An Open-Source Tool}} for {{Quick XRD Phase Identification}} and {{Refinement Automation}}.
\newblock \emph{Advanced Theory and Simulations}, 5\penalty0 (5):\penalty0 2100588, 2022.
\newblock \doi{10.1002/adts.202100588}.

\bibitem[Baratta et~al.(2023)Baratta, Cimino, Longo, Solina, and Verteramo]{Baratta2023}
A.~Baratta, A.~Cimino, F.~Longo, V.~Solina, and S.~Verteramo.
\newblock {The Impact of ESG Practices in Industry with a Focus on Carbon Emissions: Insights and Future Perspectives}.
\newblock \emph{Sustainability}, 15:\penalty0 6685, 2023.
\newblock \doi{10.3390/su15086685}.

\bibitem[Barjasteh et~al.(2023)Barjasteh, Vossoughi, Bagherzadeh, and Pooshang~Bagheri]{Barjasteh_COD_MOF_2023}
M.~Barjasteh, M.~Vossoughi, M.~Bagherzadeh, and K.~Pooshang~Bagheri.
\newblock {MIL-100(Fe)} a potent adsorbent of {Dacarbazine}: {Experimental} and molecular docking simulation.
\newblock \emph{Chemical Engineering Journal}, 452:\penalty0 138987, 2023.
\newblock ISSN 1385-8947.
\newblock \doi{10.1016/j.cej.2022.138987}.

\bibitem[Beltr{\'a}n et~al.(2023)Beltr{\'a}n, Hospital, Gelp{\'i}, and Orozco]{Beltran2024}
D.~Beltr{\'a}n, A.~Hospital, J.~L. Gelp{\'i}, and M.~Orozco.
\newblock A new paradigm for molecular dynamics databases: the {{COVID-19}} database, the legacy of a titanic community effort.
\newblock \emph{Nucleic Acids Research}, 52\penalty0 (D1):\penalty0 D393--D403, Nov. 2023.
\newblock \doi{10.1093/nar/gkad991}.

\bibitem[Bernstein et~al.(2016)Bernstein, Bollinger, Brown, Gra{\v{z}}ulis, Hester, McMahon, Spadaccini, Westbrook, and Westrip]{cif_2016}
H.~J. Bernstein, J.~C. Bollinger, I.~D. Brown, S.~Gra{\v{z}}ulis, J.~R. Hester, B.~McMahon, N.~Spadaccini, J.~D. Westbrook, and S.~P. Westrip.
\newblock Specification of the crystallographic information file format, version 2.0.
\newblock \emph{Journal of Applied Crystallography}, 49\penalty0 (1):\penalty0 277--284, 2016.
\newblock \doi{10.1107/S1600576715021871}.

\bibitem[Brown et~al.(2020)Brown, Mann, Ryder, Subbiah, Kaplan, Dhariwal, Neelakantan, Shyam, Sastry, Askell, Agarwal, {Herbert-Voss}, Krueger, Henighan, Child, Ramesh, Ziegler, Wu, Winter, Hesse, Chen, Sigler, Litwin, Gray, Chess, Clark, Berner, McCandlish, Radford, Sutskever, and Amodei]{Brown2020}
T.~B. Brown, B.~Mann, N.~Ryder, M.~Subbiah, J.~Kaplan, P.~Dhariwal, A.~Neelakantan, P.~Shyam, G.~Sastry, A.~Askell, S.~Agarwal, A.~{Herbert-Voss}, G.~Krueger, T.~Henighan, R.~Child, A.~Ramesh, D.~M. Ziegler, J.~Wu, C.~Winter, C.~Hesse, M.~Chen, E.~Sigler, M.~Litwin, S.~Gray, B.~Chess, J.~Clark, C.~Berner, S.~McCandlish, A.~Radford, I.~Sutskever, and D.~Amodei.
\newblock Language {{Models}} are {{Few-Shot Learners}}.
\newblock \emph{arXiv}, \penalty0 (arXiv:2005.14165), July 2020.
\newblock \doi{10.48550/arXiv.2005.14165}.

\bibitem[Campi et~al.(2023)Campi, Mounet, Gibertini, Pizzi, and Marzari]{mc2d_2}
D.~Campi, N.~Mounet, M.~Gibertini, G.~Pizzi, and N.~Marzari.
\newblock Expansion of the {Materials Cloud 2D Database}.
\newblock \emph{{ACS} Nano}, 17\penalty0 (12):\penalty0 11268--11278, 2023.
\newblock \doi{10.1021/acsnano.2c11510}.

\bibitem[Castelli et~al.(2021)Castelli, Arismendi-Arrieta, Bhowmik, Cekic-Laskovic, Clark, Dominko, Flores, Flowers, Frederiksen, Friis, Grimaud, Hansen, Hardwick, Hermansson, K{\"{o}}niger, Lauritzen, Cras, Li, Lyonnard, Lorrmann, Marzari, Niedzicki, Pizzi, Rahmanian, Stein, Uhrin, Wenzel, Winter, W{\"{o}}lke, and Vegge]{BIG-MAP_2021}
I.~E. Castelli, D.~J. Arismendi-Arrieta, A.~Bhowmik, I.~Cekic-Laskovic, S.~Clark, R.~Dominko, E.~Flores, J.~Flowers, K.~U. Frederiksen, J.~Friis, A.~Grimaud, K.~V. Hansen, L.~J. Hardwick, K.~Hermansson, L.~K{\"{o}}niger, H.~Lauritzen, F.~L. Cras, H.~Li, S.~Lyonnard, H.~Lorrmann, N.~Marzari, L.~Niedzicki, G.~Pizzi, F.~Rahmanian, H.~Stein, M.~Uhrin, W.~Wenzel, M.~Winter, C.~W{\"{o}}lke, and T.~Vegge.
\newblock Data management plans: the importance of data management in the {BIG-MAP} project.
\newblock \emph{Batteries \& Supercaps}, 4:\penalty0 1803--1812, 2021.
\newblock \doi{10.1002/batt.202100117}.

\bibitem[Chan et~al.(2021)Chan, Hutchison, and Morris]{Chan_CODRingPuckering_2021}
L.~Chan, G.~R. Hutchison, and G.~M. Morris.
\newblock Understanding ring puckering in small molecules and cyclic peptides.
\newblock \emph{Journal of Chemical Information and Modeling}, 61\penalty0 (2):\penalty0 743--755, 2021.
\newblock \doi{10.1021/acs.jcim.0c01144}.

\bibitem[Chen and Ong(2022)]{Chen2022b}
C.~Chen and S.~P. Ong.
\newblock A universal graph deep learning interatomic potential for the periodic table.
\newblock \emph{Nature Computational Science}, 2\penalty0 (11):\penalty0 718--728, 2022.
\newblock \doi{10.1038/s43588-022-00349-3}.

\bibitem[Chen et~al.(2019)Chen, Ye, Zuo, Zheng, and Ong]{Chen2019}
C.~Chen, W.~Ye, Y.~Zuo, C.~Zheng, and S.~P. Ong.
\newblock Graph {{Networks}} as a {{Universal Machine Learning Framework}} for {{Molecules}} and {{Crystals}}.
\newblock \emph{Chemistry of Materials}, 31\penalty0 (9):\penalty0 3564--3572, 2019.
\newblock \doi{10.1021/acs.chemmater.9b01294}.

\bibitem[Choudhary and DeCost(2021)]{Choudhary2021a}
K.~Choudhary and B.~DeCost.
\newblock Atomistic {{Line Graph Neural Network}} for improved materials property predictions.
\newblock \emph{npj Computational Materials}, 7\penalty0 (1):\penalty0 1--8, 2021.
\newblock \doi{10.1038/s41524-021-00650-1}.

\bibitem[Choudhary and Garrity(2022)]{Choudhary_COD_superconductors_2022}
K.~Choudhary and K.~Garrity.
\newblock Designing high-${T}_c$ superconductors with {BCS}-inspired screening, density functional theory, and deep-learning.
\newblock \emph{npj Computational Materials}, 8\penalty0 (1):\penalty0 244, 2022.
\newblock ISSN 2057-3960.
\newblock \doi{10.1038/s41524-022-00933-1}.

\bibitem[Choudhary et~al.(2020)Choudhary, Garrity, Reid, DeCost, Biacchi, Hight~Walker, Trautt, Hattrick-Simpers, Kusne, Centrone, et~al.]{choudhary2020joint}
K.~Choudhary, K.~F. Garrity, A.~C. Reid, B.~DeCost, A.~J. Biacchi, A.~R. Hight~Walker, Z.~Trautt, J.~Hattrick-Simpers, A.~G. Kusne, A.~Centrone, et~al.
\newblock The joint automated repository for various integrated simulations (jarvis) for data-driven materials design.
\newblock \emph{npj computational materials}, 6\penalty0 (1):\penalty0 173, 2020.
\newblock \doi{10.1038/s41524-020-00440-1}.

\bibitem[Clark et~al.(2005)Clark, Segall, Pickard, Hasnip, Probert, Refson, and Payne]{Clark2009}
S.~J. Clark, M.~D. Segall, C.~J. Pickard, P.~J. Hasnip, M.~J. Probert, K.~Refson, and M.~C. Payne.
\newblock First principles methods using {CASTEP}.
\newblock \emph{Zeitschrift fuer Kristallographie}, 220\penalty0 (5-6):\penalty0 567--570, 2005.
\newblock \doi{10.1524/zkri.220.5.567.65075}.

\bibitem[Company(2022)]{ChemicalsReport2022}
T.~B.~R. Company.
\newblock Chemicals global market report 2022, 2022.
\newblock URL \url{https://www.researchandmarkets.com/reports/5598260/chemicals-global-market-report-2022}.

\bibitem[Corporation(2022)]{PipelinePilot2022}
D.~S.~A. Corporation.
\newblock {BIOVIA Pipeline Pilot}.
\newblock \url{https://www.3ds.com/products-services/biovia/products/data-science/pipeline-pilot/}, 2022.

\bibitem[{C}rystallography {D}omain {O}ntology()]{crystallographydomainontology}
{C}rystallography {D}omain {O}ntology.
\newblock {C}rystallography {D}omain {O}ntology.
\newblock \url{https://github.com/emmo-repo/domain-crystallography}, 2019.

\bibitem[Davidsson et~al.(2021)Davidsson, Iv{\'{a}}dy, Armiento, and Abrikosov]{ADAQ}
J.~Davidsson, V.~Iv{\'{a}}dy, R.~Armiento, and I.~A. Abrikosov.
\newblock {ADAQ:} automatic workflows for magneto-optical properties of point defects in semiconductors.
\newblock \emph{Computer Physics Communications}, 269:\penalty0 108091, 2021.
\newblock ISSN 0010-4655.
\newblock \doi{10.1016/j.cpc.2021.108091}.

\bibitem[De~Breuck et~al.(2021{\natexlab{a}})De~Breuck, Evans, and Rignanese]{DeBreuck2021}
P.-P. De~Breuck, M.~L. Evans, and G.-M. Rignanese.
\newblock Robust model benchmarking and bias-imbalance in data-driven materials science: a case study on {{MODNet}}.
\newblock \emph{Journal of Physics: Condensed Matter}, 33\penalty0 (40):\penalty0 404002, 2021{\natexlab{a}}.
\newblock \doi{10.1088/1361-648x/ac1280}.

\bibitem[De~Breuck et~al.(2021{\natexlab{b}})De~Breuck, Hautier, and Rignanese]{DeBreuck2021a}
P.-P. De~Breuck, G.~Hautier, and G.-M. Rignanese.
\newblock Materials property prediction for limited datasets enabled by feature selection and joint learning with {{MODNet}}.
\newblock \emph{npj Computational Materials}, 7\penalty0 (1):\penalty0 1--8, 2021{\natexlab{b}}.
\newblock \doi{10.1038/s41524-021-00552-2}.

\bibitem[Divilov et~al.(2024)Divilov, Eckert, Hicks, Oses, Toher, Friedrich, Esters, Mehl, Zettel, Lederer, Zurek, Maria, Brenner, Campilongo, Filipovic, Fahrenholtz, Ryan, DeSalle, Crealese, Wolfe, Calzolari, and Curtarolo]{DEED}
S.~Divilov, H.~Eckert, D.~Hicks, C.~Oses, C.~Toher, R.~Friedrich, M.~Esters, M.~J. Mehl, A.~C. Zettel, Y.~Lederer, E.~Zurek, J.-P. Maria, D.~W. Brenner, X.~Campilongo, S.~Filipovic, W.~G. Fahrenholtz, C.~J. Ryan, C.~M. DeSalle, R.~J. Crealese, D.~E. Wolfe, A.~Calzolari, and S.~Curtarolo.
\newblock {Disordered enthalpy-entropy descriptor for high-entropy ceramics discovery}.
\newblock \emph{Nature}, 625:\penalty0 66--73, 2024.
\newblock \doi{10.1038/s41586-023-06786-y}.

\bibitem[Draxl and Scheffler(2018)]{NOMAD_repository1}
C.~Draxl and M.~Scheffler.
\newblock {NOMAD}: The {FAIR} concept for big data-driven materials.
\newblock \emph{MRS Bulletin}, 43:\penalty0 676--682, 2018.
\newblock \doi{10.1557/mrs.2018.208}.

\bibitem[Draxl and Scheffler(2019)]{NOMAD_repository2}
C.~Draxl and M.~Scheffler.
\newblock The {NOMAD} laboratory: from data sharing to artificial intelligence.
\newblock \emph{Journal of Physics: Materials}, 2\penalty0 (3):\penalty0 036001, may 2019.
\newblock \doi{10.1088/2515-7639/ab13bb}.

\bibitem[Elementary {M}ultiperspective {M}aterial {O}ntology {(EMMO)}()]{EMMO}
Elementary {M}ultiperspective {M}aterial {O}ntology {(EMMO)}.
\newblock Elementary {M}ultiperspective {M}aterial {O}ntology {(EMMO)}.
\newblock \url{https://github.com/emmo-repo/EMMO}, 2019.

\bibitem[Enkovaara et~al.(2010)Enkovaara, Rostgaard, Mortensen, Chen, Du{\l}ak, Ferrighi, Gavnholt, Glinsvad, Haikola, Hansen, Kristoffersen, Kuisma, Larsen, Lehtovaara, Ljungberg, Lopez-Acevedo, Moses, Ojanen, Olsen, Petzold, Romero, Stausholm-M{\o}ller, Strange, Tritsaris, Vanin, Walter, Hammer, H{\"a}kkinen, Madsen, Nieminen, N{\o}rskov, Puska, Rantala, Schi{\o}tz, Thygesen, and Jacobsen]{GPAW}
J.~Enkovaara, C.~Rostgaard, J.~J. Mortensen, J.~Chen, M.~Du{\l}ak, L.~Ferrighi, J.~Gavnholt, C.~Glinsvad, V.~Haikola, H.~A. Hansen, H.~H. Kristoffersen, M.~Kuisma, A.~H. Larsen, L.~Lehtovaara, M.~Ljungberg, O.~Lopez-Acevedo, P.~G. Moses, J.~Ojanen, T.~Olsen, V.~Petzold, N.~A. Romero, J.~Stausholm-M{\o}ller, M.~Strange, G.~A. Tritsaris, M.~Vanin, M.~Walter, B.~Hammer, H.~H{\"a}kkinen, G.~K.~H. Madsen, R.~M. Nieminen, J.~K. N{\o}rskov, M.~Puska, T.~T. Rantala, J.~Schi{\o}tz, K.~S. Thygesen, and K.~W. Jacobsen.
\newblock Electronic structure calculations with {GPAW}: a real-space implementation of the projector augmented-wave method.
\newblock \emph{J. Phys.: Condens. Matter}, 22\penalty0 (25):\penalty0 253202, 2010.
\newblock \doi{10.1088/0953-8984/22/25/253202}.

\bibitem[Esters et~al.(2023)Esters, Oses, Divilov, Eckert, Friedrich, Hicks, Mehl, Rose, Smolyanyuk, Calzolari, Campilongo, Toher, and Curtarolo]{afloworg2023}
M.~Esters, C.~Oses, S.~Divilov, H.~Eckert, R.~Friedrich, D.~Hicks, M.~J. Mehl, F.~Rose, A.~Smolyanyuk, A.~Calzolari, X.~Campilongo, C.~Toher, and S.~Curtarolo.
\newblock {aflow.org}: A web ecosystem of databases, software and tools.
\newblock \emph{Comput.\ Mater.\ Sci.}, 216:\penalty0 111808, 2023.
\newblock \doi{10.1016/j.commatsci.2022.111808}.

\bibitem[Evans and Morris(2020)]{Evans2020}
M.~L. Evans and A.~J. Morris.
\newblock {\emph{matador}}: a {{Python}} library for analysing, curating and performing high-throughput density-functional theory calculations.
\newblock \emph{Journal of Open Source Software}, 5\penalty0 (54):\penalty0 2563, 2020.
\newblock \doi{10.21105/joss.02563}.

\bibitem[Evans et~al.(2021)Evans, Andersen, Dwaraknath, Scheidgen, Fekete, and Winston]{Evans2021}
M.~L. Evans, C.~W. Andersen, S.~Dwaraknath, M.~Scheidgen, {\'A}.~Fekete, and D.~Winston.
\newblock {\emph{optimade-python-tools}}: a {{Python}} library for serving and consuming materials data via {{OPTIMADE APIs}}.
\newblock \emph{Journal of Open Source Software}, 6\penalty0 (65):\penalty0 3458, 2021.
\newblock \doi{10.21105/joss.03458}.

\bibitem[{FAIRsharing.org: OPTIMADE; Open Databases Integration for Materials Design}()]{fairsharing}
{FAIRsharing.org: OPTIMADE; Open Databases Integration for Materials Design}.
\newblock {FAIRsharing.org: OPTIMADE; Open Databases Integration for Materials Design}.
\newblock DOI: \url{10.25504/FAIRsharing.xvfqAC}, 2020.

\bibitem[Ford et~al.(2019)Ford, Hicks, Oses, Toher, and Curtarolo]{aflowbmg2019}
D.~C. Ford, D.~Hicks, C.~Oses, C.~Toher, and S.~Curtarolo.
\newblock Metallic glasses for biodegradable implants.
\newblock \emph{Acta\ Mater.}, 176:\penalty0 297--305, 2019.
\newblock \doi{10.1016/j.actamat.2019.07.008}.

\bibitem[Fuentes-Cobas et~al.(2017)Fuentes-Cobas, Chateigner, Fuentes-Montero, Pepponi, and Grazulis]{MPOD2017}
L.~Fuentes-Cobas, D.~Chateigner, M.~Fuentes-Montero, G.~Pepponi, and S.~Grazulis.
\newblock The representation of coupling interactions in the {Material} {Properties} {Open} {Database} ({MPOD}).
\newblock \emph{Advances in Applied Ceramics}, 116\penalty0 (8):\penalty0 428--433, 2017.
\newblock \doi{10.1080/17436753.2017.1343782}.

\bibitem[Ghiringhelli et~al.(2023)Ghiringhelli, Baldauf, Bereau, Brockhauser, Carbogno, Chamanara, Cozzini, Curtarolo, Draxl, Dwaraknath, Fekete, Kermode, Koch, K{\"u}hbach, Ladines, Lambrix, Himmer, Levchenko, Oliveira, Michalchuk, Miller, Onat, Pavone, Pizzi, Regler, Rignanese, Schaarschmidt, Scheidgen, Schneidewind, Sheveleva, Su, Usvyat, Valsson, W{\"o}ll, and Scheffler]{Ghiringhelli2023}
L.~M. Ghiringhelli, C.~Baldauf, T.~Bereau, S.~Brockhauser, C.~Carbogno, J.~Chamanara, S.~Cozzini, S.~Curtarolo, C.~Draxl, S.~Dwaraknath, {\'A}.~Fekete, J.~Kermode, C.~T. Koch, M.~K{\"u}hbach, A.~N. Ladines, P.~Lambrix, M.-O. Himmer, S.~V. Levchenko, M.~Oliveira, A.~Michalchuk, R.~E. Miller, B.~Onat, P.~Pavone, G.~Pizzi, B.~Regler, G.-M. Rignanese, J.~Schaarschmidt, M.~Scheidgen, A.~Schneidewind, T.~Sheveleva, C.~Su, D.~Usvyat, O.~Valsson, C.~W{\"o}ll, and M.~Scheffler.
\newblock Shared metadata for data-centric materials science.
\newblock \emph{Scientific Data}, 10\penalty0 (1):\penalty0 626, 2023.
\newblock \doi{10.1038/s41597-023-02501-8}.

\bibitem[Giannozzi et~al.(2009)Giannozzi, Baroni, Bonini, Calandra, Car, Cavazzoni, Ceresoli, Chiarotti, Cococcioni, Dabo, Dal~Corso, de~Gironcoli, Fabris, Fratesi, Gebauer, Gerstmann, Gougoussis, Kokalj, Lazzeri, Martin-Samos, Marzari, Mauri, Mazzarello, Paolini, Pasquarello, Paulatto, Sbraccia, Scandolo, Sclauzero, Seitsonen, Smogunov, Umari, Wentzcovitch, and R.M.]{Giannozzi2009}
P.~Giannozzi, S.~Baroni, N.~Bonini, M.~Calandra, R.~Car, C.~Cavazzoni, D.~Ceresoli, G.~Chiarotti, M.~Cococcioni, I.~Dabo, A.~Dal~Corso, S.~de~Gironcoli, S.~Fabris, G.~Fratesi, R.~Gebauer, U.~Gerstmann, C.~Gougoussis, A.~Kokalj, M.~Lazzeri, L.~Martin-Samos, N.~Marzari, F.~Mauri, R.~Mazzarello, S.~Paolini, A.~Pasquarello, L.~Paulatto, C.~Sbraccia, S.~Scandolo, G.~Sclauzero, A.~Seitsonen, A.~Smogunov, P.~Umari, R.~Wentzcovitch, and W.~R.M.
\newblock {Quantum ESPRESSO}: a modular and open-source software project for quantum simulations of materials.
\newblock \emph{Journal of Physics: Condensed Matter}, 21\penalty0 (39):\penalty0 395502, 2009.
\newblock \doi{10.1088/0953-8984/21/39/395502}.

\bibitem[Giannozzi et~al.(2017)Giannozzi, Andreussi, Brumme, Bunau, Nardelli, Calandra, Car, Cavazzoni, Ceresoli, Cococcioni, Colonna, Carnimeo, Corso, de~Gironcoli, Delugas, DiStasio, Ferretti, Floris, Fratesi, Fugallo, Gebauer, Gerstmann, Giustino, Gorni, Jia, Kawamura, Ko, Kokalj, K{\"{u}}{\c{c}}{\"{u}}kbenli, Lazzeri, Marsili, Marzari, Mauri, Nguyen, Nguyen, de-la Roza, Paulatto, Ponc{\'{e}}, Rocca, Sabatini, Santra, Schlipf, Seitsonen, Smogunov, Timrov, Thonhauser, Umari, Vast, Wu, and Baroni]{Giannozzi2017}
P.~Giannozzi, O.~Andreussi, T.~Brumme, O.~Bunau, M.~B. Nardelli, M.~Calandra, R.~Car, C.~Cavazzoni, D.~Ceresoli, M.~Cococcioni, N.~Colonna, I.~Carnimeo, A.~D. Corso, S.~de~Gironcoli, P.~Delugas, R.~A. DiStasio, A.~Ferretti, A.~Floris, G.~Fratesi, G.~Fugallo, R.~Gebauer, U.~Gerstmann, F.~Giustino, T.~Gorni, J.~Jia, M.~Kawamura, H.-Y. Ko, A.~Kokalj, E.~K{\"{u}}{\c{c}}{\"{u}}kbenli, M.~Lazzeri, M.~Marsili, N.~Marzari, F.~Mauri, N.~L. Nguyen, H.-V. Nguyen, A.~O. de-la Roza, L.~Paulatto, S.~Ponc{\'{e}}, D.~Rocca, R.~Sabatini, B.~Santra, M.~Schlipf, A.~P. Seitsonen, A.~Smogunov, I.~Timrov, T.~Thonhauser, P.~Umari, N.~Vast, X.~Wu, and S.~Baroni.
\newblock Advanced capabilities for materials modelling with {Quantum ESPRESSO}.
\newblock \emph{Journal of Physics: Condensed Matter}, 29\penalty0 (46):\penalty0 465901, 2017.
\newblock \doi{10.1088/1361-648X/aa8f79}.

\bibitem[Gjerding et~al.(2021)Gjerding, Taghizadeh, Rasmussen, Ali, Bertoldo, Deilmann, Kn{\o}sgaard, Kruse, Larsen, Manti, et~al.]{gjerding2021recent}
M.~N. Gjerding, A.~Taghizadeh, A.~Rasmussen, S.~Ali, F.~Bertoldo, T.~Deilmann, N.~R. Kn{\o}sgaard, M.~Kruse, A.~H. Larsen, S.~Manti, et~al.
\newblock Recent progress of the computational {2D} materials database ({C2DB}).
\newblock \emph{2D Materials}, 8\penalty0 (4):\penalty0 044002, 2021.
\newblock \doi{10.1088/2053-1583/ac1059}.

\bibitem[Gomzi et~al.(2021)Gomzi, \v{S}api\'{c}, and Vidak]{Gomzi_CODff_2021}
V.~Gomzi, I.~M. \v{S}api\'{c}, and A.~Vidak.
\newblock {ReaxFF} force field development and application for toluene adsorption on {MnMO}\textsubscript{x} ({M}={C}u, {F}e, {N}i) catalysts.
\newblock \emph{The Journal of Physical Chemistry A}, 125\penalty0 (50):\penalty0 10649--10656, 2021.
\newblock \doi{10.1021/acs.jpca.1c06939}.

\bibitem[Goodall et~al.(2022)Goodall, Parackal, Faber, Armiento, and Lee]{Goodall2022}
R.~E.~A. Goodall, A.~S. Parackal, F.~A. Faber, R.~Armiento, and A.~A. Lee.
\newblock Rapid discovery of stable materials by coordinate-free coarse graining.
\newblock \emph{Science Advances}, 8\penalty0 (30):\penalty0 eabn4117, 2022.
\newblock \doi{10.1126/sciadv.abn4117}.

\bibitem[Grassano et~al.(2023)Grassano, Campi, Marrazzo, and Marzari]{Grassano2023}
D.~Grassano, D.~Campi, A.~Marrazzo, and N.~Marzari.
\newblock Complementary screening for quantum spin {Hall} insulators in two-dimensional exfoliable materials.
\newblock \emph{Phys. Rev. Mater.}, 7:\penalty0 094202, 2023.
\newblock \doi{10.1103/PhysRevMaterials.7.094202}.

\bibitem[Grassano et~al.(2024)Grassano, Marzari, and Campi]{Grassano2023a}
D.~Grassano, N.~Marzari, and D.~Campi.
\newblock High-throughput screening of weyl semimetals.
\newblock \emph{Phys. Rev. Mater.}, 8:\penalty0 024201, Feb 2024.
\newblock \doi{10.1103/PhysRevMaterials.8.024201}.
\newblock URL \url{https://link.aps.org/doi/10.1103/PhysRevMaterials.8.024201}.

\bibitem[Gra{\v{z}}ulis et~al.(2012)Gra{\v{z}}ulis, Da{\v{s}}kevi{\v{c}}, Merkys, Chateigner, Lutterotti, Quir{\'{o}}s, Serebryanaya, Moeck, Downs, and Bail]{Grazulis_COD_2012}
S.~Gra{\v{z}}ulis, A.~Da{\v{s}}kevi{\v{c}}, A.~Merkys, D.~Chateigner, L.~Lutterotti, M.~Quir{\'{o}}s, N.~R. Serebryanaya, P.~Moeck, R.~T. Downs, and A.~L. Bail.
\newblock {C}rystallography {O}pen {D}atabase ({COD}): an open-access collection of crystal structures and platform for world-wide collaboration.
\newblock \emph{Nucleic Acids Res.}, 40:\penalty0 D420--D427, 2012.
\newblock \doi{10.1093/nar/gkr900}.

\bibitem[Haastrup et~al.(2018)Haastrup, Strange, Pandey, Deilmann, Schmidt, Hinsche, Gjerding, Torelli, Larsen, Riis-Jensen, et~al.]{haastrup2018computational}
S.~Haastrup, M.~Strange, M.~Pandey, T.~Deilmann, P.~S. Schmidt, N.~F. Hinsche, M.~N. Gjerding, D.~Torelli, P.~M. Larsen, A.~C. Riis-Jensen, et~al.
\newblock {The Computational 2D Materials Database: high-throughput modeling and discovery of atomically thin crystals}.
\newblock \emph{2D Materials}, 5\penalty0 (4):\penalty0 042002, 2018.
\newblock \doi{10.1088/2053-1583/aacfc1}.

\bibitem[Hall et~al.(1991)Hall, Allen, and Brown]{Hall_IUCr_CIF_1991}
S.~R. Hall, F.~H. Allen, and I.~D. Brown.
\newblock The crystallographic information file ({CIF}): a new standard archive file for crystallography.
\newblock \emph{Acta Crystallographica Section A}, 47:\penalty0 655--685, 1991.
\newblock \doi{10.1107/S010876739101067X}.

\bibitem[Harper et~al.(2020)Harper, Evans, Darby, Karasulu, Ko{\c c}er, Nelson, and Morris]{Harper2020}
A.~F. Harper, M.~L. Evans, J.~P. Darby, B.~Karasulu, C.~P. Ko{\c c}er, J.~R. Nelson, and A.~J. Morris.
\newblock Ab initio {{Structure Prediction Methods}} for {{Battery Materials}}: {{A}} review of recent computational efforts to predict the atomic level structure and bonding in materials for rechargeable batteries.
\newblock \emph{Johnson Matthey Technology Review}, 64\penalty0 (2):\penalty0 103--118, 2020.
\newblock \doi{10.1595/205651320x15742491027978}.

\bibitem[He et~al.(2020)He, Yao, Hegde, Naghavi, Shen, Bushick, and Wolverton]{He_OQMD_Thermoelectrics}
J.~He, Z.~Yao, V.~I. Hegde, S.~S. Naghavi, J.~Shen, K.~M. Bushick, and C.~Wolverton.
\newblock Computational discovery of stable heteroanionic oxychalcogenides {$ABX$O} ({$A$}, {$B$}=metals; {$X$}={S}, {Se}, and {Te}) and their potential applications.
\newblock \emph{Chemistry of Materials}, 32\penalty0 (19):\penalty0 8229--8242, 2020.
\newblock ISSN 0897-4756.
\newblock \doi{10.1021/acs.chemmater.0c01902}.

\bibitem[Henkel et~al.(2023)Henkel, Li, Grandhi, Vivo, and Rinke]{Henkel2023}
P.~Henkel, J.~Li, G.~K. Grandhi, P.~Vivo, and P.~Rinke.
\newblock Screening {{Mixed-Metal Sn}}{\textsubscript{2}}{{M}}({{III}}){{Ch}}{\textsubscript{2}}{{X}}{\textsubscript{3}} {{Chalcohalides}} for {{Photovoltaic Applications}}.
\newblock \emph{Chemistry of Materials}, 35\penalty0 (18):\penalty0 7761--7769, 2023.
\newblock \doi{10.1021/acs.chemmater.3c01629}.

\bibitem[Hicks et~al.(2019)Hicks, Mehl, Gossett, Toher, Levy, Hanson, Hart, and Curtarolo]{AFLOW_PROTO2}
D.~Hicks, M.~J. Mehl, E.~Gossett, C.~Toher, O.~Levy, R.~M. Hanson, G.~L.~W. Hart, and S.~Curtarolo.
\newblock {The AFLOW Library of Crystallographic Prototypes: Part 2}.
\newblock \emph{Comp. Mat. Sci.}, 161:\penalty0 S1--S1011, 2019.
\newblock \doi{10.1016/j.commatsci.2018.10.043}.

\bibitem[Hicks et~al.(2021)Hicks, Mehl, Esters, Oses, Levy, Hart, Toher, and Curtarolo]{AFLOW_PROTO3}
D.~Hicks, M.~J. Mehl, M.~Esters, C.~Oses, O.~Levy, G.~L.~W. Hart, C.~Toher, and S.~Curtarolo.
\newblock {The AFLOW Library of Crystallographic Prototypes: Part 3}.
\newblock \emph{Comp. Mat. Sci.}, 199:\penalty0 110450, 2021.
\newblock \doi{10.1016/j.commatsci.2021.110450}.

\bibitem[Himanen et~al.(2019)Himanen, Geurts, Foster, and Rinke]{Himanen2019}
L.~Himanen, A.~Geurts, A.~S. Foster, and P.~Rinke.
\newblock Data-{Driven} {Materials} {Science}: {Status}, {Challenges}, and {Perspectives}.
\newblock \emph{Adv. Sci.}, 6\penalty0 (21):\penalty0 1900808, 2019.
\newblock \doi{10.1002/advs.201900808}.

\bibitem[Hospital et~al.(2016)Hospital, Andrio, Cugnasco, Codo, Becerra, Dans, Battistini, Torrers, Go{\~{n}}i, Orozco, and Gelpi]{Hospital2015}
A.~Hospital, P.~Andrio, C.~Cugnasco, L.~Codo, Y.~Becerra, P.~D. Dans, F.~Battistini, J.~Torrers, R.~Go{\~{n}}i, M.~Orozco, and J.~L. Gelpi.
\newblock {{BIGNASim}}: a {{NoSQL}} database structure and analysis portal for nucleic acids simulation data.
\newblock \emph{Nucleic Acids Research}, 44:\penalty0 D272--D278, 2016.
\newblock \doi{10.1093/nar/gkv1301}.

\bibitem[Hospital et~al.(2020)Hospital, Battistini, Soliva, Gelpi, and Orozco]{Hospital2020}
A.~Hospital, F.~Battistini, R.~Soliva, J.~L. Gelpi, and M.~Orozco.
\newblock Surviving the deluge of biosimulation data.
\newblock \emph{WIREs Computational Molecular Science}, 10:\penalty0 e1449, 2020.
\newblock \doi{10.1002/wcms.1449}.

\bibitem[Huber et~al.(2020)Huber, Zoupanos, Uhrin, Talirz, Kahle, H{\"{a}}uselmann, Gresch, M{\"{u}}ller, Yakutovich, Andersen, Ramirez, Adorf, Gargiulo, Kumbhar, Passaro, Johnston, Merkys, Cepellotti, Mounet, Marzari, Kozinsky, and Pizzi]{AiiDA2}
S.~P. Huber, S.~Zoupanos, M.~Uhrin, L.~Talirz, L.~Kahle, R.~H{\"{a}}uselmann, D.~Gresch, T.~M{\"{u}}ller, A.~V. Yakutovich, C.~W. Andersen, F.~F. Ramirez, C.~S. Adorf, F.~Gargiulo, S.~Kumbhar, E.~Passaro, C.~Johnston, A.~Merkys, A.~Cepellotti, N.~Mounet, N.~Marzari, B.~Kozinsky, and G.~Pizzi.
\newblock {AiiDA} 1.0, a scalable computational infrastructure for automated reproducible workflows and data provenance.
\newblock \emph{Scientific Data}, 7\penalty0 (1):\penalty0 300, 2020.
\newblock \doi{10.1038/s41597-020-00638-4}.

\bibitem[Im et~al.(2022)Im, Shang, Smith, Krajewski, Lichtenstein, Sun, Bocklund, Liu, and Kim]{Im2022ThermodynamicModelingNdBi}
S.~Im, S.~L. Shang, N.~D. Smith, A.~M. Krajewski, T.~Lichtenstein, H.~Sun, B.~J. Bocklund, Z.~K. Liu, and H.~Kim.
\newblock {Thermodynamic properties of the Nd-Bi system via emf measurements, DFT calculations, machine learning, and CALPHAD modeling}.
\newblock \emph{Acta Materialia}, 223:\penalty0 117448, 2022.
\newblock \doi{10.1016/J.ACTAMAT.2021.117448}.

\bibitem[Isayev et~al.(2015)Isayev, Fourches, Muratov, Oses, Rasch, Tropsha, and Curtarolo]{aflowcartography2015}
O.~Isayev, D.~Fourches, E.~N. Muratov, C.~Oses, K.~Rasch, A.~Tropsha, and S.~Curtarolo.
\newblock Materials cartography: Representing and mining materials space using structural and electronic fingerprints.
\newblock \emph{Chem.\ Mater.}, 27\penalty0 (3):\penalty0 735--743, 2015.
\newblock \doi{10.1021/cm503507h}.

\bibitem[{IUCr}(2023)]{IUCr_quality_criteria_2023}
{IUCr}.
\newblock Data requirements for structures.
\newblock \url{https://journals.iucr.org/c/services/cif/reqdata.html}, 2023.

\bibitem[Iwasaki et~al.(2017)Iwasaki, Kusne, and Takeuchi]{Iwasaki2017}
Y.~Iwasaki, A.~G. Kusne, and I.~Takeuchi.
\newblock Comparison of dissimilarity measures for cluster analysis of {{X-ray}} diffraction data from combinatorial libraries.
\newblock \emph{npj Computational Materials}, 3\penalty0 (1):\penalty0 1--9, 2017.
\newblock \doi{10.1038/s41524-017-0006-2}.

\bibitem[Jablonka et~al.(2023)Jablonka, Ai, {Al-Feghali}, Badhwar, Bocarsly, Bran, Bringuier, Brinson, Choudhary, Circi, Cox, de~Jong, Evans, Gastellu, Genzling, Gil, Gupta, Hong, Imran, Kruschwitz, Labarre, L{\'a}la, Liu, Ma, Majumdar, Merz, Moitessier, Moubarak, Mouri{\~n}o, Pelkie, Pieler, Ramos, Rankovi{\'c}, Rodriques, Sanders, Schwaller, Schwarting, Shi, Smit, Smith, Herck, V{\"o}lker, Ward, Warren, Weiser, Zhang, Zhang, Zia, Scourtas, Schmidt, Foster, White, and Blaiszik]{Jablonka2023b}
K.~M. Jablonka, Q.~Ai, A.~{Al-Feghali}, S.~Badhwar, J.~D. Bocarsly, A.~M. Bran, S.~Bringuier, L.~C. Brinson, K.~Choudhary, D.~Circi, S.~Cox, W.~A. de~Jong, M.~L. Evans, N.~Gastellu, J.~Genzling, M.~V. Gil, A.~K. Gupta, Z.~Hong, A.~Imran, S.~Kruschwitz, A.~Labarre, J.~L{\'a}la, T.~Liu, S.~Ma, S.~Majumdar, G.~W. Merz, N.~Moitessier, E.~Moubarak, B.~Mouri{\~n}o, B.~Pelkie, M.~Pieler, M.~C. Ramos, B.~Rankovi{\'c}, S.~G. Rodriques, J.~N. Sanders, P.~Schwaller, M.~Schwarting, J.~Shi, B.~Smit, B.~E. Smith, J.~V. Herck, C.~V{\"o}lker, L.~Ward, S.~Warren, B.~Weiser, S.~Zhang, X.~Zhang, G.~A. Zia, A.~Scourtas, K.~J. Schmidt, I.~Foster, A.~D. White, and B.~Blaiszik.
\newblock 14 examples of how {{LLMs}} can transform materials science and chemistry: a reflection on a large language model hackathon.
\newblock \emph{Digital Discovery}, 2\penalty0 (5):\penalty0 1233--1250, Oct. 2023.
\newblock \doi{10.1039/d3dd00113j}.

\bibitem[Jain et~al.(2013)Jain, Ong, Hautier, Chen, Richards, Dacek, Cholia, Gunter, Skinner, Ceder, and Persson]{Materials_Project}
A.~Jain, S.~P. Ong, G.~Hautier, W.~Chen, W.~D. Richards, S.~Dacek, S.~Cholia, D.~Gunter, D.~Skinner, G.~Ceder, and K.~A. Persson.
\newblock {Commentary: The Materials Project: A materials genome approach to accelerating materials innovation}.
\newblock \emph{APL Mater.}, 1\penalty0 (1):\penalty0 011002, 2013.
\newblock \doi{10.1063/1.4812323}.

\bibitem[Kirklin et~al.(2016)Kirklin, Saal, Hegde, and Wolverton]{Kirklin_OQMD_alloy}
S.~Kirklin, J.~E. Saal, V.~I. Hegde, and C.~Wolverton.
\newblock High-throughput computational search for strengthening precipitates in alloys.
\newblock \emph{Acta Materialia}, 102:\penalty0 125--135, 2016.
\newblock ISSN 1359-6454.
\newblock \doi{10.1016/j.actamat.2015.09.016}.

\bibitem[Krajewski et~al.(2022)Krajewski, Siegel, Xu, and Liu]{Krajewski2022ExtensibleNetworks}
A.~M. Krajewski, J.~W. Siegel, J.~Xu, and Z.~K. Liu.
\newblock {Extensible Structure-Informed Prediction of Formation Energy with improved accuracy and usability employing neural networks}.
\newblock \emph{Computational Materials Science}, 208:\penalty0 111254, 2022.
\newblock \doi{10.1016/J.COMMATSCI.2022.111254}.

\bibitem[Krajewski et~al.(2024)Krajewski, Siegel, and Liu]{Krajewski2024pySIPFENN}
A.~M. Krajewski, J.~W. Siegel, and Z.-K. Liu.
\newblock Efficient structure-informed featurization and property prediction of ordered, dilute, and random atomic structures.
\newblock \emph{arXiv}, 4 2024.
\newblock \doi{10.48550/arXiv.2404.02849}.

\bibitem[Kresse and Furthm{\"{u}}ller(1996)]{Kresse1996}
G.~Kresse and J.~Furthm{\"{u}}ller.
\newblock Efficient iterative schemes for ab initio total-energy calculations using a plane-wave basis set.
\newblock \emph{Phys Rev B Condens Matter}, 54\penalty0 (16):\penalty0 11169--11186, 1996.
\newblock \doi{10.1103/physrevb.54.11169}.

\bibitem[Kusne et~al.(2020)Kusne, Yu, Wu, Zhang, {Hattrick-Simpers}, {DeCost}, Sarker, Oses, Toher, Curtarolo, Davydov, Agarwal, Bendersky, Li, Mehta, and Takeuchi]{aflowcameo2020}
A.~G. Kusne, H.~Yu, C.~Wu, H.~Zhang, J.~{Hattrick-Simpers}, B.~{DeCost}, S.~Sarker, C.~Oses, C.~Toher, S.~Curtarolo, A.~V. Davydov, R.~Agarwal, L.~A. Bendersky, M.~Li, A.~Mehta, and I.~Takeuchi.
\newblock On-the-fly closed-loop materials discovery via bayesian active learning.
\newblock \emph{Nat.\ Commun.}, 11:\penalty0 5966, 2020.
\newblock \doi{10.1038/s41467-020-19597-w}.

\bibitem[Larsen et~al.(2017)Larsen, Mortensen, Blomqvist, Castelli, Christensen, Du{\l}ak, Friis, Groves, Hammer, Hargus, Hermes, Jennings, Jensen, Kermode, Kitchin, Kolsbjerg, Kubal, Kaasbjerg, Lysgaard, Maronsson, Maxson, Olsen, Pastewka, Peterson, Rostgaard, Schi{\o}tz, Sch{\"{u}}tt, Strange, Thygesen, Vegge, Vilhelmsen, Walter, Zeng, and Jacobsen]{ase}
A.~H. Larsen, J.~J. Mortensen, J.~Blomqvist, I.~E. Castelli, R.~Christensen, M.~Du{\l}ak, J.~Friis, M.~N. Groves, B.~Hammer, C.~Hargus, E.~D. Hermes, P.~C. Jennings, P.~B. Jensen, J.~Kermode, J.~R. Kitchin, E.~L. Kolsbjerg, J.~Kubal, K.~Kaasbjerg, S.~Lysgaard, J.~B. Maronsson, T.~Maxson, T.~Olsen, L.~Pastewka, A.~Peterson, C.~Rostgaard, J.~Schi{\o}tz, O.~Sch{\"{u}}tt, M.~Strange, K.~S. Thygesen, T.~Vegge, L.~Vilhelmsen, M.~Walter, Z.~Zeng, and K.~W. Jacobsen.
\newblock {The atomic simulation environment{\textemdash}a {Python} library for working with atoms}.
\newblock \emph{Journal of Physics: Condensed Matter}, 29\penalty0 (27):\penalty0 273002, 2017.
\newblock \doi{10.1088/1361-648x/aa680e}.

\bibitem[Law et~al.(2023)Law, Pandey, Gorai, and St.~John]{Law2023}
J.~N. Law, S.~Pandey, P.~Gorai, and P.~C. St.~John.
\newblock Upper-{{Bound Energy Minimization}} to {{Search}} for {{Stable Functional Materials}} with {{Graph Neural Networks}}.
\newblock \emph{JACS Au}, 3\penalty0 (1):\penalty0 113--123, 2023.
\newblock \doi{10.1021/jacsau.2c00540}.

\bibitem[Lev{\"{a}}m{\"{a}}ki et~al.(2023)Lev{\"{a}}m{\"{a}}ki, Bock, Sangiovanni, Johnson, Tasn{\'{a}}di, Armiento, and Abrikosov]{HADB_2023}
H.~Lev{\"{a}}m{\"{a}}ki, F.~Bock, D.~G. Sangiovanni, L.~J.~S. Johnson, F.~Tasn{\'{a}}di, R.~Armiento, and I.~A. Abrikosov.
\newblock {HADB: A materials-property database for hard-coating alloys}.
\newblock \emph{Thin Solid Films}, 766:\penalty0 139627, 2023.
\newblock \doi{10.1016/j.tsf.2022.139627}.

\bibitem[Li et~al.(2024)Li, Hartig, Armiento, and Lambrix]{Li2024}
H.~Li, O.~Hartig, R.~Armiento, and P.~Lambrix.
\newblock {Ontology-based GraphQL server generation for data access and data integration}.
\newblock \emph{Semantic Web}, page 1–37, Jan. 2024.
\newblock ISSN 1570-0844.
\newblock \doi{10.3233/sw-233550}.

\bibitem[Liu et~al.(2022)Liu, De~Breuck, Wang, and Rignanese]{Liu2022}
X.~Liu, P.-P. De~Breuck, L.~Wang, and G.-M. Rignanese.
\newblock A simple denoising approach to exploit multi-fidelity data for machine learning materials properties.
\newblock \emph{npj Computational Materials}, 8:\penalty0 233, 2022.
\newblock \doi{10.1038/s41524-022-00925-1}.

\bibitem[Liu(2014)]{Liu2014}
Z.~Liu.
\newblock Perspective on materials genome®.
\newblock \emph{Chinese Science Bulletin}, 59:\penalty0 1619, 2014.
\newblock \doi{10.1007/s11434-013-0072-x}.

\bibitem[Lyngby and Thygesen(2022)]{lyngby2022data}
P.~Lyngby and K.~S. Thygesen.
\newblock Data-driven discovery of {2D} materials by deep generative models.
\newblock \emph{npj Computational Materials}, 8\penalty0 (1):\penalty0 232, 2022.
\newblock \doi{10.1038/s41524-022-00923-3}.

\bibitem[Marrazzo et~al.(2018)Marrazzo, Gibertini, Campi, Mounet, and Marzari]{Marrazzo2018}
A.~Marrazzo, M.~Gibertini, D.~Campi, N.~Mounet, and N.~Marzari.
\newblock Prediction of a large-gap and switchable {Kane-Mele} quantum spin {Hall} insulator.
\newblock \emph{Phys. Rev. Lett.}, 120:\penalty0 117701, 2018.
\newblock \doi{10.1103/PhysRevLett.120.117701}.

\bibitem[Marrazzo et~al.(2019)Marrazzo, Gibertini, Campi, Mounet, and Marzari]{Marrazzo2019}
A.~Marrazzo, M.~Gibertini, D.~Campi, N.~Mounet, and N.~Marzari.
\newblock Relative abundance of $\mathbb{Z}_2$ topological order in exfoliable two-dimensional insulators.
\newblock \emph{Nano Letters}, 19\penalty0 (12):\penalty0 8431--8440, 2019.
\newblock \doi{10.1021/acs.nanolett.9b02689}.

\bibitem[MatCloud()]{MatCloudURL}
MatCloud.
\newblock Matcloud.
\newblock \url{http://matcloud.com.cn}.
\newblock accessed 2024.

\bibitem[{Medina-Smith} et~al.(2021){Medina-Smith}, Becker, Plante, Bartolo, Dima, Warren, and Hanisch]{Medina-Smith2021}
A.~{Medina-Smith}, C.~A. Becker, R.~L. Plante, L.~M. Bartolo, A.~Dima, J.~A. Warren, and R.~J. Hanisch.
\newblock A {{Controlled Vocabulary}} and {{Metadata Schema}} for {{Materials Science Data Discovery}}.
\newblock \emph{Data Science Journal}, 20\penalty0 (1):\penalty0 18, 2021.
\newblock \doi{10.5334/dsj-2021-018}.

\bibitem[Mehl et~al.(2017)Mehl, Hicks, Toher, Levy, Hanson, Hart, and Curtarolo]{AFLOW_PROTO1}
M.~J. Mehl, D.~Hicks, C.~Toher, O.~Levy, R.~M. Hanson, G.~L.~W. Hart, and S.~Curtarolo.
\newblock {The AFLOW Library of Crystallographic Prototypes: Part 1}.
\newblock \emph{Comp. Mat. Sci.}, 136:\penalty0 S1--S828, 2017.
\newblock \doi{10.1016/j.commatsci.2017.01.017}.

\bibitem[Mendenhall et~al.(2020)Mendenhall, Brown, Kothiwale, and Meiler]{Mendenhall2020}
J.~Mendenhall, B.~P. Brown, S.~Kothiwale, and J.~Meiler.
\newblock {BCL::Conf: Improved Open-Source Knowledge-Based Conformation Sampling Using the Crystallography Open Database}.
\newblock \emph{Journal of Chemical Information and Modeling}, 61\penalty0 (1):\penalty0 189--201, 2020.
\newblock \doi{10.1021/acs.jcim.0c01140}.

\bibitem[Merchant et~al.(2023)Merchant, Batzner, Schoenholz, Aykol, Cheon, and Cubuk]{Merchant2023}
A.~Merchant, S.~Batzner, S.~S. Schoenholz, M.~Aykol, G.~Cheon, and E.~D. Cubuk.
\newblock Scaling deep learning for materials discovery.
\newblock \emph{Nature}, 624\penalty0 (7990):\penalty0 80--85, Dec. 2023.
\newblock ISSN 1476-4687.
\newblock \doi{10.1038/s41586-023-06735-9}.

\bibitem[Mortensen et~al.(2024)]{mortensen2023gpaw}
J.~J. Mortensen et~al.
\newblock {GPAW: An open Python package for electronic structure calculations}.
\newblock \emph{The Journal of Chemical Physics}, 160\penalty0 (9):\penalty0 092503, 03 2024.
\newblock \doi{10.1063/5.0182685}.

\bibitem[Mounet et~al.(2018)Mounet, Gibertini, Schwaller, Campi, Merkys, Marrazzo, Sohier, Castelli, Cepellotti, Pizzi, and Marzari]{mc2d_1}
N.~Mounet, M.~Gibertini, P.~Schwaller, D.~Campi, A.~Merkys, A.~Marrazzo, T.~Sohier, I.~E. Castelli, A.~Cepellotti, G.~Pizzi, and N.~Marzari.
\newblock Two-dimensional materials from high-throughput computational exfoliation of experimentally known compounds.
\newblock \emph{Nature Nanotechnology}, 13\penalty0 (3):\penalty0 246--252, 2018.
\newblock ISSN 1748-3395.
\newblock \doi{10.1038/s41565-017-0035-5}.

\bibitem[Moustafa et~al.(2022)Moustafa, Larsen, Gjerding, Mortensen, Thygesen, and Jacobsen]{moustafa2022computational}
H.~Moustafa, P.~M. Larsen, M.~N. Gjerding, J.~J. Mortensen, K.~S. Thygesen, and K.~W. Jacobsen.
\newblock Computational exfoliation of atomically thin one-dimensional materials with application to {Majorana} bound states.
\newblock \emph{Physical Review Materials}, 6\penalty0 (6):\penalty0 064202, 2022.
\newblock \doi{10.1103/PhysRevMaterials.6.064202}.

\bibitem[Nyshadham et~al.(2017)Nyshadham, Oses, Hansen, Takeuchi, Curtarolo, and Hart]{aflowsuperalloys2017}
C.~Nyshadham, C.~Oses, J.~E. Hansen, I.~Takeuchi, S.~Curtarolo, and G.~L.~W. Hart.
\newblock A computational high-throughput search for new ternary superalloys.
\newblock \emph{Acta\ Mater.}, 122:\penalty0 438--447, 2017.
\newblock \doi{10.1016/j.actamat.2016.09.017}.

\bibitem[Ong et~al.(2013)Ong, Richards, Jain, Hautier, Kocher, Cholia, Gunter, Chevrier, Persson, and Ceder]{Ong_pymatgen_2013}
S.~P. Ong, W.~D. Richards, A.~Jain, G.~Hautier, M.~Kocher, S.~Cholia, D.~Gunter, V.~L. Chevrier, K.~A. Persson, and G.~Ceder.
\newblock {Python} {Materials} {Genomics} (pymatgen): A robust, open-source {Python} library for materials analysis.
\newblock \emph{Comput. Mater. Sci.}, 68:\penalty0 314--319, 2013.
\newblock \doi{10.1016/j.commatsci.2012.10.028}.

\bibitem[Ontology driven {O}pen {T}ranslation {E}nvironment {(OntoTrans)}()]{OntoTrans}
Ontology driven {O}pen {T}ranslation {E}nvironment {(OntoTrans)}.
\newblock Ontology driven {O}pen {T}ranslation {E}nvironment {(OntoTrans)}.
\newblock \url{https://ontotrans.eu}, 2020.
\newblock accessed 2023.

\bibitem[Oses et~al.(2023)Oses, Esters, Hicks, Divilov, Eckert, Friedrich, Mehl, Smolyanyuk, Campilongo, {van de Walle}, Schroers, Kusne, Takeuchi, Zurek, {Buongiorno Nardelli}, Fornari, Lederer, Levy, Toher, and Curtarolo]{aflowpp2023}
C.~Oses, M.~Esters, D.~Hicks, S.~Divilov, H.~Eckert, R.~Friedrich, M.~J. Mehl, A.~Smolyanyuk, X.~Campilongo, A.~{van de Walle}, J.~Schroers, A.~G. Kusne, I.~Takeuchi, E.~Zurek, M.~{Buongiorno Nardelli}, M.~Fornari, Y.~Lederer, O.~Levy, C.~Toher, and S.~Curtarolo.
\newblock {aflow++}: A {C++} framework for autonomous materials design.
\newblock \emph{Comput.\ Mater.\ Sci.}, 217:\penalty0 111889, 2023.
\newblock \doi{10.1016/j.commatsci.2022.111889}.

\bibitem[Ozaki et~al.(2020)Ozaki, Suzuki, Hawai, Saito, Onishi, and Ono]{Ozaki2020}
Y.~Ozaki, Y.~Suzuki, T.~Hawai, K.~Saito, M.~Onishi, and K.~Ono.
\newblock Automated crystal structure analysis based on blackbox optimisation.
\newblock \emph{npj Computational Materials}, 6\penalty0 (1):\penalty0 1--7, 2020.
\newblock \doi{10.1038/s41524-020-0330-9}.

\bibitem[Pakdel et~al.(2024)Pakdel, Rasmussen, Taghizadeh, Kruse, Olsen, and Thygesen]{pakdel2023emergent}
S.~Pakdel, A.~Rasmussen, A.~Taghizadeh, M.~Kruse, T.~Olsen, and K.~S. Thygesen.
\newblock High-throughput computational stacking reveals emergent properties in natural van der {{Waals}} bilayers.
\newblock \emph{Nature Communications}, 15\penalty0 (1):\penalty0 932, Jan. 2024.
\newblock ISSN 2041-1723.
\newblock \doi{10.1038/s41467-024-45003-w}.

\bibitem[{Pauling File}()]{PaulingFile_2024}
{Pauling File}.
\newblock {Pauling File}.
\newblock \url{https://paulingfile.com}, 2024.

\bibitem[Pedregosa et~al.(2011)Pedregosa, Varoquaux, Gramfort, Michel, Thirion, Grisel, Blondel, Prettenhofer, Weiss, Dubourg, Vanderplas, Passos, Cournapeau, Brucher, Perrot, and Duchesnay]{scikit-learn}
F.~Pedregosa, G.~Varoquaux, A.~Gramfort, V.~Michel, B.~Thirion, O.~Grisel, M.~Blondel, P.~Prettenhofer, R.~Weiss, V.~Dubourg, J.~Vanderplas, A.~Passos, D.~Cournapeau, M.~Brucher, M.~Perrot, and E.~Duchesnay.
\newblock Scikit-learn: Machine learning in {P}ython.
\newblock \emph{Journal of Machine Learning Research}, 12:\penalty0 2825--2830, 2011.

\bibitem[Pepponi et~al.(2012)Pepponi, Gra{\v{z}}ulis, and Chateigner]{MPOD2012}
G.~Pepponi, S.~Gra{\v{z}}ulis, and D.~Chateigner.
\newblock {MPOD}: {A} {Material} {Property} {Open} {Database} linked to structural information.
\newblock \emph{Nuclear Instruments and Methods in Physics Research Section B: Beam Interactions with Materials and Atoms}, 284:\penalty0 10--14, 2012.
\newblock \doi{https://doi.org/10.1016/j.nimb.2011.08.070}.

\bibitem[Perim et~al.(2016)Perim, Lee, Liu, Toher, Gong, {Li}, Simmons, Levy, Vlassak, Schroers, and Curtarolo]{aflowbmg2016}
E.~Perim, D.~Lee, Y.~Liu, C.~Toher, P.~Gong, Y.~{Li}, W.~N. Simmons, O.~Levy, J.~J. Vlassak, J.~Schroers, and S.~Curtarolo.
\newblock Spectral descriptors for bulk metallic glasses based on the thermodynamics of competing crystalline phases.
\newblock \emph{Nat.\ Commun.}, 7:\penalty0 12315, 2016.
\newblock \doi{10.1038/ncomms12315}.

\bibitem[Pezoa et~al.(2016)Pezoa, Reutter, Suarez, Ugarte, and Vrgo{\v{c}}]{jsonschema}
F.~Pezoa, J.~L. Reutter, F.~Suarez, M.~Ugarte, and D.~Vrgo{\v{c}}.
\newblock {Foundations of JSON schema}.
\newblock In \emph{Proceedings of the 25th International Conference on World Wide Web}, pages 263--273. International World Wide Web Conferences Steering Committee, 2016.
\newblock \doi{10.1145/2872427.2883029}.

\bibitem[Pizzi et~al.(2016)Pizzi, Cepellotti, Sabatini, Marzari, and Kozinsky]{AiiDA}
G.~Pizzi, A.~Cepellotti, R.~Sabatini, N.~Marzari, and B.~Kozinsky.
\newblock {AiiDA}: automated interactive infrastructure and database for computational science.
\newblock \emph{Comput. Mater. Sci.}, 111:\penalty0 218--230, 2016.
\newblock \doi{10.1016/j.commatsci.2015.09.013}.

\bibitem[Pizzi et~al.(2021)Pizzi, Milana, Ferrari, Marzari, and Gibertini]{Pizzi_COD_2D_2021}
G.~Pizzi, S.~Milana, A.~C. Ferrari, N.~Marzari, and M.~Gibertini.
\newblock Shear and breathing modes of layered materials.
\newblock \emph{ACS Nano}, 15\penalty0 (8):\penalty0 12509--12534, 2021.
\newblock \doi{10.1021/acsnano.0c10672}.

\bibitem[Plante et~al.(2021)Plante, Becker, {Medina-Smith}, Brady, Dima, Long, Bartolo, Warren, and Hanisch]{Plante2021}
R.~L. Plante, C.~A. Becker, A.~{Medina-Smith}, K.~Brady, A.~Dima, B.~Long, L.~M. Bartolo, J.~A. Warren, and R.~J. Hanisch.
\newblock Implementing a {{Registry Federation}} for {{Materials Science Data Discovery}}.
\newblock \emph{Data Science Journal}, 20\penalty0 (1):\penalty0 15, 2021.
\newblock \doi{10.5334/dsj-2021-015}.

\bibitem[Qiao et~al.(2023)Qiao, Pizzi, and Marzari]{Qiao2023}
J.~Qiao, G.~Pizzi, and N.~Marzari.
\newblock Projectability disentanglement for accurate and automated electronic-structure {Hamiltonians}.
\newblock \emph{npj Computational Materials}, 9\penalty0 (1):\penalty0 208, 2023.
\newblock \doi{10.1038/s41524-023-01146-w}.

\bibitem[{Reyes Tirado} et~al.(2018){Reyes Tirado}, {Perrin Toinin}, and Dunand]{ReyesTirado_superalloys_ActaMat_2018}
F.~L. {Reyes Tirado}, J.~{Perrin Toinin}, and D.~C. Dunand.
\newblock $\gamma+\gamma^{\prime}$ microstructures in the {Co}-{Ta}-{V} and {Co}-{Nb}-{V} ternary systems.
\newblock \emph{Acta\ Mater.}, 151:\penalty0 137--148, 2018.
\newblock \doi{10.1016/j.actamat.2018.03.057}.

\bibitem[Rose et~al.(2017)Rose, Toher, Gossett, Oses, {Buongiorno Nardelli}, Fornari, and Curtarolo]{AFLUX}
F.~Rose, C.~Toher, E.~Gossett, C.~Oses, M.~{Buongiorno Nardelli}, M.~Fornari, and S.~Curtarolo.
\newblock {AFLUX}: The {LUX} materials search {API} for the {AFLOW} data repositories.
\newblock \emph{Comput. Mater. Sci.}, 137:\penalty0 362--370, 2017.
\newblock \doi{10.1016/j.commatsci.2017.04.036}.

\bibitem[Saal et~al.(2013)Saal, Kirklin, Aykol, Meredig, and Wolverton]{Saal_OQMD2013}
J.~E. Saal, S.~Kirklin, M.~Aykol, B.~Meredig, and C.~Wolverton.
\newblock {Materials Design and Discovery with High-Throughput Density Functional Theory: The Open Quantum Materials Database (OQMD)}.
\newblock \emph{JOM}, 65\penalty0 (11):\penalty0 1501--1509, 2013.
\newblock ISSN 1543-1851.
\newblock \doi{10.1007/s11837-013-0755-4}.

\bibitem[Sanvito et~al.(2017)Sanvito, Oses, Xue, Tiwari, {\v{Z}}ic, Archer, Tozman, Venkatesan, Coey, and Curtarolo]{aflowmagnets2017}
S.~Sanvito, C.~Oses, J.~Xue, A.~Tiwari, M.~{\v{Z}}ic, T.~Archer, P.~Tozman, M.~Venkatesan, J.~M.~D. Coey, and S.~Curtarolo.
\newblock Accelerated discovery of new magnets in the {H}eusler alloy family.
\newblock \emph{Sci.\ Adv.}, 3\penalty0 (4):\penalty0 e1602241, 2017.
\newblock \doi{10.1126/sciadv.1602241}.

\bibitem[Sarker et~al.(2018)Sarker, Harrington, Toher, Oses, Samiee, Maria, Brenner, Vecchio, and Curtarolo]{aflowhec2018}
P.~Sarker, T.~Harrington, C.~Toher, C.~Oses, M.~Samiee, J.-P. Maria, D.~W. Brenner, K.~S. Vecchio, and S.~Curtarolo.
\newblock High-entropy high-hardness metal carbides discovered by entropy descriptors.
\newblock \emph{Nat.\ Commun.}, 9\penalty0 (1):\penalty0 4980, 2018.
\newblock \doi{10.1038/s41467-018-07160-7}.

\bibitem[Sbail{\`{o}} et~al.(2022)Sbail{\`{o}}, Fekete, Ghiringhelli, and Scheffler]{NOMAD_toolkit2022}
L.~Sbail{\`{o}}, {\'{A}}.~Fekete, L.~M. Ghiringhelli, and M.~Scheffler.
\newblock The {NOMAD} artificial-intelligence toolkit: turning materials-science data into knowledge and understanding.
\newblock \emph{npj Computational Materials}, 8\penalty0 (1):\penalty0 250, 2022.
\newblock \doi{10.1038/s41524-022-00935-z}.

\bibitem[Scheffler et~al.(2022)Scheffler, Aeschlimann, Albrecht, Bereau, Bungartz, Felser, Greiner, Gro{\ss}, Koch, Kremer, et~al.]{NOMAD_fair2022}
M.~Scheffler, M.~Aeschlimann, M.~Albrecht, T.~Bereau, H.-J. Bungartz, C.~Felser, M.~Greiner, A.~Gro{\ss}, C.~T. Koch, K.~Kremer, et~al.
\newblock {FAIR} data enabling new horizons for materials research.
\newblock \emph{Nature}, 604\penalty0 (7907):\penalty0 635--642, 2022.
\newblock \doi{10.1038/s41586-022-04501-x}.

\bibitem[Scheidgen et~al.(2023)Scheidgen, Himanen, Ladines, Sikter, Nakhaee, {{\'{A}}}d{\'{a}}m Fekete, Chang, Golparvar, M{\'{a}}rquez, Brockhauser, Br{\"{u}}ckner, Ghiringhelli, Dietrich, Lehmberg, Denell, Albino, N{\"{a}}sstr{\"{o}}m, Shabih, Dobener, K{\"{u}}hbach, Mozumder, Rudzinski, Daelman, Pizarro, Kuban, Salazar, Ondra{\v{c}}ka, Bungartz, and Draxl]{NOMAD_2023}
M.~Scheidgen, L.~Himanen, A.~N. Ladines, D.~Sikter, M.~Nakhaee, {{\'{A}}}d{\'{a}}m Fekete, T.~Chang, A.~Golparvar, J.~A. M{\'{a}}rquez, S.~Brockhauser, S.~Br{\"{u}}ckner, L.~M. Ghiringhelli, F.~Dietrich, D.~Lehmberg, T.~Denell, A.~Albino, H.~N{\"{a}}sstr{\"{o}}m, S.~Shabih, F.~Dobener, M.~K{\"{u}}hbach, R.~Mozumder, J.~F. Rudzinski, N.~Daelman, J.~M. Pizarro, M.~Kuban, C.~Salazar, P.~Ondra{\v{c}}ka, H.-J. Bungartz, and C.~Draxl.
\newblock {NOMAD: A distributed web-based platform for managing materials science research data}.
\newblock \emph{Journal of Open Source Software}, 8\penalty0 (90):\penalty0 5388, 2023.
\newblock \doi{10.21105/joss.05388}.

\bibitem[Schmidt et~al.(2017)Schmidt, Shi, Borlido, Chen, Botti, and Marques]{schmidt2017}
J.~Schmidt, J.~Shi, P.~Borlido, L.~Chen, S.~Botti, and M.~A.~L. Marques.
\newblock Predicting the thermodynamic stability of solids combining density functional theory and machine learning.
\newblock \emph{Chem. Mater.}, 29\penalty0 (12):\penalty0 5090--5103, 2017.
\newblock \doi{10.1021/acs.chemmater.7b00156}.

\bibitem[Schmidt et~al.(2018)Schmidt, Chen, Botti, and Marques]{jonathan2018}
J.~Schmidt, L.~Chen, S.~Botti, and M.~A.~L. Marques.
\newblock Predicting the stability of ternary intermetallics with density functional theory and machine learning.
\newblock \emph{J. Chem. Phys.}, 148\penalty0 (24):\penalty0 241728, 2018.
\newblock \doi{10.1063/1.5020223}.

\bibitem[Schmidt et~al.(2021{\natexlab{a}})Schmidt, Pettersson, Verdozzi, Botti, and Marques]{CGAT}
J.~Schmidt, L.~Pettersson, C.~Verdozzi, S.~Botti, and M.~A.~L. Marques.
\newblock Crystal graph attention networks for the prediction of stable materials.
\newblock \emph{Sci. Adv.}, 7\penalty0 (49):\penalty0 eabi7948, 2021{\natexlab{a}}.
\newblock \doi{10.1126/sciadv.abi7948}.

\bibitem[Schmidt et~al.(2021{\natexlab{b}})Schmidt, Wang, Cerqueira, Botti, and Marques]{scan_dataset}
J.~Schmidt, H.-C. Wang, T.~F.~T. Cerqueira, S.~Botti, and M.~A.~L. Marques.
\newblock {A new dataset of 175k stable and metastable materials calculated with the PBEsol and SCAN functionals}.
\newblock \emph{Sci. Data}, 12\penalty0 (1):\penalty0 64, 2021{\natexlab{b}}.
\newblock \doi{10.1038/s41597-022-01177-w}.

\bibitem[Schmidt et~al.(2023{\natexlab{a}})Schmidt, Hoffmann, Wang, Borlido, Carri{\c{c}}o, Cerqueira, Botti, and Marques]{CGATHT}
J.~Schmidt, N.~Hoffmann, H.-C. Wang, P.~Borlido, P.~J. M.~A. Carri{\c{c}}o, T.~F.~T. Cerqueira, S.~Botti, and M.~A.~L. Marques.
\newblock Machine-learning-assisted determination of the global zero-temperature phase diagram of materials.
\newblock \emph{Adv. Mater.}, 35\penalty0 (22):\penalty0 2210788, 2023{\natexlab{a}}.
\newblock \doi{10.1002/adma.202210788}.

\bibitem[Schmidt et~al.(2023{\natexlab{b}})Schmidt, Wang, Schmidt, and Marques]{garnet}
J.~Schmidt, H.-C. Wang, G.~Schmidt, and M.~A. Marques.
\newblock Machine learning guided high-throughput search of non-oxide garnets.
\newblock \emph{npj Computational Materials}, 9\penalty0 (1):\penalty0 63, 2023{\natexlab{b}}.
\newblock \doi{10.1038/s41524-023-01009-4}.

\bibitem[Schwarz et~al.(2022)Schwarz, Uhlig, Lindner, Lampke, Wagner, and Seyller]{Schwarz_CODalloys_2022}
H.~Schwarz, T.~Uhlig, T.~Lindner, T.~Lampke, G.~Wagner, and T.~Seyller.
\newblock {Hardness Enhancement in {CoCrFeNi\textsubscript{1-x}(WC)\textsubscript{x}} High-Entropy Alloy Thin Films Synthesised by Magnetron Co-Sputtering}.
\newblock \emph{Coatings}, 12\penalty0 (2):\penalty0 269, 2022.
\newblock \doi{10.3390/coatings12020269}.

\bibitem[Shang et~al.(2021)Shang, Sun, Pan, Wang, Krajewski, Banu, Li, and Liu]{Shang2021FormingAlFeJoints}
S.-L. Shang, H.~Sun, B.~Pan, Y.~Wang, A.~M. Krajewski, M.~Banu, J.~Li, and Z.-K. Liu.
\newblock {Forming Mechanism of Equilibrium and Non-equilibrium Metallurgical Phases in Dissimilar Materials: Illustrated With Aluminum/steel (Al-Fe) Joints}.
\newblock \emph{Scientific Reports}, 11:\penalty0 24251, 2021.
\newblock \doi{10.1038/s41598-021-03578-0}.

\bibitem[Shen et~al.(2022)Shen, Griesemer, Gopakumar, Baldassarri, Saal, Aykol, Hegde, and Wolverton]{Shen_ReflectionOQMD}
J.~Shen, S.~D. Griesemer, A.~Gopakumar, B.~Baldassarri, J.~E. Saal, M.~Aykol, V.~I. Hegde, and C.~Wolverton.
\newblock Reflections on one million compounds in the open quantum materials database {(OQMD)}.
\newblock \emph{Journal of Physics: Materials}, 5\penalty0 (3):\penalty0 031001, 2022.
\newblock ISSN 2515-7639.
\newblock \doi{10.1088/2515-7639/ac7ba9}.

\bibitem[{SMARTS - A Language for Describing Molecular Patterns}()]{smarts}
{SMARTS - A Language for Describing Molecular Patterns}.
\newblock {SMARTS - A Language for Describing Molecular Patterns}.
\newblock \url{https://www.daylight.com/dayhtml/doc/theory/theory.smarts.html}.
\newblock accessed 2024.

\bibitem[Sohier et~al.(2018)Sohier, Campi, Marzari, and Gibertini]{Sohier2018}
T.~Sohier, D.~Campi, N.~Marzari, and M.~Gibertini.
\newblock Mobility of two-dimensional materials from first principles in an accurate and automated framework.
\newblock \emph{Phys. Rev. Mater.}, 2:\penalty0 114010, 2018.
\newblock \doi{10.1103/PhysRevMaterials.2.114010}.

\bibitem[Sohier et~al.(2019)Sohier, Gibertini, Campi, Pizzi, and Marzari]{Sohier2019}
T.~Sohier, M.~Gibertini, D.~Campi, G.~Pizzi, and N.~Marzari.
\newblock Valley-engineering mobilities in two-dimensional materials.
\newblock \emph{Nano Letters}, 19\penalty0 (6):\penalty0 3723--3729, 2019.
\newblock \doi{10.1021/acs.nanolett.9b00865}.

\bibitem[Stanev et~al.(2018)Stanev, Oses, Kusne, Rodriguez, Paglione, Curtarolo, and Takeuchi]{aflowsc2018}
V.~Stanev, C.~Oses, A.~G. Kusne, E.~Rodriguez, J.~Paglione, S.~Curtarolo, and I.~Takeuchi.
\newblock Machine learning modeling of superconducting critical temperature.
\newblock \emph{npj\ Comput.\ Mater.}, 4:\penalty0 29, 2018.
\newblock \doi{10.1038/s41524-018-0085-8}.

\bibitem[Suh et~al.(2020)Suh, Fare, Warren, and Pyzer-Knapp]{Suh2020}
C.~Suh, C.~Fare, J.~Warren, and E.~Pyzer-Knapp.
\newblock Evolving the materials genome: How machine learning is fueling the next generation of materials discovery.
\newblock \emph{Annual Review of Materials Research}, 50:\penalty0 1--25, 2020.
\newblock \doi{10.1146/annurev-matsci-082019-105100}.

\bibitem[{SVELTE}()]{svelte_2024}
{SVELTE}.
\newblock {SVELTE}.
\newblock \url{https://svelte.dev}, 2024.

\bibitem[Talirz et~al.(2020)Talirz, Kumbhar, Passaro, Yakutovich, Granata, Gargiulo, Borelli, Uhrin, Huber, Zoupanos, Adorf, Andersen, Sch{\"{u}}tt, Pignedoli, Passerone, VandeVondele, Schulthess, Smit, Pizzi, and Marzari]{MaterialsCloud}
L.~Talirz, S.~Kumbhar, E.~Passaro, A.~V. Yakutovich, V.~Granata, F.~Gargiulo, M.~Borelli, M.~Uhrin, S.~P. Huber, S.~Zoupanos, C.~S. Adorf, C.~W. Andersen, O.~Sch{\"{u}}tt, C.~A. Pignedoli, D.~Passerone, J.~VandeVondele, T.~C. Schulthess, B.~Smit, G.~Pizzi, and N.~Marzari.
\newblock Materials {Cloud}, a platform for open computational science.
\newblock \emph{Scientific Data}, 7\penalty0 (1):\penalty0 299, 2020.
\newblock \doi{10.1038/s41597-020-00637-5}.

\bibitem[The {M}arket{P}lace {P}roject()]{MarketPlace}
The {M}arket{P}lace {P}roject.
\newblock The {M}arket{P}lace {P}roject.
\newblock \url{https://materials-marketplace.eu}, 2018.
\newblock accessed 2023.

\bibitem[{The OPTIMADE Developers}(accessed 2023{\natexlab{a}})]{OPTIMADE_github}
{The OPTIMADE Developers}.
\newblock {Open Database Integration for Materials Design (OPTIMADE)}.
\newblock \url{https://github.com/Materials-Consortia/OPTIMADE}, accessed 2023{\natexlab{a}}.

\bibitem[{The OPTIMADE Developers}(accessed 2023{\natexlab{b}})]{OPTIMADE_providers_dashboard}
{The OPTIMADE Developers}.
\newblock {OPTIMADE Providers Dashboard}.
\newblock \url{https://www.optimade.org/providers-dashboard/}, accessed 2023{\natexlab{b}}.

\bibitem[{The OPTIMADE Developers}(accessed 2023{\natexlab{c}})]{OPTIMADE_providers_list}
{The OPTIMADE Developers}.
\newblock {OPTIMADE Providers List}.
\newblock \url{https://providers.optimade.org}, accessed 2023{\natexlab{c}}.

\bibitem[Toby and Von~Dreele(2013)]{Toby2013}
B.~H. Toby and R.~B. Von~Dreele.
\newblock {{GSAS-II}}: the genesis of a modern open-source all purpose crystallography software package.
\newblock \emph{Journal of Applied Crystallography}, 46\penalty0 (2):\penalty0 544--549, 2013.
\newblock \doi{10.1107/S0021889813003531}.

\bibitem[Uhrin et~al.(2021)Uhrin, Huber, Yu, Marzari, and Pizzi]{AiiDA3}
M.~Uhrin, S.~P. Huber, J.~Yu, N.~Marzari, and G.~Pizzi.
\newblock {Workflows in AiiDA: Engineering a high-throughput, event-based engine for robust and modular computational workflows}.
\newblock \emph{Computational Materials Science}, 187:\penalty0 110086, 2021.
\newblock \doi{10.1016/j.commatsci.2020.110086}.

\bibitem[Vahdat et~al.(2022)Vahdat, Agrawal, and Pizzi]{Vahdat2022}
M.~T. Vahdat, K.~V. Agrawal, and G.~Pizzi.
\newblock Machine-learning accelerated identification of exfoliable two-dimensional materials.
\newblock \emph{Machine Learning: Science and Technology}, 3\penalty0 (4):\penalty0 045014, 2022.
\newblock \doi{10.1088/2632-2153/ac9bca}.

\bibitem[{van~Roekeghem} et~al.(2016){van~Roekeghem}, Carrete, Oses, Curtarolo, and Mingo]{aflowte2016}
A.~{van~Roekeghem}, J.~Carrete, C.~Oses, S.~Curtarolo, and N.~Mingo.
\newblock High-throughput computation of thermal conductivity of high-temperature solid phases: The case of oxide and fluoride perovskites.
\newblock \emph{Phys.\ Rev.\ X}, 6:\penalty0 041061, 2016.
\newblock \doi{10.1103/PhysRevX.6.041061}.

\bibitem[Wang et~al.(2022{\natexlab{a}})Wang, Fan, and Yue]{CGCNN_2022}
B.~Wang, Q.~Fan, and Y.~Yue.
\newblock Study of crystal properties based on attention mechanism and crystal graph convolutional neural network.
\newblock \emph{J. Phys.: Condens. Matter}, 34:\penalty0 195901, 2022{\natexlab{a}}.
\newblock \doi{10.1088/1361-648X/ac5705}.

\bibitem[Wang et~al.(2021)Wang, Botti, and Marques]{Wang2021b}
H.-C. Wang, S.~Botti, and M.~A.~L. Marques.
\newblock Predicting stable crystalline compounds using chemical similarity.
\newblock \emph{npj Computational Materials}, 7\penalty0 (1):\penalty0 1--9, 2021.
\newblock \doi{10.1038/s41524-020-00481-6}.

\bibitem[Wang et~al.(2022{\natexlab{b}})Wang, Zhang, Th{\'{e}}, and Yu]{Wang_COD_model_validation_2022}
T.~Wang, K.~Zhang, J.~Th{\'{e}}, and H.~Yu.
\newblock Accurate prediction of band gap of materials using stacking machine learning model.
\newblock \emph{Computational Materials Science}, 201:\penalty0 110899, 2022{\natexlab{b}}.
\newblock ISSN 0927-0256.
\newblock \doi{10.1016/j.commatsci.2021.110899}.

\bibitem[Wang et~al.(2023{\natexlab{a}})Wang, Gong, Evans, Yan, Wang, Miao, Zheng, Rignanese, and Wang]{WangElectrides2023}
Z.~Wang, Y.~Gong, M.~L. Evans, Y.~Yan, S.~Wang, N.~Miao, R.~Zheng, G.-M. Rignanese, and J.~Wang.
\newblock Machine {{Learning-Accelerated Discovery}} of {{A}}{\textsubscript{2}}{{BC}}{\textsubscript{2}} {{Ternary Electrides}} with {{Diverse Anionic Electron Densities}}.
\newblock \emph{Journal of the American Chemical Society}, 145\penalty0 (48):\penalty0 26412--26424, Dec. 2023{\natexlab{a}}.
\newblock ISSN 0002-7863.
\newblock \doi{10.1021/jacs.3c10538}.

\bibitem[Wang et~al.(2023{\natexlab{b}})Wang, Gong, Evans, Yan, Wang, Miao, Zheng, Rignanese, and Wang]{wangElectridesMCA2023}
Z.~Wang, Y.~Gong, M.~L. Evans, Y.~Yan, S.~Wang, N.~Miao, R.~Zheng, G.-M. Rignanese, and J.~Wang.
\newblock Machine learning-accelerated discovery of {{A}}{$_{2}$}{{BC}}{$_2$} ternary electrides with diverse anionic electron densities.
\newblock \emph{{Materials Cloud Archive}}, {2023.181}:\penalty0 26412--26424, 2023{\natexlab{b}}.
\newblock \doi{10.24435/materialscloud:c8-gy}.

\bibitem[Weininger(1988)]{Weininger1988}
D.~Weininger.
\newblock {{SMILES}}, a chemical language and information system. 1. {{Introduction}} to methodology and encoding rules.
\newblock \emph{Journal of Chemical Information and Computer Sciences}, 28\penalty0 (1):\penalty0 31--36, Feb. 1988.
\newblock \doi{10.1021/ci00057a005}.

\bibitem[Wines et~al.(2023)Wines, Gurunathan, Garrity, DeCost, Biacchi, Tavazza, and Choudhary]{winesreview}
D.~Wines, R.~Gurunathan, K.~F. Garrity, B.~DeCost, A.~J. Biacchi, F.~Tavazza, and K.~Choudhary.
\newblock {Recent progress in the JARVIS infrastructure for next-generation data-driven materials design}.
\newblock \emph{Applied Physics Reviews}, 10\penalty0 (4):\penalty0 041302, 10 2023.
\newblock \doi{10.1063/5.0159299}.

\bibitem[Xie and Grossman(2018)]{Xie2017}
T.~Xie and J.~C. Grossman.
\newblock Crystal {{Graph Convolutional Neural Networks}} for {{Accurate}} and {{Interpretable Prediction}} of {{Material Properties}}.
\newblock \emph{Physical Review Letters}, 120\penalty0 (14):\penalty0 145301, 2018.
\newblock \doi{10.1103/PhysRevLett.120.145301}.

\bibitem[Yakutovich et~al.(2021)Yakutovich, Eimre, Sch{\"u}tt, Talirz, Adorf, Andersen, Ditler, Du, Passerone, Smit, Marzari, Pizzi, and Pignedoli]{aiidalab}
A.~V. Yakutovich, K.~Eimre, O.~Sch{\"u}tt, L.~Talirz, C.~S. Adorf, C.~W. Andersen, E.~Ditler, D.~Du, D.~Passerone, B.~Smit, N.~Marzari, G.~Pizzi, and C.~A. Pignedoli.
\newblock {{AiiDAlab}} \textendash{} an ecosystem for developing, executing, and sharing scientific workflows.
\newblock \emph{Computational Materials Science}, 188:\penalty0 110165, 2021.
\newblock ISSN 0927-0256.
\newblock \doi{10.1016/j.commatsci.2020.110165}.

\bibitem[Yang et~al.(2018{\natexlab{a}})Yang, Wang, and et~al.]{MatCloud2018}
X.~Yang, Z.~Wang, and J.~S. et~al.
\newblock A high-throughput computational materials infrastructure: Present, future visions and challenges.
\newblock \emph{Chin. Phys. B}, 27\penalty0 (11):\penalty0 110301--110301, 2018{\natexlab{a}}.
\newblock \doi{10.1088/1674-1056/27/11/110301}.

\bibitem[Yang et~al.(2018{\natexlab{b}})Yang, Wang, and et~al.]{Yang2018}
X.~Yang, Z.~Wang, and J.~S. et~al.
\newblock Matcloud: A high-throughput computational infrastructure for integrated management of materials simulation, data and resources.
\newblock \emph{Computational Materials Science}, 146:\penalty0 319--333, 2018{\natexlab{b}}.
\newblock \doi{10.1016/j.commatsci.2018.01.039}.

\bibitem[Ye et~al.(2022)Ye, Lei, Aykol, and Montoya]{Ye2022}
W.~Ye, X.~Lei, M.~Aykol, and J.~H. Montoya.
\newblock Novel inorganic crystal structures predicted using autonomous simulation agents.
\newblock \emph{Scientific Data}, 9\penalty0 (1):\penalty0 302, 2022.
\newblock \doi{10.1038/s41597-022-01438-8}.

\bibitem[Zivanovic et~al.(2020)Zivanovic, Bayarri, Colizzi, Moreno, Gelpi, Soliva, Hospital, and Orozco]{Zivanovic2020}
S.~Zivanovic, G.~Bayarri, F.~Colizzi, D.~Moreno, J.~L. Gelpi, R.~Soliva, A.~Hospital, and M.~Orozco.
\newblock Bioactive conformational ensemble server and database. a public framework to speed up in silico drug discovery.
\newblock \emph{Journal of Chemical Theory and Computation}, 16:\penalty0 6586--6597, 2020.
\newblock \doi{10.1021/acs.jctc.0c00305}.

\end{thebibliography}

\end{document}